\documentclass[twocolumn,times]{aastex62}

\hypersetup{linkcolor=teal,citecolor=teal,filecolor=teal,urlcolor=teal}

\newcommand{\ciimu}{\lbrack C\ensuremath{\,\textsc{ii}}\rbrack\,158\,$\mu$m}
\newcommand{\lya}{Ly\ensuremath{\alpha}}
\newcommand{\hi}{H\ensuremath{\,\textsc{i}}}
\newcommand{\mgii}{Mg\ensuremath{\,\textsc{ii}}}

\newcommand{\civ}{C\ensuremath{\,\textsc{iv}}}
\newcommand{\ha}{H\ensuremath{\alpha}}
\newcommand{\nv}{N\ensuremath{\,\textsc{v}}}
\newcommand{\heii}{He\ensuremath{\,\textsc{ii}}}

\def\lsim{\mathrel{\rlap{\lower 3pt \hbox{$\sim$}} \raise 2.0pt \hbox{$<$}}}
\def\gsim{\mathrel{\rlap{\lower 3pt \hbox{$\sim$}} \raise 2.0pt \hbox{$>$}}}

\def\msunyr{{\rm M}_\odot\,{\rm yr}^{-1}}
\def\kms{{\rm km}\,{\rm s}^{-1}}

\def\llya{{\rm L}({\rm Ly}\alpha)}

\def\M1450{{\rm M}_{\rm 1450}}
\def\REQUIEM{\textit{REQUIEM}}

\usepackage{graphicx}
\usepackage{natbib}
\usepackage{relsize}
\usepackage{textcomp}
\usepackage{upgreek}
\usepackage{amsmath}
\usepackage{amssymb}
\usepackage{calrsfs}
\usepackage{afterpage}
\usepackage{float}
\usepackage{makecell}

\setcitestyle{notesep={}}

\shorttitle{The \textit{REQUIEM} Survey I}

\shortauthors{Farina et al.}

\begin{document}

\title{The \textit{REQUIEM} Survey I: A Search for Extended Ly--Alpha Nebular Emission Around 31 $z>5.7$ Quasars.}

\author[0000-0002-6822-2254]{Emanuele Paolo Farina}
\affiliation{Max Planck Institut f\"ur Astronomie, K\"onigstuhl 17, D-69117, Heidelberg, Germany}
\affiliation{Max Planck Institut f\"ur Astrophysik, Karl--Schwarzschild--Stra{\ss}e 1, D-85748, Garching bei M\"unchen, Germany}
\email{emanuele.paolo.farina@gmail.com}

\author[0000-0002-4770-6137]{Fabrizio Arrigoni--Battaia}
\affiliation{Max Planck Institut f\"ur Astrophysik, Karl--Schwarzschild--Stra{\ss}e 1, D-85748, Garching bei M\"unchen, Germany}

\author{Tiago Costa}
\affiliation{Max Planck Institut f\"ur Astrophysik, Karl--Schwarzschild--Stra{\ss}e 1, D-85748, Garching bei M\"unchen, Germany}

\author[0000-0003-4793-7880]{Fabian Walter}
\affiliation{Max Planck Institut f\"ur Astronomie, K\"onigstuhl 17, D-69117, Heidelberg, Germany}

\author[0000-0002-7054-4332]{Joseph F.\ Hennawi}
\affiliation{Max Planck Institut f\"ur Astronomie, K\"onigstuhl 17, D-69117, Heidelberg, Germany}
\affiliation{Department of Physics, University of California, Santa Barbara, CA 93106-9530, USA}

\author[0000-0002-0174-3362]{Alyssa B.\ Drake}
\affiliation{Max Planck Institut f\"ur Astronomie, K\"onigstuhl 17, D-69117, Heidelberg, Germany}

\author[0000-0002-2662-8803]{Roberto Decarli}
\affiliation{INAF --- Osservatorio di Astrofisica e Scienza dello Spazio di Bologna, via Gobetti 93/3, I-40129, Bologna, Italy}

\author[0000-0001-6179-7701]{Thales A.\ Gutcke}
\affiliation{Max Planck Institut f\"ur Astrophysik, Karl--Schwarzschild--Stra{\ss}e 1, D-85748, Garching bei M\"unchen, Germany}

\author[0000-0002-5941-5214]{Chiara Mazzucchelli}
\affiliation{European Southern Observatory, Alonso de C\'ordova 3107, Vitacura, Regi\'on Metropolitana, Chile}

\author{Marcel Neeleman}
\affiliation{Max Planck Institut f\"ur Astronomie, K\"onigstuhl 17, D-69117, Heidelberg, Germany}

\author[0000-0001-8471-6679]{Iskren Georgiev}
\affiliation{Max Planck Institut f\"ur Astronomie, K\"onigstuhl 17, D-69117, Heidelberg, Germany}

\author[0000-0003-2895-6218]{Anna--Christina Eilers}
\affiliation{Max Planck Institut f\"ur Astronomie, K\"onigstuhl 17, D-69117, Heidelberg, Germany}

\author[0000-0003-0821-3644]{Frederick B.\ Davies}
\affiliation{Department of Physics, University of California, Santa Barbara, CA 93106-9530, USA}

\author[0000-0002-2931-7824]{Eduardo Ba\~nados}
\affiliation{Max Planck Institut f\"ur Astronomie, K\"onigstuhl 17, D-69117, Heidelberg, Germany}

\author[0000-0003-3310-0131]{Xiaohui Fan}
\affiliation{Steward Observatory, University of Arizona, 933 N Cherry Ave, Tucson, AZ 85719, USA}

\author{Masafusa Onoue}
\affiliation{Max Planck Institut f\"ur Astronomie, K\"onigstuhl 17, D-69117, Heidelberg, Germany}

\author[0000-0002-4544-8242]{Jan--Torge Schindler}
\affiliation{Max Planck Institut f\"ur Astronomie, K\"onigstuhl 17, D-69117, Heidelberg, Germany}

\author[0000-0001-9024-8322]{Bram P.\ Venemans}
\affiliation{Max Planck Institut f\"ur Astronomie, K\"onigstuhl 17, D-69117, Heidelberg, Germany}

\author[0000-0002-7633-431X]{Feige Wang}
\affiliation{Department of Physics, University of California, Santa Barbara, CA 93106-9530, USA}

\author{Jinyi Yang}
\affiliation{Steward Observatory, University of Arizona, 933 N Cherry Ave, Tucson, AZ 85719, USA}

\author{Sebastian Rabien}
\affiliation{Max Planck Institut f\"ur Extraterrestrische Physik Gie{\ss}enbachstra{\ss}e 1, D-85748, Garching bei M\"unchen, Germany}

\author{Lorenzo Busoni}
\affiliation{INAF --- Osservatorio Astronomico di Arcetri, Largo Enrico Fermi 5, I-50125, Firenze, Italy}

\begin{abstract}
The discovery of quasars few hundred megayears after the Big Bang represents a major challenge to our understanding of black holes and galaxy formation and evolution.
Their luminosity is produced by extreme gas accretion onto black holes, which already reached masses of $M_{\rm BH}>10^9$\,M$_{\odot}$ by $z\sim6$.
Simultaneously, their host galaxies form hundreds of stars per year, using up gas in the process.
To understand which environments are able to sustain the rapid formation of these extreme sources we started a \textit{VLT}/MUSE effort aimed at characterizing the surroundings of a sample of $5.7<z<6.6$ quasars dubbed:
the Reionization Epoch QUasar InvEstigation with MUSE (\REQUIEM) survey.
We here present results of our searches for extended \lya\ halos around the first 31 targets observed as part of this program.
Reaching 5--$\sigma$ surface brightness limits of $0.1-1.1\times10^{-17}$erg\,s$^{-1}$\,cm$^{-2}$\,arcsec$^{-2}$ over a 1\,arcsec$^2$ aperture, we were able to unveil the presence of 12 \lya\ nebulae, 8 of which are newly discovered.
The detected nebulae show a variety of emission properties and morphologies with luminosities ranging from $8\times10^{42}$ to $2\times10^{44}$\,erg\,s$^{-1}$, FWHMs between 300 and 1700\,km\,s$^{-1}$, sizes $<30$\,pkpc, and redshifts consistent with those of the quasar host galaxies.
As the first statistical and homogeneous investigation of the circum--galactic medium of massive galaxies at the end of the reionization epoch, the \REQUIEM\ survey enables the study of the evolution of the cool gas surrounding quasars in the first 3\,Gyr of the Universe.
A comparison with the extended \lya\ emission observed around bright ($\M1450\lesssim-25$\,mag) quasars at intermediate redshift indicates little variations on the properties of the cool gas from $z\sim6$ to $z\sim3$ followed by a decline in the average surface brightness down to $z\sim2$.
\end{abstract}


\section{INTRODUCTION}\label{sec:introduction}

\textit{Where do the first quasars form?}
Two decades after the discovery of the first quasar at $z>6$ \citep[i.e., J1030$+$0524 at $z=6.3$,][]{Fan2001}, this question still puzzles astronomers.
Assuming a simple model where a massive black hole grows at the Eddington limit starting at a certain time $t_0$ from a seed with mass $M_{\rm BH}(t_0)=M_{\rm seed}$, the evolution of the mass with time can be expressed as:
\begin{equation}\label{eq:seed}
M_{\rm BH}(t) = M_{\rm seed}\times \exp{\left[f_{\rm Duty}(1-\eta)\frac{t-t_0}{t_{\rm Sal}}\right]}
\end{equation}
where $f_{\rm Duty}$ is the duty cycle and $\eta$ is the fraction of rest mass energy released during the accretion.
The time scale of the mass growth is set by the Salpeter time \citep[][]{Salpeter1964}: $t_{\rm Sal}=\epsilon\sigma_{\rm T}\,c\,/\,\left(4\pi\,G\,m_{\rm p}\right)=\epsilon\,450\,{\rm Myr}$, where $\sigma_{\rm T}$ is the Thomson cross--section, $m_{\rm p}$ is the proton mass, and $\epsilon$ is the radiation efficiency\footnote{The presence of helium, with a mass of $\sim4\times m_{\rm p}$ and 2 free electrons, allows a faster growth of the black holes. Considering a plasma with abundances $X=0.75$ for hydrogen and $Y=0.25$ for helium, the Salpeter time becomes $t_{\rm Sal}=\epsilon\,390\,{\rm Myr}$}.
In standard radiatively efficient accretion disks, all the energy is radiated away and it is typically assumed that $\epsilon=\eta=0.1$ \citep{Soltan1982, Tanaka2009, Davis2011, Davies2019soltan}.
\autoref{eq:seed} implies that, for instance, a $10^{2}$\,M$_\odot$ remnant of a PoP~III star at $z=30$ needs to accrete at the Eddington limit for its entire life ($f_{\rm Duty}=1$) to reach a black hole mass $>10^{9}$\,M$_\odot$ at $z\sim6$, as observed in quasars \citep[e.g.,][]{Mortlock2011, Derosa2011, Derosa2014, Wu2015, Mazzucchelli2017, Banados2018, Shen2019, Reed2019, Pons2019}.
In addition, investigations at mm and sub--mm wavelengths revealed that also the host--galaxies of these first quasars are vigorously growing mass, with star formation rates ${\rm SFR}\gg100\,\msunyr$ \citep[e.g.,][]{Walter2009, Wang2013, Willott2015, Willott2017, Decarli2018, Venemans2012, Venemans2016, Venemans2018, Kim2019, Shao2019, Wang2019J0100, Yang2018Lens}.

To comprehend how these first quasars form and grow it is important to understand where they are hosted.
\citet{Efstathiou1988} first proposed that, in the current $\Lambda$CDM paradigm of galaxy formation \citep[e.g.,][]{White1978}, only rare high peaks in the density field contain enough gas to build--up the black hole and star mass (taking into account mass losses due to supernova--driven winds) of high--redshift quasars.
This scenario is supported by cosmological hydrodynamic simulations \citep[e.g.,][]{Sijacki2009, Costa2014} and analytical arguments \citep[e.g.,][]{Volonteri2006} showing that only the small fraction of black holes that, by $z\sim6$, are hosted by $\gtrsim10^{12}$\,M$_\odot$ dark matter halos can grow efficiently into a population of quasars with masses and accretion rates matching current observational constraints \citep[but see discussion in][]{Fanidakis2013}.
To compensate for the rapid gas consumption, the host--galaxies need a continuous replenishment of fresh fuel provided by filamentary streams of $T=10^4-10^5$\,K pristine gas from the intergalactic medium (IGM) and/or by mergers with gas rich halos \citep[e.g.,][]{Keres2005, Keres2009a, Yoo2004, Volonteri2005, Volonteri2010, Volonteri2012, Li2007, Keres2009a, Dekel2006, Dekel2009, Fumagalli2011, vandeVoort2012, DiMatteo2012, Habouzit2018, Mayer2019}.
Observational validations of this framework can be set by the detection of gas reservoirs and satellites in the so--called circum--galactic medium (CGM, empirically defined as the regions within a few hundreds of kiloparsecs from a galaxy) of high--redshift quasars.

Historically, information on the CGM has been provided by absorption signatures imprinted on background sightlines.
This revealed the presence of halos of cool and enriched gas extending to $\sim200$\,pkpc from high--redshift galaxies \citep[e.g.,][]{Steidel1994, Bahcall1969, Chen2008, Chen2010a, Chen2010b, Gauthier2010, Nielsen2013a, Nielsen2013b, Churchill2013, Werk2016, Tumlinson2017}.
In particular, this technique applied on close projected quasar pairs revealed that  intermediate redshift quasars are surrounded by massive ($>10^{10}$\,M$_\odot$), metal rich ($Z\gtrsim0.1 Z_\odot$), and cool ($T\sim10^{4}$\,K) gas reservoirs \citep[e.g.,][]{Bowen2006, Hennawi2006, Hennawi2007, Decarli2009, Prochaska2009, Prochaska2013a, Prochaska2013b, Farina2013, Farina2014, Johnson2015, Lau2016, Lau2018}.
However, the rapid drop in the number density of bright background sources with redshift, make the absorption studies to lose effectiveness at $z\gtrsim4$.

A promising way to push investigation of the CGM of quasars up to the epoch of reionization is to probe the cool gas in emission.
The strong flux of UV photons radiating from the AGN can be reprocessed in the hydrogen \lya\ line at 1215.7\,\AA\ \citep{Lyman1906, Millikan1920} by the surrounding gas, giving rise to an extended ``fuzz'' of fluorescent \lya\ emission \citep[e.g.,][]{Rees1988, Haiman2001CGM, Alam2002}.
Several pioneering efforts have been performed to reveal such halos in the vicinity of $z\sim2-4$ quasars \citep[e.g.,][]{Heckman1991a, Heckman1991b, Christensen2006, North2012, Hennawi2013, Roche2014, Herenz2015, Arrigoni2016, Arrigoni2019}.
This led to the general consensus that $10-50$\,kpc nebulae are (almost) ubiquitous around intermediate redshift quasars, and that a few objects (typically associated with galaxy overdensities) are surrounded by giant \lya\ nebulae with sizes $>300$\,kpc, i.e. larger than the expected virial radius for such systems \citep[e.g.,][]{Cantalupo2014, Martin2014, Hennawi2015, Cai2017}.

A change of gear in these searches was driven by the recent development of the new generation of sensitive integral field spectrographs (IFS) on 10--m class telescopes, i.e. the Multi--Unit Spectroscopic Explorer \citep[MUSE;][]{Bacon2010} on the ESO/\textit{VLT} and the \textit{Keck} Cosmic Web Imager \citep[KCWI;][]{Morrissey2012, Morrissey2018} on the \textit{Keck\,II} telescope.
These instruments have been successfully exploited to map the diffuse gas in the CGM of hundreds of intermediate redshift galaxies \citep[e.g.,][]{Wisotzki2016, Leclercq2017, Erb2018} and quasars \citep[e.g.,][]{Martin2014, Husband2015, Borisova2016, Fumagalli2016, Arrigoni2018, Arrigoni2019, Arrigoni2019Filament, Ginolfi2018, Lusso2019, Cai2019}.
The picture emerging is that the cool gas around  $z\sim2-4$ radio--quiet quasars has a quiescent kinematics and it is likely to be constituted by a population of compact (with sizes of $\lesssim50$\,pc) dense ($n_{\rm H}\gtrsim1$\,cm$^{-3}$) clouds that are optically thin to the quasar radiation \citep[e.g.,][]{Hennawi2013, Hennawi2015, Arrigoni2015HeII, Cantalupo2017, Cantalupo2019}.

However, by $z=4$ the Universe is already 1.5\,Gyr old and a population of massive, quiescent galaxies is already in place \citep[e.g.,][]{Straatman2014, Straatman2016}.
To probe the first stages of galaxy formation it is thus necessary to push these studies to $z\gtrsim6$.
To date, extended \lya\ halos have been reported only for a handful of $z\sim6$ quasars exploiting different techniques: narrow--band imaging \citep[][]{Goto2009, Decarli2012, Momose2018}, long--slit spectroscopy \citep[][]{Willott2011, Goto2012, Roche2014}, and IFS \citep[][]{Farina2017, Drake2019}.
This small sample showed that the first quasars can be surrounded by extended nebulae with luminosities up to $\llya\sim10^{44}$\,erg\,s$^{-1}$ and sizes $\lesssim40$\,pkpc.
However, a detailed interpretation of these results is hampered by the small number statistic and by the heterogeneity of the data.

To overcome these limitations, we started the Reionization Epoch QUasar InvEstigation with MUSE (\REQUIEM) survey aimed at performing a statistical and homogeneous census of the close environment of the first quasars.
In this Paper, we report results from the investigation of the first 31 $5.7<z<6.6$ quasars part of this ongoing program (including the re--analysis of MUSE data from \citealt{Farina2017} and \citealt{Drake2019}), focusing our attention on the properties of the extended \lya\ halos as tracer of the gas reservoirs able to fuel the activity of the first quasars.
We defer the analysis of the close galactic environment of these systems to a future paper.

To summarize, the analysis of the MUSE observations (see \autoref{sec:observations}) of the 31~targets presented in \autoref{sec:sample} with the procedure described in \autoref{sec:psfsubtraction} led to the discovery of 12 extended \lya\ nebulae above a surface brightness limit of ${\rm SB}_{\rm Ly\alpha}\sim{\rm few}\times10^{18}$\,erg\,s$^{-1}$\,cm$^{-2}$\,arcsec$^{-2}$ ($\sim40\%$ of the cases, see \autoref{sec:discussion}).
In \autoref{sec:discussion} we report on the attributes of the detected halos, we compare them with the properties of the quasar host galaxies and of the central supermassive black holes, and we test for possible signatures of CGM evolution down to $z\sim3$.
Finally, a summary is given in \autoref{sec:conclusions}.

Throughout this paper we assume a concordance cos\-mo\-lo\-gy with H$_0$=70\,km\,s$^{-1}$\,Mpc$^{-1}$, $\Omega_{\rm M}$=0.3, and \mbox{$\Omega_\Lambda$=1-$\Omega_{\rm M}$=0.7}. 
In this cosmology, at $z$=6.2 (the average redshift of our sample) the Universe is 0.877\,Gyr old, and an angular scale of $\theta$$=$1\arcsec\ corresponds a proper transverse separation of~5.6\,kpc.
We remind the reader that MUSE is able to cover the \lya\ line up to redshift $z\sim6.6$ at a spectral resolution of $R=\lambda/\Delta\lambda\sim3500$ at $\lambda\sim9000$\,\AA\ with a spatial sampling of $0\farcs2\times0\farcs2$ (corresponding to 1.1\,pkpc$\times$1.1\,pkpc at $z=6$) over a $\sim1$\,arcmin$^2$ field--of--view.

\section{SAMPLE SELECTION}\label{sec:sample}

\begin{figure}[tb]
\begin{center}
\includegraphics[width=0.98\columnwidth]{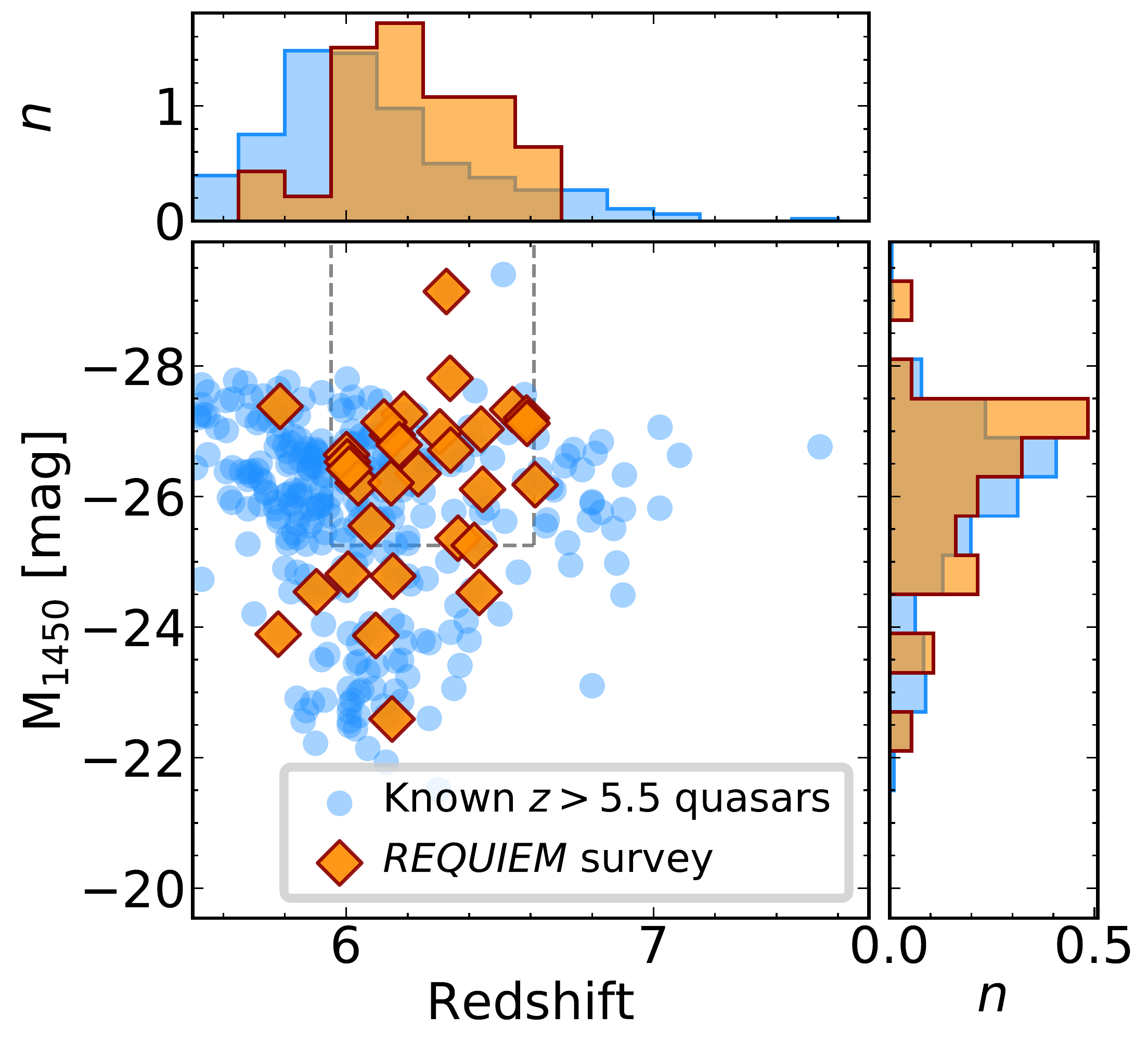}
\caption{
Distribution of all $z>5.5$ quasars known to date in the redshift vs. absolute magnitude plane at 1450\,\AA\  (light blue circles and histograms).
Orange diamonds and histograms mark targets from our survey.
Histograms are normalized by the total number of targets and by the bin size (with steps $0.15$ in redshift and of $0.6$\,mag in absolute magnitude).
The limits in luminosity and redshift of our \textit{core sample} (see \autoref{sec:sample}) are plotted as gray dashed lines.
The five quasars outside these boundaries are: J0129$-$0035 at $z=5.78$, J1044$-$0125 at $z=5.78$, J2228$+$0110 at $z=5.90$, J0055$+$0146 at $z=6.01$, J2216$-$0016 at $z=6.10$, J2219$+$0102 at $z=6.15$, J2229$+$1457 at $z=6.15$, and J2318$-$3113 at $z=6.44$.
A 2D Kolmogorov--Smirnov test \citep[][]{Fasano1987} performed with bootstrap re--sampling of the parent dataset of the $5.95<z<6.62$ and $\M1450<-25.25$\,mag quasars does not refute the null hypothesis that our \textit{core sample} has a different distribution as the parent dataset distribution ($p-{\rm value}\gtrsim0.2$).
}\label{fig:sample}
\end{center}
\end{figure}

Our sample consists of 31~quasars in the redshift range $5.77<z<6.62$ located in the southern sky \citep[e.g.,][]{Fan2001, Fan2003, Fan2006, Willott2007, Willott2010, Venemans2013, Jiang2016, Banados2016, Reed2017, Mazzucchelli2017, Matsuoka2018, Wang2018, Yang2019Sample}.
This includes all available MUSE observations of $z>5.7$ quasars present in the ESO Archive at the time of writing (Aug.~2019).
These quasars have an average redshift of $\langle z \rangle=6.22$ and an average absolute magnitude of $\langle \M1450 \rangle=-26.9$\,mag (see \autoref{tab:sample} and \autoref{fig:sample}).
Among these, only J2228$+$0110 is a confirmed radio--loud quasar \citep[considering radio--loud quasars as having $R$$=$$f_{\nu,\,5{\rm GHz}}$$/$$f_{\nu,\,4400{\rm Ang.}}$$>$$10$,][, in prep.]{Kellermann1989, Banados2015radio}.

In the following we will refer to the entire dataset as our \textit{full sample}, and to the subset of 23 quasars with $\M1450<-25.25$\,mag and $5.95<z<6.62$ as our \textit{core sample}.
This well defined sub--sample is highly representative of the \mbox{high--$z$} population of luminous quasars (see \autoref{fig:sample}) and largely overlaps with the survey of dust continuum and \ciimu\ fine--structure emission lines in $z>6$ quasar host--galaxies using the Atacama Large Millimeter Array (ALMA) presented in \citet{Decarli2017, Decarli2018} and \citet{Venemans2018}.

\begin{deluxetable*}{lcccccccc}[tbp]
\tablecaption{Quasars observed with MUSE in decreasing redshift order\label{tab:sample}}
\tablecolumns{9}
\tablenum{1}
\tablewidth{0pt}
\tablehead{
\colhead{ID}   & \colhead{RA}      & \colhead{Dec.}    & \colhead{Redshift} & \colhead{M$_{\rm1450}$} & \colhead{Prog. ID.}  & \colhead{Exp. Time} & \colhead{Image Quality} & \colhead{SB$_{5\,\sigma}^{1}$} \\
\colhead{}     & \colhead{(J2000)} & \colhead{(J2000)} & \colhead{}         & \colhead{(mag)}         &                      & \colhead{(sec.)}    & \colhead{(\arcsec)}     & \colhead{(erg\,s$^{-1}$\,cm$^{-2}$\,arcsec$^{-2}$)}
}
\startdata
J0305$-$3150 & 03:05:16.916 & $-$31:50:55.90 & 6.6145$\pm$0.0001\tablenotemark{a}       & $-$26.12 &     \phm{0}094.B-0893 & \phm{1}8640. & 0.53 & 0.29$\times$10$^{-17}$ \\
P323$+$12    & 21:32:33.191 & $+$12:17:55.26 & 6.5881$\pm$0.0003\tablenotemark{b}       & $-$27.06 &           0101.A-0656 & \phm{1}2964. & 0.85 & 0.48$\times$10$^{-17}$ \\
P231$-$20    & 15:26:37.841 & $-$20:50:00.66 & 6.5864$\pm$0.0005\phm{\tablenotemark{o}} & $-$27.14 &     \phm{0}099.A-0682 &       11856. & 0.63 & 0.30$\times$10$^{-17}$ \\
P036$+$03    & 02:26:01.876 & $+$03:02:59.39 & 6.5412$\pm$0.0018\tablenotemark{c}       & $-$27.28 &           0101.A-0656 & \phm{1}2964. & 0.61 & 0.33$\times$10$^{-17}$ \\
J2318$-$3113 & 23:18:18.351 & $-$31:13:46.35 & 6.4435$\pm$0.0004\phm{\tablenotemark{o}} & $-$26.06 &           0101.A-0656 & \phm{1}2964. & 0.65 & 0.54$\times$10$^{-17}$ \\
P183$+$05    & 12:12:26.981 & $+$05:05:33.49 & 6.4386$\pm$0.0004\phm{\tablenotemark{o}} & $-$26.98 &     \phm{0}099.A-0682 & \phm{1}2964. & 0.62 & 0.92$\times$10$^{-17}$ \\
J0210$-$0456 & 02:10:13.190 & $-$04:56:20.90 & 6.4323$\pm$0.0005\tablenotemark{d}       & $-$24.47 &           0103.A-0562 & \phm{1}2964. & 1.24 & 0.26$\times$10$^{-17}$ \\
J2329$-$0301 & 23:29:08.275 & $-$03:01:58.80 & 6.4164$\pm$0.0008\tablenotemark{e}       & $-$25.19 &     \phm{01}60.A-9321 & \phm{1}7170. & 0.65 & 0.18$\times$10$^{-17}$ \\
J1152$+$0055 & 11:52:21.269 & $+$00:55:36.69 & 6.3643$\pm$0.0005\phm{\tablenotemark{f}} & $-$25.30 &           0103.A-0562 & \phm{1}2964. & 1.18 & 0.90$\times$10$^{-17}$ \\
J2211$-$3206 & 22:11:12.391 & $-$32:06:12.94 & 6.3394$\pm$0.0010\phm{\tablenotemark{o}} & $-$26.66 &           0101.A-0656 & \phm{1}2964. & 0.73 & 1.49$\times$10$^{-17}$ \\
J0142$-$3327 & 01:42:43.727 & $-$33:27:45.47 & 6.3379$\pm$0.0004\phm{\tablenotemark{o}} & $-$27.76 &           0101.A-0656 & \phm{1}2964. & 0.71 & 1.11$\times$10$^{-17}$ \\
J0100$+$2802 & 01:00:13.027 & $+$28:02:25.84 & 6.3258$\pm$0.0010\tablenotemark{g}       & $-$29.09 &           0101.A-0656 & \phm{1}2964. & 1.29 & 1.13$\times$10$^{-17}$ \\
J1030$+$0524 & 10:30:27.098 & $+$05:24:55.00 & 6.3000$\pm$0.0002\tablenotemark{h}       & $-$26.93 &     \phm{0}095.A-0714 &       23152. & 0.51 & 0.08$\times$10$^{-17}$ \\
P308$-$21    & 20:32:09.996 & $-$21:14:02.31 & 6.2341$\pm$0.0005\phm{\tablenotemark{o}} & $-$26.29 &     \phm{0}099.A-0682 &       17784. & 0.77 & 0.26$\times$10$^{-17}$ \\
P065$-$26    & 04:21:38.052 & $-$26:57:15.60 & 6.1877$\pm$0.0005\phm{\tablenotemark{o}} & $-$27.21 &           0101.A-0656 & \phm{1}2964. & 0.68 & 0.25$\times$10$^{-17}$ \\
P359$-$06    & 23:56:32.455 & $-$06:22:59.26 & 6.1722$\pm$0.0004\phm{\tablenotemark{o}} & $-$26.74 &           0101.A-0656 & \phm{1}2964. & 0.58 & 0.28$\times$10$^{-17}$ \\
J2229$+$1457 & 22:29:01.649 & $+$14:57:08.99 & 6.1517$\pm$0.0005\tablenotemark{i}       & $-$24.72 &           0103.A-0562 & \phm{1}2964. & 0.54 & 0.27$\times$10$^{-17}$ \\
P217$-$16    & 14:28:21.394 & $-$16:02:43.29 & 6.1498$\pm$0.0011\phm{\tablenotemark{o}} & $-$26.89 &           0101.A-0656 & \phm{1}2964. & 0.90 & 0.32$\times$10$^{-17}$ \\
J2219$+$0102 & 22:19:17.217 & $+$01:02:48.90 & 6.1492$\pm$0.0002\tablenotemark{e}       & $-$22.54 &           0103.A-0562 & \phm{1}2964. & 0.69 & 0.48$\times$10$^{-17}$ \\
J2318$-$3029 & 23:18:33.100 & $-$30:29:33.37 & 6.1458$\pm$0.0005\phm{\tablenotemark{o}} & $-$26.16 &           0101.A-0656 & \phm{1}2964. & 0.73 & 0.30$\times$10$^{-17}$ \\
J1509$-$1749 & 15:09:41.778 & $-$17:49:26.80 & 6.1225$\pm$0.0007\phm{\tablenotemark{o}} & $-$27.09 &           0101.A-0656 & \phm{1}2964. & 0.88 & 0.46$\times$10$^{-17}$ \\
J2216$-$0016 & 22:16:44.473 & $-$00:16:50.10 & 6.0962$\pm$0.0003\tablenotemark{l}       & $-$23.82 &           0103.A-0562 & \phm{1}2964. & 1.12 & 0.48$\times$10$^{-17}$ \\
J2100$-$1715 & 21:00:54.616 & $-$17:15:22.50 & 6.0812$\pm$0.0005\phm{\tablenotemark{o}} & $-$25.50 &     \phm{0}297.A-5054 &       13338. & 0.67 & 0.23$\times$10$^{-17}$ \\
J2054$-$0005 & 20:54:06.481 & $-$00:05:14.80 & 6.0391$\pm$0.0001\tablenotemark{m}       & $-$26.15 &           0101.A-0656 & \phm{1}3869. & 0.81 & 0.24$\times$10$^{-17}$ \\
P340$-$18    & 22:40:48.997 & $-$18:39:43.81 & 6.01$\pm$0.05\tablenotemark{n}           & $-$26.36 &           0101.A-0656 & \phm{1}2964. & 0.55 & 0.26$\times$10$^{-17}$ \\
J0055$+$0146 & 00:55:02.910 & $+$01:46:18.30 & 6.0060$\pm$0.0008\tablenotemark{i}       & $-$24.76 &           0103.A-0562 & \phm{1}2964. & 0.75 & 0.27$\times$10$^{-17}$ \\
P009$-$10    & 00:38:56.522 & $-$10:25:53.90 & 6.0039$\pm$0.0004\phm{\tablenotemark{o}} & $-$26.50 &           0101.A-0656 & \phm{1}2964. & 0.67 & 0.27$\times$10$^{-17}$ \\
P007$+$04    & 00:28:06.560 & $+$04:57:25.68 & 6.0008$\pm$0.0004\phm{\tablenotemark{o}} & $-$26.59 &           0101.A-0656 & \phm{1}2964. & 1.19 & 0.35$\times$10$^{-17}$ \\
J2228$+$0110 & 22:28:43.535 & $+$01:10:32.20 & 5.9030$\pm$0.0002\tablenotemark{o}       & $-$24.47 &     \phm{0}095.B-0419 &       40950. & 0.61 & 0.11$\times$10$^{-17}$ \\
J1044$-$0125 & 10:44:33.042 & $-$01:25:02.20 & 5.7847$\pm$0.0007\tablenotemark{l}       & $-$27.32 &           0103.A-0562 & \phm{1}2964. & 0.94 & 0.32$\times$10$^{-17}$ \\
J0129$-$0035 & 01:29:58.510 & $-$00:35:39.70 & 5.7787$\pm$0.0001\tablenotemark{m}       & $-$23.83 &           0103.A-0562 & \phm{1}2964. & 1.19 & 0.26$\times$10$^{-17}$ \\
\enddata
\tablerefs{Unless otherwise specified, we report systemic redshifts measured from the \ciimu\ emission lines by \citet{Decarli2018}.
\tablenotetext{a}{\footnotesize{\ciimu\ redshift from \citet{Venemans2013}.}}\vspace{-0.1cm}
\tablenotetext{b}{\footnotesize{\ciimu\ redshift from \citet{Mazzucchelli2017}.}}\vspace{-0.1cm}
\tablenotetext{c}{\footnotesize{\ciimu\ redshift from \citet{Banados2015}.}}\vspace{-0.1cm}
\tablenotetext{d}{\footnotesize{\ciimu\ redshift from \citet{Willott2013}.}}\vspace{-0.1cm}
\tablenotetext{e}{\footnotesize{\ciimu\ redshift from \citet{Willott2017}.}}\vspace{-0.1cm}
\tablenotetext{f}{\footnotesize{\citet{Izumi2018} report a slightly different (but consistent within the error) \ciimu\ redshift for J1152$+$0055: $z=6.3637\pm0.0005$.}}\vspace{-0.1cm}
\tablenotetext{g}{\footnotesize{\ciimu\ redshift from \citet{Wang2016}.}}\vspace{-0.1cm}
\tablenotetext{h}{\footnotesize{Redshift derived by \citet{Derosa2011} from the fit of the \mgii\ broad emission line.}}\vspace{-0.1cm}
\tablenotetext{i}{\footnotesize{\ciimu\ redshift from \citet{Willott2015}.}}\vspace{-0.1cm}
\tablenotetext{l}{\footnotesize{\ciimu\ redshift from \citet{Izumi2018}.}}\vspace{-0.1cm}
\tablenotetext{m}{\footnotesize{\ciimu\ redshift from \citet{Wang2013}.}}\vspace{-0.1cm}
\tablenotetext{n}{\footnotesize{The \ciimu\ emission of P340$-$18 was not detected in 8\,minutes ALMA integration by \citet{Decarli2018}. We report the redshift inferred from the observed optical spectrum by \citet{Banados2016}.}}\vspace{-0.1cm}
\tablenotetext{o}{\footnotesize{Redshift derived by \citet{Roche2014} from the measurement of the \lya\ line.}}
}
\tablecomments{
Seeing and 5--$\sigma$ surface brightness limits have been estimated on pseudo--narrow--band images obtained collapsing 5 wavelength channels (for a total of 6.25\,\AA) at the expected location of the \lya\ emission of the quasars.}
\end{deluxetable*}

\subsection{Notes on Individual Objects}\label{sec:notes}

\paragraph{J0305$-$3150} 
\citet{Farina2017} reported the presence of a faint nebular emission extending $\sim9$\,pkpc toward the southwest of the quasar.
In addition, the presence of a \lya\ emitter (LAE) at a projected separation of 12.5\,kpc suggests that J0305$-$3150 is tracing an overdensity of galaxies.
This hypothesis is corroborated by recent high--resolution \textit{ALMA} imaging that revealed the presence of three \ciimu\ emitters located within $\sim40$\,kpc and $\sim1000$\,$\kms$ from the quasar \citep{Venemans2019}.
These observations also showed the complex morphology of the host--galaxy, possibly due to interactions with nearby galaxies \citep{Venemans2019}.
\citet{Ota2018}, using deep narrow--band imaging obtained with the Subaru Telescope Suprime--Cam, reported an LAE number density comparable with the background.
However, the displacement between the location of the redshifted \lya\ emission and wavelengths with high response of the NB921 filter used \citep[see Fig.\,2 in][]{Ota2018} may have hindered the detection of galaxies associated with the quasar.

\paragraph{P231$-$20} 
\textit{ALMA} observations of this quasar revealed the presence of a massive \ciimu\ bright galaxy in its immediate vicinity \citep[with a projected separation of 13.8\,kpc and a velocity difference of 591\,$\kms$,][]{Decarli2017}.
A sensitive search for the rest--frame UV emission from this companion galaxy is presented in \citet[][]{Mazzucchelli2019}.
An additional weaker \ciimu\ emitter has been identified by \citet{Neeleman2019} 14\,kpc south--southeast of the quasar.
Deep MUSE observations already revealed the presence of a $\sim18$\,pkpc \lya\ nebular emission around this quasar \citep[][]{Drake2019}.

\paragraph{P183$+$05} 
For this quasar \citet{Banados2019} reported the presence of a proximate damped \lya\ absorption system (pDLA) located at $z$=6.40392 (1400\,km\,s$^{-1}$ away from the quasar host galaxy), making this system the highest redshift pDLA known to date.
It shows an \hi\ column density of N$_{\rm HI}$=10$^{20.77\pm0.25}$\,cm$^{-2}$ and relative chemical abundances typical of an high redshift low--mass galaxy.
The pDLA can act as a coronagraphs and, by blocking its light, it allows one to perform sensitive searches for extended emission associated to the background quasar \citep[e.g.,][]{Hennawi2009}.
The galaxy originating the pDLA is not detected as a \lya\ line \textit{down--the--barrel} in the MUSE quasar's spectrum (see \autoref{fig:allspec} in \autoref{app:allspec}).
However, it could be located at a larger impact parameter \citep[e.g.,][]{Neeleman2016, Neeleman2017, Neeleman2018, DOdorico2018}.
The possibility to detected the galaxy in the full MUSE datacube, both in emission and as a shadow against the extended background \lya\ halo, will be explored in a future paper of this series (Farina et al., in prep.).

\paragraph{J2329$-$0301} 
The \lya\ halo of this quasar has been the subject of several studies \citep[][]{Goto2009, Goto2012, Willott2011, Momose2018, Drake2019}.
\citet{Goto2017} reported the complete absence of LAEs down to a narrow--band magnitude of NB906=25.4\,mag (at 50\% completeness) in the entire field--of--view of the Subaru Telescope Suprime--Cam ($\sim$200\,cMpc$^2$).

\paragraph{J0100$+$2802} 
With $\M1450=-29.09$\,mag, J0100$+$2802 is the brightest (unlensed) quasar known at $z>6$ \citep{Wu2015}.
Sub--arcsecond resolution observations of the \ciimu\ and \mbox{CO} emission lines suggest that the host galaxy has a dynamical mass of only $\sim1.9\times10^{11}$\,M$_\odot$ \citep[][]{Wang2019J0100}.
Given this high luminosity, its proximity zone appears to be small [$R_{\rm p}=(7.12\pm0.13)$\,pMpc], implying that this quasar is relatively young, with a quasar age of $t_{\rm QSO}\sim10^5$\,years \citep{Eilers2017, Davies2019}.

\paragraph{J1030$+$0524} 
Deep broad band optical and near--IR investigation evidenced an overdensity of Lyman--Break galaxies in the field of this quasar \citep[][]{Morselli2014, Balmaverde2017, Decarli2019J1030}.
Searches for the presence of \lya\ extended emission around this target has already been investigated with sensitive \textit{HST} observations by \citet{Decarli2012} and with MUSE by \citet[][]{Drake2019}.

\paragraph{P308$-$21} 
The \ciimu\ emission line of this quasar host--galaxy is displaced by $\sim$25\,kpc and shows an enormous velocity gradient extending across more than 1000\,$\kms$ \citep{Decarli2017}.
High--resolution \textit{ALMA} and \textit{HST} observations revealed that the host--galaxy emission is split into (at least) three distinct components.
The observed gas morphology and kinematics is consistent with the close interaction of a single satellite with the quasar \citep[][]{Decarli2019}.
Deep \textit{Chandra} observations the companion galaxy might contain a heavily--obscured AGN \citep[][]{Connor2019}.
A direct comparison of our new MUSE data with the \textit{ALMA} \ciimu\ and dust maps will be presented in a forthcoming paper (Farina et al. in prep.).

\paragraph{J2229$+$1457} 
With a size of only $R_{\rm p}=(0.45\pm0.14)$\,pMpc, the proximity zone of this object is the smallest among the 31 $5.8\lesssim z \lesssim6.5$ quasars investigated by \citet[][]{Eilers2017}.
This suggests a short quasar age ($t_{\rm QSO}\lesssim10^5$\,years) for this object \citep{Eilers2017, Davies2019}.

\paragraph{J2219$+$0102} 
This is the faintest target in our survey.
Despite the low luminosity of the accretion disk, the host galaxy is undergoing a powerful starburst detected at mm--wavelengths (with an inferred star--formation rate of ${\rm SFR}\sim250$\,M$_\odot$\,yr$^{-1}$) and appears to be resolved with a size of 2--3\,kpc \citep[][]{Willott2017}.

\paragraph{J2216$-$0016} 
The rest--frame UV spectrum of this faint quasar shows a \nv\ broad absorption line \citep[][]{Matsuoka2016}.
The structure of the \ciimu\ line appears to be complex, suggesting the presence of a companion galaxy merging with the quasar host--galaxy \citep[][]{Izumi2018}

\paragraph{J2100$-$1715} 
\citet{Decarli2017} reported the presence of a \ciimu\ bright companion located at a projected separation of 60.7\,kpc and with a velocity difference of $-$41\,$\kms$ from the quasar's host galaxy.
The search for the \lya\ emission arising from this companion in the MUSE data is presented in \citet[][]{Mazzucchelli2019}.
\citet{Drake2019} reported the absence of extended \lya\ emission around this quasar.

\paragraph{P007$+$04} 
The broad \lya\ line of this quasar is truncated by the presence of a pDLA (see \autoref{fig:allspec} in \autoref{app:allspec}).
The analysis of the absorbing gas generating this feature and the search for its rest--frame UV counterpart will be presented in a future paper of this series (Farina et al., in prep.).

\paragraph{J2228$+$0110} 
A faint, extended \lya\ emission has been detected by \citet{Roche2014} in deep long--slit spectroscopic observations of this faint radio--loud quasar \citep[with radio--loudness: $R$$\sim$$60$,][]{Banados2015radio}.
The presence of the halo was confirmed by \citet{Drake2019} with MUSE observations.

\paragraph{J1044$-$0125} 
\textit{ALMA} 0\farcs2 resolution observations of the \ciimu\ fine structure line showed evidence of turbulent gas kinematics in the host galaxy and revealed the possible presence of a faint companion galaxy located at a separation of 4.9\,kpc \citep{WangRan2019}.

\section{OBSERVATIONS AND DATA REDUCTION}\label{sec:observations}

Observations of the quasars in our sample have been collected with the MUSE instrument on the \textit{VLT} telescope \textit{YEPUN} as a part of the ESO programs: 60.A-9321(A, Science Verification), 094.B-0893(A, PI: Venemans), 095.B-0419(A, PI: Roche), 095.A-0714(A, PI: Karman), 099.A-0682(A, PI: Farina), 0101.A-0656(A, PI: Farina), 0103.A-0562(A, PI: Farina), and 297.A-5054(A, PI: Decarli).
Typically, the total time on target was $\sim$50\,min, divided into two exposures of 1482\,s differentiated by a $<$5$^{\prime\prime}$ shift and a 90\,degree rotation.
For eight targets, longer integrations have been acquired (ranging from 65 to 680\,min) and the shift and rotation pattern was repeated several times (see \autoref{tab:sample})

Data reduction was performed as in \citet{Farina2017} using the \textsc{MUSE Data Reduction Software} version 2.6 \citep[][]{Weilbacher2012, Weilbacher2014} complemented by our own set of custom built routines.
Basic steps are summarized in the following.
Individual exposures were bias subtracted, corrected for flat field and illumination, and calibrated in wavelength and flux.
We then subtracted the sky emission and re--sampled the data onto a $0\farcs2$$\times$$0\farcs$$2$$\times$1.25\,\AA\ grid\footnote{Cosmic rays could have an impact on the final quality of the cubes when only two exposures have been collected.
Their rejection is performed by the pipeline in the post--processing of the data considering a sigma rejection factor of \texttt{crsigma}=15.}.
White light images were then created and used to estimate the relative offsets between different exposures of a single target.
From these images we also determined the relative flux scaling between exposures by performing force photometry on sources in the field.
Finally, we average--combined the exposures into a single cube.
Residual illumination patterns were removed using the \textsc{Zurich Atmosphere Purge} (\textsc{ZAP}) software \citep[version 2.0][]{Soto2016}, setting the number of eigenspectra (\texttt{nevals}) to~3 and masking sources detected in the white light images.
This procedure, however, comes with the price of possibly removing some astronomical flux from the cubes.
In the following, we will present results from the ``cleaned'' datacubes.
However, we also double checked for extended emission in the data prior to the use of \textsc{ZAP}.
To take voxel--to--voxel correlations into account, that can result in an underestimation of the noise calculated by the pipeline, we rescaled the variance datacube to match the measured variance of the background \citep[see e.g.,][]{Bacon2015, Borisova2016, Farina2017, Arrigoni2019}.
The astrometry solution was refined by matching sources with the Pan--STARRS1 (PS1) data archive \citep[][]{Chambers2016, Flewelling2016} or with other available surveys if the field was not covered by the PS1 footprint.
We corrected for reddening towards the quasar location using $E(B-V)$ values from \citet{Schlafly2011} and assuming R$_V$=3.1 \citep[e.g.,][]{Cardelli1989, Fitzpatrick1999}.
Absolute flux calibration was obtained matching the $z$--band photometry of sources in the field with PS1 and/or with the Dark Energy Camera Legacy Survey\footnote{\url{http://legacysurvey.org/}}.
In \autoref{tab:sample} we report the \mbox{5--$\sigma$} surface brightness limits estimated over a 1\,arcsec$^{2}$ aperture after collapsing 5 wavelength slices that were centered at the expected position of the \lya\ line shifted to the systemic redshift of the quasar (SB$_{5\,\sigma, {\rm Ly\alpha}}^{1}$).
These range from SB$_{5\,\sigma, {\rm Ly\alpha}}^{1}$=0.1 to 1.1$\times$10$^{-17}$erg\,s$^{-1}$\,cm$^{-2}$\,arcsec$^{-2}$ depending on exposure times, sky conditions, and on the redshift of the quasar.
Postage stamps of the quasar vicinities and quasar spectra are shown in \autoref{app:allspec}.

\section{SEARCHING FOR EXTENDED EMISSION}\label{sec:psfsubtraction}

An accurate PSF subtraction is necessary to recover the faint signal of the diffuse \lya\ emission emerging from the PSF wings of the bright unresolved nuclear component.
The steps we executed on each datacube to accomplish this goal are summarized in the following:

\begin{enumerate}
\item We removed possible foreground objects located in close proximity to the quasar.
To perform this step, we first collapsed the datacube along wavelengths blueward of the redshifted \lya\ line location.
Due to the Gunn--Peterson effect, the resulting image is virtually free of any object with a redshift consistent with or larger than the quasar's one.
For each source detected in this image we extracted the emission over an aperture 3~times larger than the effective radius.
We used this as an empirical model of the object's light profile\footnote{
By construction, we are following the average star--light emission profile of a galaxy.
However, nebular line emission can extend on larger scales and thus is not well reproduced by our empirical model.
We also stress that the expected improvement in seeing with wavelength ($\propto\lambda^{-0.2}$) has a negligible impact in the spectral range we are considering.
}.
This model was then propagated through the datacube by rescaling it to the flux of source measured at each wavelength channel.
Finally, all these models were combined together and subtracted from the datacube.
\item An empirical model of the PSF was directly created from the quasar light by summing up spectral regions virtually free of any extended emission, i.e. $>2500\,\kms$ from the wavelength of the \lya\ line redshifted to the quasar's systemic redshift\footnote{In \citet{Farina2017} we showed that PSF models created from nearby stars and directly from the quasar itself provide similar results in terms of detecting extended emission.}.
For this procedure, we excluded all channels where the background noise was increased by the presence of bright sky emission lines.
\item In each wavelength layer, the PSF model was rescaled to match the quasar flux measured within a radius of 2~spatial pixels, assuming that the unresolved emission of the AGN dominates within this region.
\item Following a similar procedure as in, e.g., \citet{Hennawi2013, Arrigoni2015, Farina2017} we created a smoothed $\upchi_{x,y,\lambda}$ cube defined as:
\begin{equation}
\begin{aligned}
&{\rm SMOOTH}\left[\upchi_{x,y,\lambda}\right]=\\
& =\frac{{\rm CONVOL}\left[{\rm DATA_{x,y,\lambda}}-{\rm MODEL_{x,y,\lambda}}\right]}{\sqrt{{\rm CONVOL}^2\left[ \sigma^2_{x,y,\lambda} \right]}}
\end{aligned}
\end{equation}
where ${\rm DATA_{x,y,\lambda}}$ is the datacube, ${\rm MODEL_{x,y,\lambda}}$ is the PSF model created in the step above, and $\sigma_{x,y,\lambda}$ is the square root of the variance datacube.
The operation ${\rm CONVOL}$ is a convolution with a 3D Gaussian kernel with $\sigma_{\rm spat}$=0\farcs2 in each spatial direction and $\sigma_{\rm spec}$=2.50\,\AA\ in the spectral direction, and ${\rm CONVOL}^2$ denotes a convolution with the square of the smoothing kernel used in ${\rm CONVOL}$.
\item To identify significant extended emission, we then ran a friends--of--friends algorithm that connects voxels that have $S/N$$>$2 in the ${\rm SMOOTH}\left[\upchi_{x,y,\lambda}\right]$ cube.
We chose a linking length of 2\,voxels (in both spatial and spectral directions).
Voxels located within the effective radius of a removed foreground source were excluded\footnote{The empirical procedure used to remove foreground sources intrinsically conceals information (possibly) present at their center.
We thus decided to mask these regions to avoid false detections and/or bias estimates of the halo properties.
However, this may result in an underestimate of the total halo emission.}.
Additionally, voxels contaminated by instrumental artifacts were also excluded.
We consider a group identified by the FoF as a halo associated with the quasar if, at the same time:
\textit{(i)} there is at least one voxel with $S/N>2$ within a radius of 1\,arcsec in the spatial direction and within $\pm250$\,km\,s$^{-1}$ in the spectral direction from the expected location of the \lya\ emission of the quasar;
\textit{(ii)} it contains more than 300 connected voxels\footnote{As a rule of thumb, for a spatially unresolved source, 300\,voxels correspond to a cylinder with a base of 1.5\,arcsec$^{2}$ and an height of 160\,km\,s$^{-1}$.} (this was empirically derived to avoid contaminations from cosmic rays and/or instrument artifacts not fully removed by the pipeline);
and \textit{(iii)} it spans more than two consecutive channels in the spectral direction.
\item If a halo is detected, we created a 3--dimensional mask containing all connected voxels (${\rm MASK_{x,y,\lambda}}$) and used it to extract information from the ${\rm DATA_{x,y,\lambda}}$\-$-$\-${\rm MODEL_{x,y,\lambda}}$ cube.
\end{enumerate}

In \autoref{fig:psfsub} we show the results of this procedure applied to the \REQUIEM\ survey dataset.
For each object we plot a 11\arcsec$\times$11\arcsec\ (roughly 60\,pkpc$\times$60\,pkpc at $z=6$) pseudo--narrow--band image centered at the quasar location.
The spectral region of the cube defining each narrow--band image was set by the minimum ($\lambda_{\rm min}^{\rm mask}$) and maximum ($\lambda_{\rm Max}^{\rm mask}$) wavelengths covered by ${\rm MASK_{x,y,\lambda}}$ (see \autoref{tab:mask}).
The black contours highlight regions where significant (as described above) extended emission was detected.

In summary, we report the presence of 12 \lya\ nebulae around $z>5.7$ quasars, 8 of which are newly discovered.
In the following, we describe the procedure used to extract physical information about each detected nebula.

\begin{figure*}[tbp]
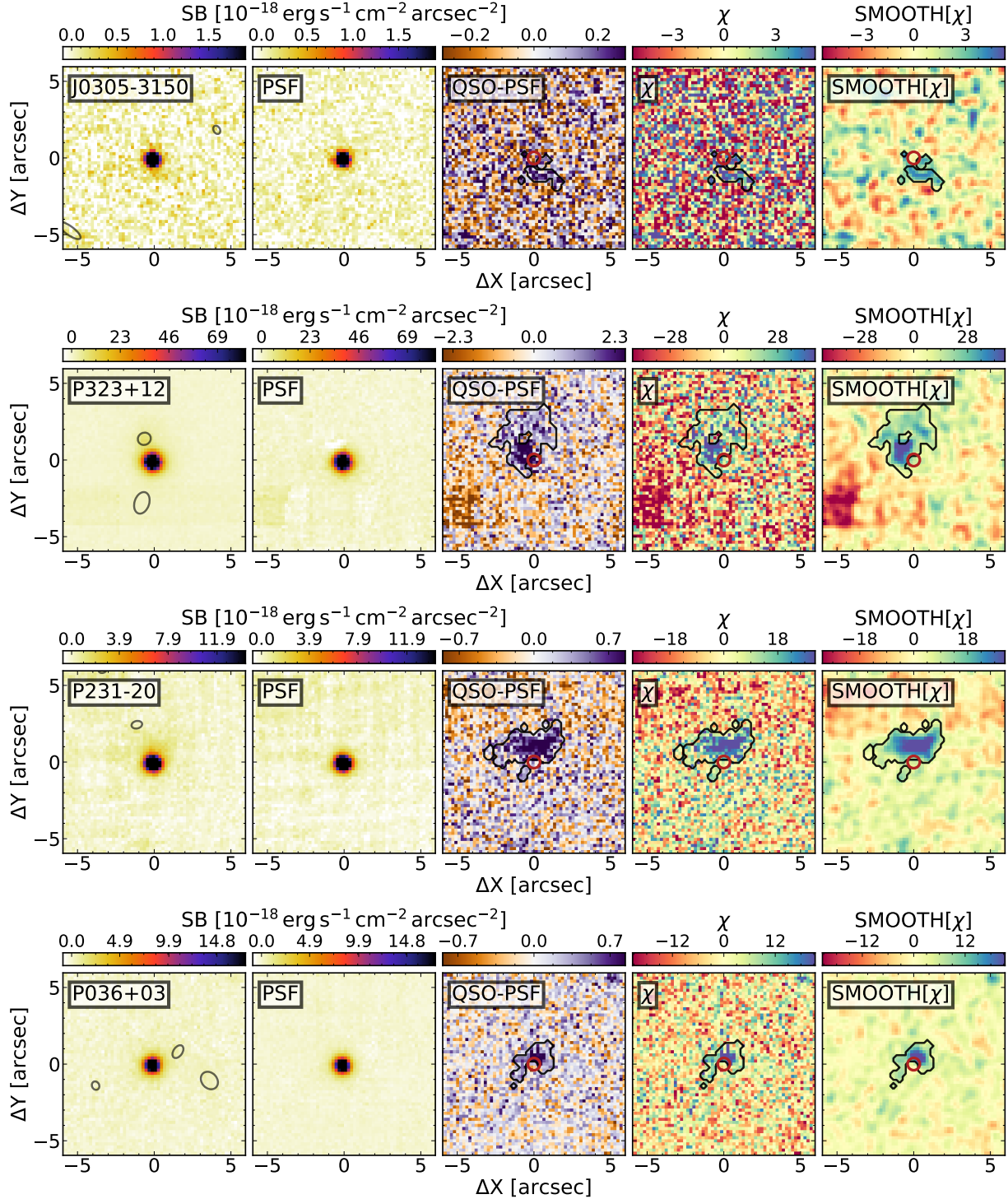

\begin{center}
\includegraphics[width=0.90\textwidth]{{{psfSub_J0305m3150_z6.6145}}}
\includegraphics[width=0.90\textwidth]{{{psfSub_P323p12_z6.5881}}}
\includegraphics[width=0.90\textwidth]{{{psfSub_P231m20_z6.5864}}}
\includegraphics[width=0.90\textwidth]{{{psfSub_P036p03_z6.5412}}}
\caption{Results from the PSF--subtraction procedure described in \autoref{sec:psfsubtraction}.
The different panels show the pseudo--narrow--band images obtained by collapsing (from left to right) the
${\rm DATA_{x,y,\lambda}}$,
${\rm MODEL_{x,y,\lambda}}$,
${\rm DATA_{x,y,\lambda}}-{\rm MODEL_{x,y,\lambda}}$,
$\upchi_{x,y,\lambda}$, and
${\rm SMOOTH}\left[\upchi_{x,y,\lambda}\right]$
cubes in the wavelength range where extended emission was detected.
If no significant nebular emission is present, the collapsed region is between $-$500 and $+$500\,$\kms$.
Black contours are constructed by collapsing ${\rm MASK_{x,y,\lambda}}$ along the velocity axis.
These are the regions used to extract the spectra of the halos in \autoref{sec:spechalo}.
In the left--most panel, the location of the detected (and removed) foreground sources are marked as gray ellipses.
Note that saturation spikes from nearby bright stars (e.g. in the bottom left corner of the quasar P323$+$12, or on the top of J1509$-$1749) and/or instrumental artifacts such as IFU--to--IFU edge effects (e.g. for the quasars P183$+$05 and P217$-$16) have been masked during the halo identification procedure but are still shown here.}\label{fig:psfsub}
\end{center}
\end{figure*}

\addtocounter{figure}{-1}

\begin{figure*}[tbp]
\begin{center}
\includegraphics[width=0.90\textwidth]{{{psfSub_J2318m3113_z6.4435}}}
\includegraphics[width=0.90\textwidth]{{{psfSub_P183p05_z6.4386}}}
\includegraphics[width=0.90\textwidth]{{{psfSub_J0210m0456_z6.4323}}}
\includegraphics[width=0.90\textwidth]{{{psfSub_J2329m0301_z6.4164}}}
\includegraphics[width=0.90\textwidth]{{{psfSub_J1152p0055_z6.3637}}}
\caption{continued.}
\end{center}
\end{figure*}

\addtocounter{figure}{-1}

\begin{figure*}[tbp]
\begin{center}
\includegraphics[width=0.90\textwidth]{{{psfSub_J2211m3206_z6.3394}}}
\includegraphics[width=0.90\textwidth]{{{psfSub_J0142m3327_z6.3379}}}
\includegraphics[width=0.90\textwidth]{{{psfSub_J0100p2802_z6.3258}}}
\includegraphics[width=0.90\textwidth]{{{psfSub_J1030p0524_z6.3}}}
\includegraphics[width=0.90\textwidth]{{{psfSub_P308m21_z6.2341}}}
\caption{continued.}
\end{center}
\end{figure*}

\addtocounter{figure}{-1}

\begin{figure*}[tbp]
\begin{center}
\includegraphics[width=0.90\textwidth]{{{psfSub_P065m26_z6.1877}}}
\includegraphics[width=0.90\textwidth]{{{psfSub_P359m06_z6.1722}}}
\includegraphics[width=0.90\textwidth]{{{psfSub_J2229p1457_z6.1517}}}
\includegraphics[width=0.90\textwidth]{{{psfSub_P217m16_z6.1498}}}
\includegraphics[width=0.90\textwidth]{{{psfSub_J2219p0102_z6.1492}}}
\caption{continued.}
\end{center}
\end{figure*}

\addtocounter{figure}{-1}

\begin{figure*}[tbp]
\begin{center}
\includegraphics[width=0.90\textwidth]{{{psfSub_J2318m3029_z6.1458}}}
\includegraphics[width=0.90\textwidth]{{{psfSub_J1509m1749_z6.1225}}}
\includegraphics[width=0.90\textwidth]{{{psfSub_J2216m0016_z6.0962}}}
\includegraphics[width=0.90\textwidth]{{{psfSub_J2100m1715_z6.0812}}}
\includegraphics[width=0.90\textwidth]{{{psfSub_J2054m0005_z6.0391}}}
\caption{continued.}
\end{center}
\end{figure*}

\addtocounter{figure}{-1}

\begin{figure*}[tbp]
\begin{center}
\includegraphics[width=0.90\textwidth]{{{psfSub_P009m10_z6.0039}}}
\includegraphics[width=0.90\textwidth]{{{psfSub_P340m18_z6.01}}}
\includegraphics[width=0.90\textwidth]{{{psfSub_J0055p0146_z6.006}}}
\includegraphics[width=0.90\textwidth]{{{psfSub_P007p04_z6.0008}}}
\includegraphics[width=0.90\textwidth]{{{psfSub_J2228p0110_z5.903}}}
\caption{continued.}
\end{center}
\end{figure*}

\addtocounter{figure}{-1}

\begin{figure*}[tbp]
\begin{center}
\includegraphics[width=0.90\textwidth]{{{psfSub_J1044m0125_z5.7847}}}
\includegraphics[width=0.90\textwidth]{{{psfSub_J0129m0035_z5.7787}}}
\caption{continued.}
\end{center}
\end{figure*}

\subsection{Spectra of the Extended Emission}\label{sec:spechalo}

We extract the nebular emission spectrum using a 2D mask obtained by collapsing ${\rm MASK_{x,y,\lambda}}$ along the spectral axis.
The construction of a halo mask described in the previous section is instrumental in obtaining the highest signal--to--noise spectrum of a detected halo.
However, given that this procedure is based on a fix cut in signal--to--noise per voxel, it inevitably results in a loss of information at larger radii.
For each halo we thus also extract a spectrum from the circular aperture with radius equal to the distance between the quasar and the most distant significant voxel detected in the collapsed ${\rm MASK_{x,y,\lambda}}$ ($d_{\rm QSO}^{\rm mask}$, see \autoref{tab:mask}).
Spectra extracted over the collapsed mask and over the circular aperture (plotted in red and in yellow in \autoref{fig:spechalo}) have similar shapes, but the latter shows a systematically higher flux density at each wavelength.

We estimate the central wavelength ($\lambda_{\rm c}$) as a non--parametric flux--and--error--weighted mean of the emission between $\lambda_{\rm min}^{\rm mask}$ and $\lambda_{\rm Max}^{\rm mask}$ (see \autoref{tab:mask} and \autoref{fig:spechalo}), i.e. without assuming any particular shape for the \lya\ line.
While in \autoref{tab:spechalo} we only report measurements from the \textit{masked} spectrum, we point out that the central wavelengths measured from the 2D masks and from the circular aperture extraction are consistent within the errors, with an average difference of only ($-15\pm37$)\,km\,s$^{-1}$.
In order to reduce the effects of noise spikes, the FWHMs of the nebular emission were estimated after smoothing spectra extracted from the 2D masks with a Gaussian kernel of $\sigma=2.5$\,\AA.
The derived FHWMs are shown as gray horizontal bars in \autoref{fig:spechalo} and listed in \autoref{tab:spechalo}.
Finally, total fluxes were calculated by integrating the spectra extracted over the circular apertures between $\lambda_{\rm min}^{\rm mask}$ and $\lambda_{\rm Max}^{\rm mask}$.
If a nebula was not detected, we extracted the spectrum over a circular aperture with a fixed radius of 20\,pkpc (corresponding to 3\farcs5 at $z=6$).
From this, we derived the 1--$\sigma$ detection limit as $\sigma^{2}_{\rm Lim}=\sum^{+500\,{\rm km}\,{\rm s}^{-1}}_{-500\,{\rm km}\,{\rm s}^{-1}}\sigma_\lambda^{2}/N_{\lambda}$, where $\sigma^2_\lambda$ is the variance at each wavelength and $N_{\lambda}$ is the number of spectral pixels in the $\pm500$\,km\,s$^{-1}$ stretch from the quasar's systemic redshift.

\begin{figure*}[p]
\begin{center}
\includegraphics[width=0.90\columnwidth]{{{SpecHalo_J0305m3150_z6.6145}}} \qquad\qquad
\includegraphics[width=0.90\columnwidth]{{{SpecHalo_P323p12_z6.5881}}}    \par\vspace{0.1cm}
\includegraphics[width=0.90\columnwidth]{{{SpecHalo_P231m20_z6.5864}}}    \qquad\qquad
\includegraphics[width=0.90\columnwidth]{{{SpecHalo_P036p03_z6.5412}}}    \par\vspace{0.1cm}
\includegraphics[width=0.90\columnwidth]{{{SpecHalo_J2318m3113_z6.4435}}} \qquad\qquad
\includegraphics[width=0.90\columnwidth]{{{SpecHalo_P183p05_z6.4386}}}    \par\vspace{0.1cm}
\includegraphics[width=0.90\columnwidth]{{{SpecHalo_J0210m0456_z6.4323}}} \qquad\qquad
\includegraphics[width=0.90\columnwidth]{{{SpecHalo_J2329m0301_z6.4164}}} \par\vspace{0.1cm}
\includegraphics[width=0.90\columnwidth]{{{SpecHalo_J1152p0055_z6.3637}}} \qquad\qquad
\includegraphics[width=0.90\columnwidth]{{{SpecHalo_J2211m3206_z6.3394}}} 
\caption{Atlas of spectra and surface brightness profiles of extended emission detected in the \REQUIEM\ survey.
\textit{Left--hand panel} --- Spectrum of the nebular emission extracted over the collapsed halo mask as described in \autoref{sec:spechalo} (red shaded histogram) with corresponding 1--$\sigma$ error (gray histogram).
The spectrum extracted over a circular aperture of radius $d_{\rm QSO}$ is also shown as a yellow histogram.
The vertical dotted lines mark the wavelength range where the extended emission was detected and the location of its flux--weighted centroid.
In the cases where no halo was detected, the spectrum extracted from a circular aperture of radius 20\,pkpc is plotted.
\textit{Right--hand panel} --- Surface brightness profile extracted over circular annuli evenly spaced in logarithmic space (purple points).
The formal 1-- and 2--$\sigma$ error in surface brightness are plotted as orange shaded regions (see \autoref{sec:sb} for further details).
Stars on the top right corners indicate objects for which significant extended emission has been detected.
}\label{fig:spechalo}
\end{center}
\end{figure*}

\addtocounter{figure}{-1}

\begin{figure*}[p]
\begin{center}
\includegraphics[width=0.90\columnwidth]{{{SpecHalo_J0142m3327_z6.3379}}} \qquad\qquad
\includegraphics[width=0.90\columnwidth]{{{SpecHalo_J0100p2802_z6.3258}}} \par\vspace{0.1cm}
\includegraphics[width=0.90\columnwidth]{{{SpecHalo_J1030p0524_z6.3}}}    \qquad\qquad
\includegraphics[width=0.90\columnwidth]{{{SpecHalo_P308m21_z6.2341}}}    \par\vspace{0.1cm}
\includegraphics[width=0.90\columnwidth]{{{SpecHalo_P065m26_z6.1877}}}    \qquad\qquad
\includegraphics[width=0.90\columnwidth]{{{SpecHalo_P359m06_z6.1722}}}    \par\vspace{0.1cm}
\includegraphics[width=0.90\columnwidth]{{{SpecHalo_J2229p1457_z6.1517}}} \qquad\qquad
\includegraphics[width=0.90\columnwidth]{{{SpecHalo_P217m16_z6.1498}}}    \par\vspace{0.1cm}
\includegraphics[width=0.90\columnwidth]{{{SpecHalo_J2219p0102_z6.1492}}} \qquad\qquad
\includegraphics[width=0.90\columnwidth]{{{SpecHalo_J2318m3029_z6.1458}}} \par\vspace{0.1cm}
\includegraphics[width=0.90\columnwidth]{{{SpecHalo_J1509m1749_z6.1225}}} \qquad\qquad
\includegraphics[width=0.90\columnwidth]{{{SpecHalo_J2216m0016_z6.0962}}} 
\caption{continued.}
\end{center}
\end{figure*}

\addtocounter{figure}{-1}

\begin{figure*}[p]
\begin{center}
\includegraphics[width=0.90\columnwidth]{{{SpecHalo_J2100m1715_z6.0812}}} \qquad\qquad
\includegraphics[width=0.90\columnwidth]{{{SpecHalo_J2054m0005_z6.0391}}} \par\vspace{0.1cm}
\includegraphics[width=0.90\columnwidth]{{{SpecHalo_P009m10_z6.0039}}}    \qquad\qquad
\includegraphics[width=0.90\columnwidth]{{{SpecHalo_P340m18_z6.01}}}      \par\vspace{0.1cm}
\includegraphics[width=0.90\columnwidth]{{{SpecHalo_J0055p0146_z6.006}}}  \qquad\qquad
\includegraphics[width=0.90\columnwidth]{{{SpecHalo_P007p04_z6.0008}}}    \par\vspace{0.1cm}
\includegraphics[width=0.90\columnwidth]{{{SpecHalo_J2228p0110_z5.903}}}  \qquad\qquad
\includegraphics[width=0.90\columnwidth]{{{SpecHalo_J1044m0125_z5.7847}}} \par\vspace{0.1cm}
\includegraphics[width=0.90\columnwidth]{{{SpecHalo_J0129m0035_z5.7787}}} 
\caption{continued.}
\end{center}
\end{figure*}

\subsection{Surface Brightness Profiles}\label{sec:sb}

The right--hand panels of \autoref{fig:spechalo} show circularly averaged surface brightness profiles of the extended emission around each quasar.
These are extracted from pseudo--narrow band images constructed summing up spectral channels located between $-500$ and $+500$\,km\,s$^{-1}$ of the quasar's systemic redshift.
The choice of a fixed width for the entire sample facilitates a uniform comparison of both detections and non--detections of nebular emission.
In addition, this velocity range corresponds to 30\,\AA, roughly matching the width of the pseudo--narrow--band images used to extract surface brightness profiles by \citet{Arrigoni2019} and by \citet{Cai2019} for their samples of $z\sim2-3$ quasars.
However, the size of the bin selected by \citeauthor{Arrigoni2019} and by \citeauthor{Cai2019} is twice as large as ours in velocity space.
This is also much narrower than previous narrow--band studies targeting high--redshift quasars: e.g., $\sim100$\,\AA\ in \citet{Decarli2012} or $\sim160$\,\AA\ in \citet{Momose2018}.
While this choice may lead to the loss of some signal from the wings of the nebulae, it allows us to optimize the signal--to--noise ratio of the extended emission whilst being sensitive to faint emission that may be present at larger scales. 
Before extracting the profile binned in annuli with radii evenly spaced in logarithmic space, we masked regions where apparent instrumental artifacts were present and regions located within the effective radius of the removed foreground sources.
Errors associated to each bin of the surface brightness profile were estimated from the collapsed variance datacube.

In addition, from the pseudo--narrow--band images we also derive a noise independent measurement of the size of the nebulae ($d_{\rm QSO}^{\rm fix}$, see \autoref{tab:mask}).
This is the distance from a quasar where the circularly averaged surface brightness profile drops below a surface--brightness of
$(1+z)^4\times3\times10^{-18}{\rm erg\,s^{-1}\,cm^{-2}\,arcsec^{-2}}$.
This value has been chosen to have $d_{\rm QSO}^{\rm fix}\approx d_{\rm QSO}^{\rm mask}$ for data collected with the shortest exposure times.

\subsection{Moment Maps}\label{sec:mom}

In order to trace the kinematics of the detected extended emission we produced the zeroth, first, and second moment maps of the flux distribution in velocity space (see \autoref{fig:mom}).
These maps encode information about the variation of the line centroid velocity and width at different spatial locations.
To create the maps, we first smoothed each wavelength layer with a 2D Gaussian kernel with $\sigma=1$ spatial pixel.
Then we extracted the the different moments within the ${\rm MASK_{x,y,\lambda}}$ region (i.e. only voxels significantly associated with the halo are included in the maps).
Given the complex kinematics of the \lya\ emission observed around high redshift quasars \citep[e.g.][]{Martin2015, Borisova2016, Ginolfi2018, Arrigoni2018, Arrigoni2019, Drake2019} and the relatively low spectral resolution of MUSE, the moments are estimated in a non--parametric way by flux--weighting each voxel.
In other words, no assumption was made about the 3D shape of the line emitting region.

\begin{figure*}[tbp]
\begin{center}
\includegraphics[width=0.63\textwidth]{{{momMap_J0305m3150_z6.6145}}}
\includegraphics[width=0.27\textwidth]{{{3DHalo_J0305m3150_z6.6145}}} \par\vspace{0.1cm}
\includegraphics[width=0.63\textwidth]{{{momMap_P323p12_z6.5881}}}
\includegraphics[width=0.27\textwidth]{{{3DHalo_P323p12_z6.5881}}}    \par\vspace{0.1cm}
\includegraphics[width=0.63\textwidth]{{{momMap_P231m20_z6.5864}}}
\includegraphics[width=0.27\textwidth]{{{3DHalo_P231m20_z6.5864}}}    \par\vspace{0.1cm}
\includegraphics[width=0.63\textwidth]{{{momMap_P036p03_z6.5412}}}
\includegraphics[width=0.27\textwidth]{{{3DHalo_P036p03_z6.5412}}}    
\caption{Spatially resolved kinematics of the halos detected in our sample.
From left to right: integrated flux (zeroth--moment), velocity--field with respect to the flux--weighted centroid of the line emission (first--moment), and velocity dispersion (second--moment).
These maps were obtained as described in \autoref{sec:mom}.
Note that the maps are not corrected for the finite spectral resolution of MUSE.
In the wavelength range explored by our sample, this washes away any information with $\sigma\lesssim35$\,km\,s$^{-1}$.
The right--most column shows the 3D visualization of each nebula.
Here $\Delta X$ goes from East to West, $\Delta Y$ from South to North, and $\Delta V$ from green to red with respect to the centroid of the extended emission.
}\label{fig:mom}
\end{center}
\end{figure*}

\addtocounter{figure}{-1}

\begin{figure*}[tbp]
\begin{center}
\includegraphics[width=0.63\textwidth]{{{momMap_J2329m0301_z6.4164}}}
\includegraphics[width=0.27\textwidth]{{{3DHalo_J2329m0301_z6.4164}}} \par\vspace{0.1cm}
\includegraphics[width=0.63\textwidth]{{{momMap_J1030p0524_z6.3}}}
\includegraphics[width=0.27\textwidth]{{{3DHalo_J1030p0524_z6.3}}}    \par\vspace{0.1cm}
\includegraphics[width=0.63\textwidth]{{{momMap_P308m21_z6.2341}}}
\includegraphics[width=0.27\textwidth]{{{3DHalo_P308m21_z6.2341}}}    \par\vspace{0.1cm}
\includegraphics[width=0.63\textwidth]{{{momMap_P065m26_z6.1877}}}
\includegraphics[width=0.27\textwidth]{{{3DHalo_P065m26_z6.1877}}}    
\caption{continued.}
\end{center}
\end{figure*}

\addtocounter{figure}{-1}

\begin{figure*}[tbp]
\begin{center}
\includegraphics[width=0.63\textwidth]{{{momMap_P359m06_z6.1722}}}  
\includegraphics[width=0.27\textwidth]{{{3DHalo_P359m06_z6.1722}}}   \par\vspace{0.1cm}
\includegraphics[width=0.63\textwidth]{{{momMap_P009m10_z6.0039}}}
\includegraphics[width=0.27\textwidth]{{{3DHalo_P009m10_z6.0039}}}   \par\vspace{0.1cm}
\includegraphics[width=0.63\textwidth]{{{momMap_P340m18_z6.01}}}
\includegraphics[width=0.27\textwidth]{{{3DHalo_P340m18_z6.01}}}     \par\vspace{0.1cm}
\includegraphics[width=0.63\textwidth]{{{momMap_J2228p0110_z5.903}}}
\includegraphics[width=0.27\textwidth]{{{3DHalo_J2228p0110_z5.903}}} 
\caption{continued.}
\end{center}
\end{figure*}

\begin{deluxetable}{lccccc}[tb]
\tablecaption{Summary of ${\rm MASK_{x,y,\lambda}}$ properties of identified nebulae.\label{tab:mask}}
\tablewidth{\textwidth}
\tabletypesize{\footnotesize}
\tablecolumns{6}
\tablewidth{0pt}
\tablehead{
\colhead{ID}                                                             &
\colhead{$\lambda_{\rm min}^{\rm mask}$--$\lambda_{\rm max}^{\rm mask}$} &
\colhead{$d_{\rm Max}^{\rm mask}$}                                       &
\colhead{$d_{\rm QSO}^{\rm mask}$}                                       &
\colhead{$A_{\rm Halo}^{\rm mask}$}                                      &
\colhead{$d_{\rm QSO}^{\rm fix}$}                                        \\
\colhead{}                &
\colhead{(\AA)}           &
\colhead{(\arcsec/pkpc)}  &
\colhead{(\arcsec/pkpc)}  &
\colhead{(arcsec$^{2}$)}  &
\colhead{(\arcsec/pkpc)}  
}
\startdata
J0305$-$3150 & 9255.0--9265.0 & 3.1$/$16.8 & 2.5$/$13.5       & \phm{1}1.8 & 1.3$/$\phm{1}7.2 \\
P323$+$12    & 9182.5--9248.8 & 4.6$/$24.7 & 3.7$/$20.2       & \phm{1}8.6 & 3.6$/$19.7       \\
P231$-$20    & 9200.0--9253.8 & 5.1$/$27.6 & 3.1$/$16.8       & \phm{1}7.8 & 2.8$/$15.3       \\
P036$+$03    & 9150.0--9178.8 & 3.6$/$19.4 & 1.8$/$\phm{1}9.9 & \phm{1}2.8 & 1.5$/$\phm{1}8.1 \\
J2318$-$3113 & \nodata        & \nodata    & \nodata          & \nodata    & \nodata          \\
P183$+$05    & \nodata        & \nodata    & \nodata          & \nodata    & \nodata          \\
J0210$-$0456 & \nodata        & \nodata    & \nodata          & \nodata    & \nodata          \\
J2329$-$0301 & 9003.8--9036.3 & 4.0$/$22.3 & 2.0$/$11.2       & \phm{1}6.9 & 1.7$/$\phm{1}9.1 \\
J1152$+$0055 & \nodata        & \nodata    & \nodata          & \nodata    & \nodata          \\
J2211$-$3206 & \nodata        & \nodata    & \nodata          & \nodata    & \nodata          \\
J0142$-$3327 & \nodata        & \nodata    & \nodata          & \nodata    & \nodata          \\
J0100$+$2802 & \nodata        & \nodata    & \nodata          & \nodata    & \nodata          \\
J1030$+$0524 & 8870.0--8897.5 & 6.1$/$34.0 & 3.7$/$20.7       &       10.3 & 1.3$/$\phm{1}7.3 \\
P308$-$21    & 8781.3--8826.3 & 7.7$/$43.2 & 4.9$/$27.4       &       18.2 & 2.0$/$11.3       \\
P065$-$26    & 8701.3--8756.3 & 4.4$/$24.7 & 2.7$/$15.0       & \phm{1}7.2 & 1.4$/$\phm{1}8.0 \\
P359$-$06    & 8700.0--8745.0 & 3.0$/$16.9 & 1.6$/$\phm{1}9.1 & \phm{1}2.7 & 1.6$/$\phm{1}8.8 \\
J2229$+$1457 & \nodata        & \nodata    & \nodata          & \nodata    & \nodata          \\
P217$-$16    & \nodata        & \nodata    & \nodata          & \nodata    & \nodata          \\
J2219$+$0102 & \nodata        & \nodata    & \nodata          & \nodata    & \nodata          \\
J2318$-$3029 & \nodata        & \nodata    & \nodata          & \nodata    & \nodata          \\
J1509$-$1749 & \nodata        & \nodata    & \nodata          & \nodata    & \nodata          \\
J2216$-$0016 & \nodata        & \nodata    & \nodata          & \nodata    & \nodata          \\
J2100$-$1715 & \nodata        & \nodata    & \nodata          & \nodata    & \nodata          \\
J2054$-$0005 & \nodata        & \nodata    & \nodata          & \nodata    & \nodata          \\
P340$-$18    & 8470.0--8548.8 & 3.2$/$18.4 & 2.6$/$14.6       & \phm{1}3.0 & 1.8$/$10.5       \\
J0055$+$0146 & \nodata        & \nodata    & \nodata          & \nodata    & \nodata          \\
P009$-$10    & 8508.8--8521.3 & 2.7$/$15.4 & 1.5$/$\phm{1}8.6 & \phm{1}1.4 & 1.4$/$\phm{1}8.2 \\
P007$+$04    & \nodata        & \nodata    & \nodata          & \nodata    & \nodata          \\
J2228$+$0110 & 8356.3--8416.3 & 5.2$/$29.7 & 2.8$/$16.0       &       10.2 & 2.1$/$12.2       \\
J1044$+$0125 & \nodata        & \nodata    & \nodata          & \nodata    & \nodata          \\
J0129$-$0035 & \nodata        & \nodata    & \nodata          & \nodata    & \nodata          \\
\enddata
\tablecomments{For each detected nebula we report the spectral range where significant emission was detected ($\lambda_{\rm min}^{\rm mask}$--$\lambda_{\rm max}^{\rm mask}$), and, its maximum extent projected on the sky ($d_{\rm Max}^{\rm mask}$), the distance between the quasar and the furthest significant voxel ($d_{\rm QSO}^{\rm mask}$), and the total area covered by the mask ($A_{\rm Halo}^{\rm mask}$, see \autoref{sec:psfsubtraction} for further details).
In addition, we also list the distance form the quasar where the circularly averaged surface brightness profile drops below a limit of $(1+z)^4\times3\times10^{-18}{\rm erg\,s^{-1}\,cm^{-2}\,arcsec^{-2}}$ ($d_{\rm QSO}^{\rm fix}$, see \autoref{sec:sb}).
}
\end{deluxetable}

\begin{deluxetable*}{lccccc}[t]
\tablecaption{Spectral properties of the extended emission.\label{tab:spechalo}}
\tablecolumns{6}
\tablenum{3}
\tablewidth{0pt}
\tablehead{
\colhead{ID}                                     &
\colhead{$\lambda_{\rm c,{\rm Ly}\alpha}$}       &
\colhead{$z_{{\rm Ly}\alpha}$}                   &
\colhead{FWHM$_{{\rm Ly}\alpha}$}                &
\colhead{F$_{{\rm Ly}\alpha}$}                   &
\colhead{L$_{{\rm Ly}\alpha}$}                   \\
\colhead{}                                       &
\colhead{(\AA)}                                  &
\colhead{}                                       &
\colhead{($\kms$)}                               &
\colhead{($10^{-17}$\,erg\,s$^{-1}$\,cm$^{-2}$)} &
\colhead{($10^{43}$\,erg\,s$^{-1}$)}
}
\startdata
J0305$-$3150 & 9259.9$\pm$1.0 & 6.6171$\pm$0.0008 & \phm{1}325$\pm$\phm{1}85 & \phm{1}1.6$\pm$0.4 & \phm{1}0.8$\pm$0.2 \\
P323$+$12    & 9217.2$\pm$2.2 & 6.5819$\pm$0.0018 &       1385$\pm$145       &       40.5$\pm$1.2 &       20.1$\pm$0.6 \\
P231$-$20    & 9228.3$\pm$1.9 & 6.5911$\pm$0.0016 &       1180$\pm$\phm{1}85 &       22.2$\pm$0.6 &       11.0$\pm$0.3 \\
P036$+$03    & 9167.5$\pm$1.6 & 6.5411$\pm$0.0013 & \phm{1}695$\pm$\phm{1}90 & \phm{1}7.8$\pm$0.4 & \phm{1}3.8$\pm$0.2 \\
J2318$-$3113 & \nodata        & \nodata           & \nodata                  &         $<$0.4     &         $<$0.2     \\
P183$+$05    & \nodata        & \nodata           & \nodata                  &         $<$1.2     &         $<$0.6     \\
J0210$-$0456 & \nodata        & \nodata           & \nodata                  &         $<$0.5     &         $<$0.2     \\
J2329$-$0301 & 9018.8$\pm$1.7 & 6.4188$\pm$0.0014 & \phm{1}830$\pm$\phm{1}60 &       11.0$\pm$0.3 & \phm{1}5.1$\pm$0.1 \\
J1152$+$0055 & \nodata        & \nodata           & \nodata                  &         $<$2.1     &         $<$0.9     \\
J2211$-$3206 & \nodata        & \nodata           & \nodata                  &         $<$0.9     &         $<$0.4     \\
J0142$-$3327 & \nodata        & \nodata           & \nodata                  &         $<$0.6     &         $<$0.3     \\
J0100$+$2802 & \nodata        & \nodata           & \nodata                  &         $<$1.0     &         $<$0.5     \\
J1030$+$0524 & 8880.3$\pm$1.5 & 6.3048$\pm$0.0012 & \phm{1}590$\pm$120       & \phm{1}5.6$\pm$0.7 & \phm{1}2.5$\pm$0.3 \\
P308$-$21    & 8803.2$\pm$1.7 & 6.2414$\pm$0.0014 &       1020$\pm$\phm{1}60 &       20.3$\pm$0.7 & \phm{1}8.8$\pm$0.3 \\
P065$-$26    & 8729.8$\pm$2.3 & 6.1810$\pm$0.0019 &       1675$\pm$\phm{1}90 &       15.4$\pm$0.6 & \phm{1}6.6$\pm$0.2 \\
P359$-$06    & 8722.8$\pm$1.9 & 6.1753$\pm$0.0016 &       1160$\pm$330       & \phm{1}7.8$\pm$0.4 & \phm{1}3.3$\pm$0.2 \\
J2229$+$1457 & \nodata        & \nodata           & \nodata                  &         $<$0.4     &         $<$0.2     \\
P217$-$16    & \nodata        & \nodata           & \nodata                  &         $<$0.3     &         $<$0.1     \\
J2219$+$0102 & \nodata        & \nodata           & \nodata                  &         $<$0.5     &         $<$0.2     \\
J2318$-$3029 & \nodata        & \nodata           & \nodata                  &         $<$0.3     &         $<$0.1     \\
J1509$-$1749 & \nodata        & \nodata           & \nodata                  &         $<$0.4     &         $<$0.2     \\
J2216$-$0016 & \nodata        & \nodata           & \nodata                  &         $<$0.5     &         $<$0.2     \\
J2100$-$1715 & \nodata        & \nodata           & \nodata                  &         $<$0.2     &         $<$0.1     \\
J2054$-$0005 & \nodata        & \nodata           & \nodata                  &         $<$0.3     &         $<$0.1     \\
P340$-$18    & 8510.5$\pm$2.5 & 6.0007$\pm$0.0020 &       1320$\pm$155       &       18.8$\pm$0.8 & \phm{1}7.5$\pm$0.3 \\
J0055$+$0146 & \nodata        & \nodata           & \nodata                  &         $<$0.3     &         $<$0.1     \\
P009$-$10    & 8513.8$\pm$1.2 & 6.0033$\pm$0.0010 & \phm{1}395$\pm$\phm{1}60 & \phm{1}2.3$\pm$0.2 & \phm{1}0.9$\pm$0.1 \\
P007$+$04    & \nodata        & \nodata           & \nodata                  &         $<$0.5     &         $<$0.2     \\
J2228$+$0110 & 8388.0$\pm$1.7 & 5.8999$\pm$0.0014 & \phm{1}940$\pm$\phm{1}65 &       20.3$\pm$0.4 & \phm{1}7.8$\pm$0.1 \\
J1044$+$0125 & \nodata        & \nodata           & \nodata                  &         $<$0.3     &         $<$0.1     \\
J0129$-$0035 & \nodata        & \nodata           & \nodata                  &         $<$0.3     &         $<$0.1     \\
\enddata
\tablecomments{The reported central wavelengths ($\lambda_{\rm c,{\rm Ly}\alpha}$) and FWHMs (FWHM$_{{\rm Ly}\alpha}$) are derived from the spectrum of the nebular emission extracted from the collapsed ${\rm MASK_{x,y,\lambda}}$ as described in \autoref{sec:spechalo}.
The total fluxes (F$_{{\rm Ly}\alpha}$) are instead derived from spectra extracted over a circular aperture of radius $d_{\rm QSO}^{\rm mask}$.
In the cases of non--detections, 3--$\sigma$ limits on fluxes are reported.}
\end{deluxetable*}

\section{RESULTS AND DISCUSSION}\label{sec:discussion}

The analysis of the fields of the 31 quasars that constitute the \REQUIEM\ survey revealed the presence of extended \lya\ emission around $\sim$39\% of the sample (12 out of 31 targets, 11/23 considering only our \textit{core sample}).
At the face value, this detection rate is lower than the 100\% reported for $z\sim3$ quasars by \citet{Borisova2016} and \citet{Arrigoni2019}.
However, only $\sim50$\% of the \citet{Arrigoni2019} would be detected if their surface brightness limit are rescaled to compensate for the effects of the cosmological dimming (a factor of $\sim10\times$ from $z\sim3$ to $z\sim6$).
The nebulae detected at $z\sim6$ show a variety of morphologies and properties, spanning a factor of 25 in luminosity (from $8\times10^{42}$ to $2\times10^{44}$\,erg\,s$^{-1}$), have FWHM ranging from $\sim300$ to $\sim1700$\,km\,s$^{-1}$, and maximum sizes from $\sim8$ to $\sim27$\,pkpc.
In the following, we investigate the origin of this emission, relate it to the properties of the central powering source, and compare with lower redshift samples.

\subsection{Extended Halos and Quasar Host--Galaxies}

Direct detections of the stars of the host galaxy of the first quasars still elude us \citep[e.g.,][; and \autoref{sec:luci}]{Decarli2012,Mechtley2012}.
On the other hand, gas and dust in the interstellar medium are routinely detected at mm-- and sub--millimeter wavelengths.
For instance, for all but two quasars in our sample (i.e., J2228$+$0110, and P340$-$18) sensitive measurements of the \ciimu\ emission line and of the underlying far--infrared continuum have been collected \citep[see][; and references therein]{Decarli2018, Venemans2018}.
These observations provide direct insights on the properties of the  host--galaxies, including, among others, precise systemic redshifts, dynamics of the gas, and star--formation rates.
In the following, we will test for connections between these properties of quasar host--galaxies and the extended \lya\ halos where they reside.

\subsubsection{Velocity shifts with respect to the systemic redshifts}

We first estimate the velocity difference ($\Delta V_{\rm sys}$) between the flux--weighted centroid of the extended emission and the precise systemic redshift of the quasar host--galaxies provided by the \ciimu\ line.
If no measurement of the \ciimu\ emission line is available in the literature (see \autoref{tab:sample}) we consider systemic redshifts from the quasar broad \lya\ or \mgii\ emission lines \citep[including the empirical correction for \mgii\--based systemic redshifts from][]{Shen2016}.
The velocity difference is defined as:
\begin{equation}\label{eq:dv}
\Delta V_{\rm sys} = \frac{c\,\left(z_{\rm \lya}-z_{\rm sys}\right)}{1+z_{\rm sys}}
\end{equation}
where $c$ is the speed of light.
This means that a positive $\Delta V_{\rm sys}$ corresponds to a halo shifted redward of the systemic redshift.

All the detected halos have velocity shifts between $\Delta V_{\rm sys}=-500$\,km\,s$^{-1}$ and $+500$\,km\,s$^{-1}$, with an average  $\langle \Delta V_{\rm sys} \rangle = (+71\pm31)$\,km\,s$^{-1}$ and a median of $+54$\,km\,s$^{-1}$.
This value agrees with $\langle \Delta V_{\rm sys} \rangle = (+69\pm36)$\,km\,s$^{-1}$ (with a median of $+112$\,km\,s$^{-1}$) calculated taking only \ciimu\ redshifts into account (see the left--hand panel of \autoref{fig:dv}).
These small velocity differences hint at a strong connection between the extended halos and $z\sim6$ quasar host--galaxies.
Much larger shifts are reported for \lya\ nebulosities around intermediate redshift quasars.
For instance, \citet{Borisova2016} measured a median shift of $1821$\,km\,s$^{-1}$ in a sample of bright $3 \lesssim z \lesssim 4$ quasars.
Similarly, in their sample of 61 $z\sim3$ quasars, \citet{Arrigoni2019} reported a large shift between \lya\ halos and their best estimates of the quasar systemic redshifts, with a median of $782$\,km\,s$^{-1}$.
We argue that the discrepancy between intermediate and high redshift halos is related to the large intrinsic uncertainties in the \civ--based systemic redshifts used in \citeauthor{Borisova2016} and in \citeauthor{Arrigoni2019} \citep[of the order of $\sim400$\,km\,s$^{-1}$, e.g.,][]{Richards2002, Shen2016}.
Indeed, the median shift for the sample of \citet{Arrigoni2019} reduces to $144$\,km\,s$^{-1}$ when the peak of the broad \lya\ line of the $z\sim3$ quasars themselves is used as a tracer of the systemic redshift.
This matches the median shift between the halo and the \ciimu\ redshifts observed in the \REQUIEM\ sample.

\subsubsection{FWHM of the extended emission}

\begin{figure*}[tb]
\begin{center}
\includegraphics[width=0.49\textwidth]{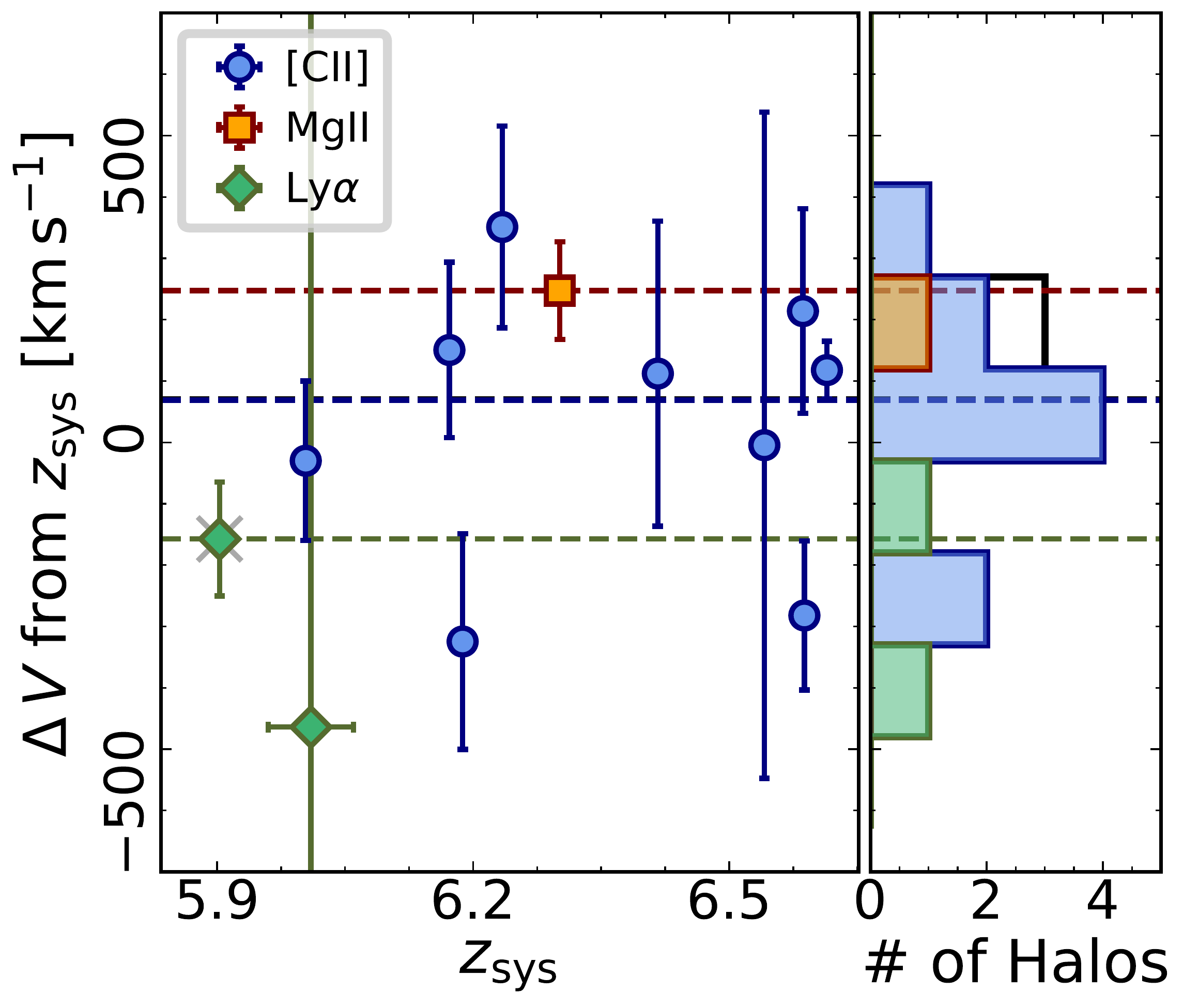}\quad
\includegraphics[width=0.49\textwidth]{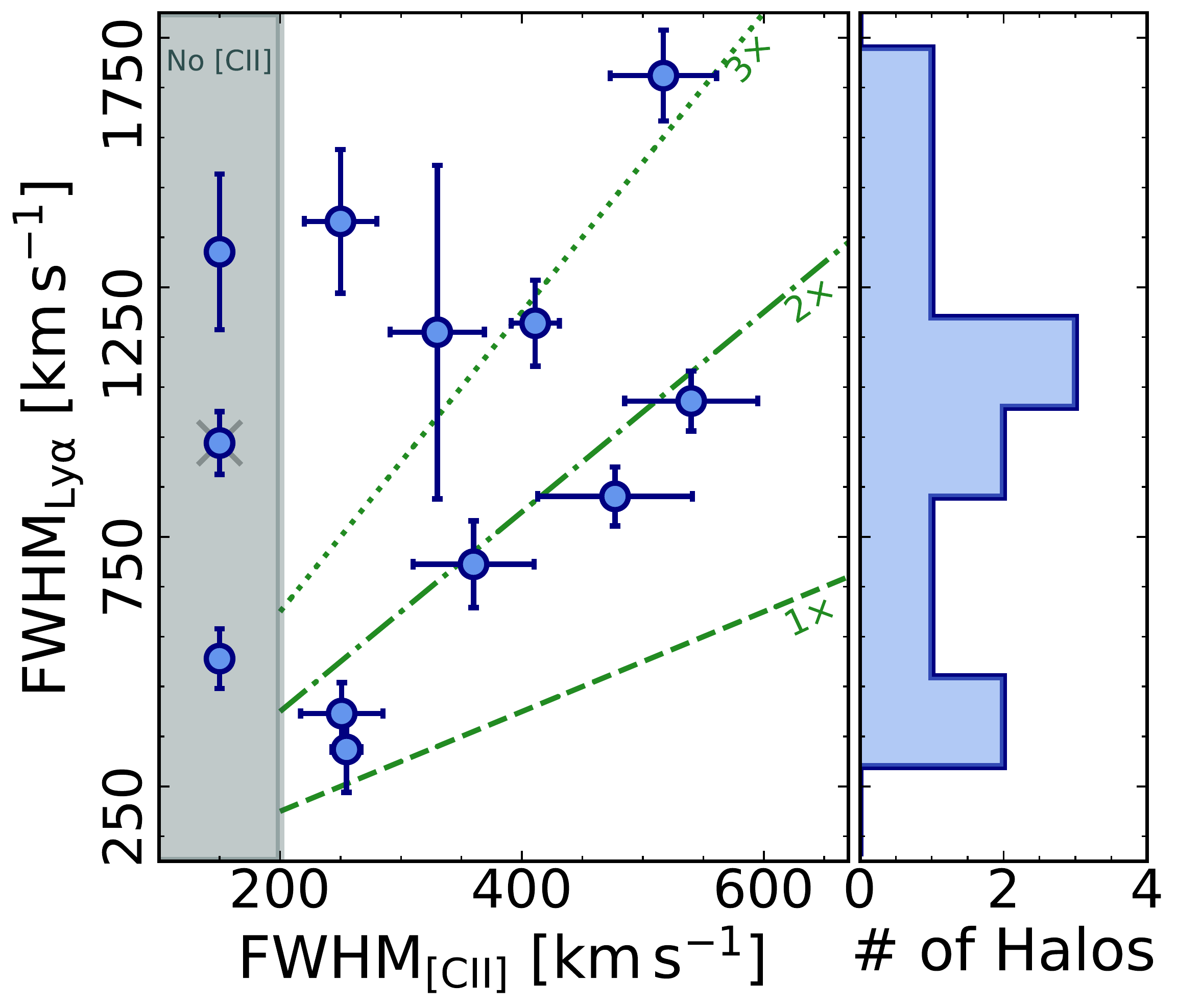}
\caption{
\textit{Left Panel ---}
Velocity difference between the flux--weighted centroid of the nebular emission and the systemic redshift of the quasar (see \autoref{eq:dv}).
Different symbols indicate different emission lines used to derive the quasar redshifts (i,e, blue points for \ciimu, orange squares for \mgii, and green diamonds for \lya, see \autoref{tab:sample} for details).
Horizontal dashed lines indicate the weighted average velocity difference of the different lines.
The error--weighted average shift calculated for the entire sample is $\langle \Delta V_{\rm sys} \rangle = (+71\pm31)$\,km\,s$^{-1}$.
This is shown as a black dashed line that overlaps with the average estimated from the \ciimu\ redshifts only.
The histograms on the left--hand side of this panel display the distribution of the shifts color--coded as on the right--hand side.
\textit{Right Panel ---}
Comparison between the FWHM of the detected extended \lya\ emission (FWHM$_{\rm Ly\alpha}$) and of the \ciimu\ line arising from the quasar's host--galaxy (FWHM$_{\rm C[II]}$).
To guide the eye, green lines mark FWHM$_{\rm Ly\alpha}$ that are 1, 2, and 3$\times$ wider than FWHM$_{\rm C[II]}$.
The wide distribution of the FWHM$_{\rm Ly\alpha}$ (ranging from $\sim250$\,km\,s$^{-1}$
to $\sim1750$\,km\,s$^{-1}$) is apparent in the histogram on the left--hand side of this panel.\newline
In both panels, gray crosses mark targets not part of our \textit{core sample} (see \autoref{sec:sample}).
}\label{fig:dv}
\end{center}
\end{figure*}

The right--hand panel of \autoref{fig:dv} presents the distribution of FWHM of the detected halos with respect to the \ciimu\ lines (FWHM$_{\rm [CII]}$).
FWHM$_{\rm Ly\alpha}$ appears to be consistently a factor $>2\times$ larger than FWHM$_{\rm [CII]}$.
Given that the \lya\ and the \ciimu\ are tracing different gas components, a different broadening of the two lines is indeed expected.
(Sub--)arcsecond investigation of \ciimu\ emission lines reveled that $z\sim6$ quasar host--galaxies are compact objects with size of a few~kiloparsecs or less \citep[e.g.][]{Wang2013, Decarli2018, Venemans2019, Neeleman2019}, while the extended emission is detected at scales of dozens of kiloparsecs (see \autoref{tab:mask}).
\textit{Zoom--in} simulations of massive $z>6$ dark--matter halos hosting quasars show that the deep potential well of stellar component dominates the kinematics in the central regions, while dark matter prevails at $\gtrsim10$\,pkpc \citep[e.g.,][]{Dubois2012, Costa2015}, giving rise to the difference between the velocities of the dense gas component traced by the \ciimu\ and of the cool gas responsible for the \lya\ emission.
A direct interpretation of this result is, however, not trivial.
The resonant nature of the \lya\ line and the turbulent motion of the gas due to interactions and feedback effects are likely to contribute to the broadening of the line emission.
Moreover, one can speculate that the presence of cool streams, often invoked to replenish the central galaxy with gas, could contribute to the larger observed FWHM \citep[][]{DiMatteo2012, DiMatteo2017, Feng2014}.
A more detailed investigation of these different possibilities will be provided in Costa et al.\ (in prep.).

\subsubsection{The SFR of the quasar host--galaxies}

It is tempting to explore the possibility that the intense starbursts observed at mm--wavelengths directly influence the powering of the extended \lya\ emission.
SFR--based \lya\ luminosities (${\rm L^{\rm SFR}_{\rm Ly\alpha}}$) are expected to follow the linear relation:
\begin{equation}\label{eq:sfr}
\frac{\rm L^{\rm SFR}_{\rm Ly\alpha}}{10^{42}\,{\rm erg\,s^{-1}}} = 1.62\frac{\rm SFR}{\rm M_\odot\,yr^{-1}},
\end{equation}
for which we assume the \ha\ calibration relation \citep[e.g.,][]{Kennicutt2012} and the case~B recombination \lya--to--\ha\ line ratio.
Star formation rates can be derived either from the \ciimu\ emission line or from rest frame far--infrared dust continuum luminosity of the quasar hosts.
\ciimu\ estimates, however, depend on the (unknown) dust metallicity, especially in the case of compact starbursts \citep[e.g.,][]{Delooze2014}.
The dust continuum, on the other hand, can be used to estimate SFRs more directly by assuming that the dust is heated by star formation \citep[i.e., considering that the quasar has a negligible contribution to the observed emission][]{Leipski2014} and that the dust spectral energy distribution is well parameterized by a modified blackbody with a (typical) temperature of T$_{\rm dust}=47$\,K and a spectral index of $\beta=1.6$ \citep[][]{Beelen2006, Barnett2015}.
Conveniently, we detected the rest frame far--infrared dust continuum significantly for all quasars in our sample except J2228$+$0110.
In the following, we will thus consider SFRs based on the dust continuum from \citet[][, and references therein]{Venemans2018}.
We stress that in high--$z$ quasar host--galaxies, SFRs derived from \ciimu\ and from far--infrared continuum correlate, albeit with a large scatter \citep[see discussion in, e.g.,][]{Decarli2018, Venemans2018}.

\begin{figure}[tb]
\begin{center}
\includegraphics[width=0.98\columnwidth]{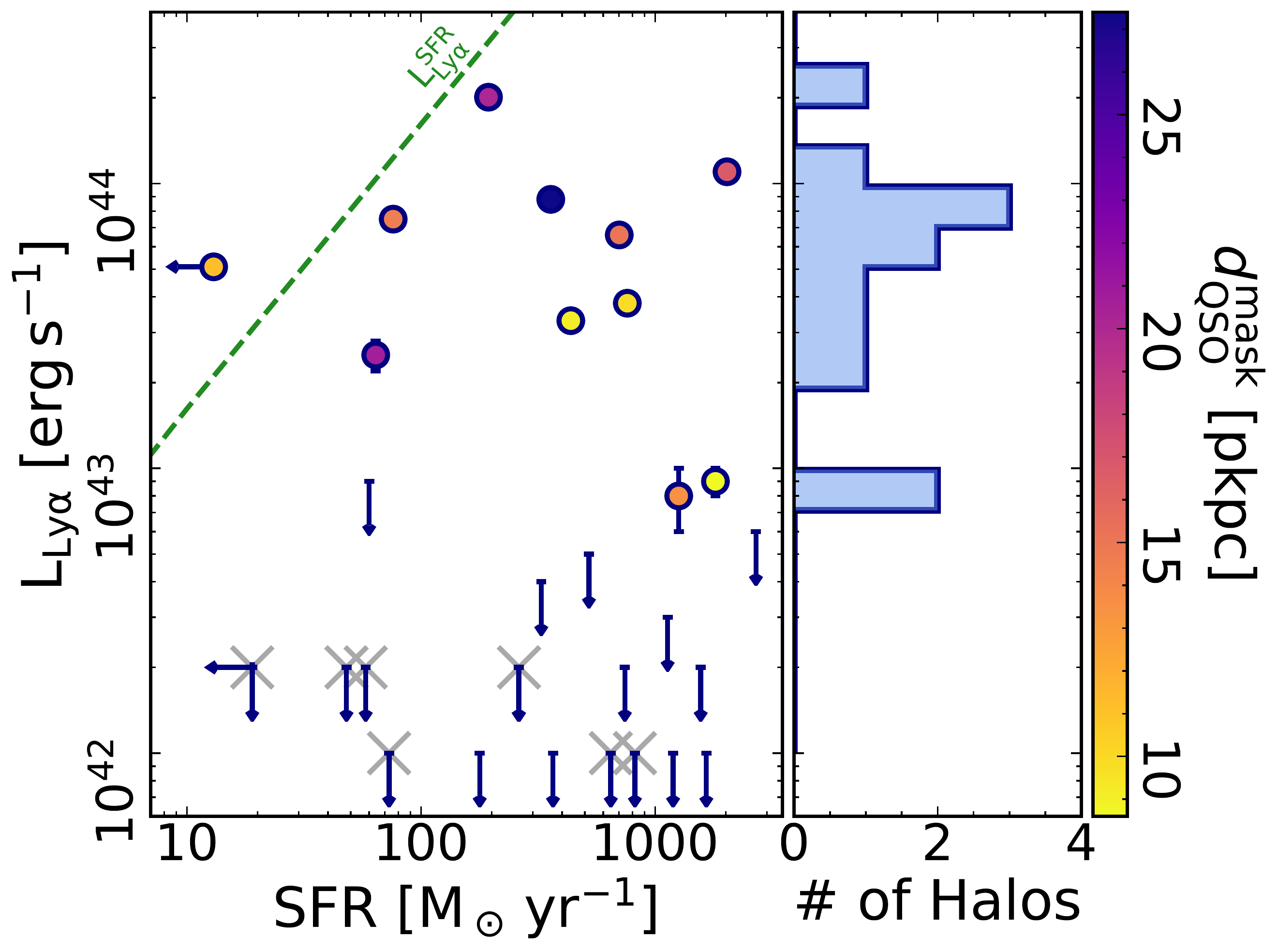}
\caption{Total \lya\ luminosity of the extended halos (L$_{\rm Ly\alpha}$, see \autoref{tab:spechalo}) versus star formation rate derived from the dust continuum emission of the quasar host--galaxies (SFR).
3--$\sigma$ upper--limits on L$_{\rm Ly\alpha}$ and on SFR are shown as downward and leftward arrows, respectively.
Different colors show the value of $d^{\rm mask}_{\rm QSO}$ (see \autoref{tab:mask}).
The green dashed line indicates the unobscured \lya\ emission expected from UV photons generated in intense starbursts, such as detected at mm--wavelengths (\autoref{eq:sfr}).
Most of the detected halos have L$_{\rm Ly\alpha}$ below this prediction.
Gray crosses represent quasars that are not part of our \textit{core sample} (see \autoref{sec:sample}).
}\label{fig:sfr}
\end{center}
\end{figure}

\autoref{fig:sfr} shows that \lya\ luminosities of the extended emission are broadly independent of the SFRs of the quasar host--galaxies and are typically well below the expectation based on \autoref{eq:sfr} (shown as a green dashed line).
The resonant nature of the \lya\ line with the large mass in dust present in the host--galaxies \citep[M$_{\rm dust}=10^7-10^9$\,M$_\odot$,][]{Venemans2018} are possible processes responsible for the suppression of the \lya\ emission \citep[e.g.][; see also \citealt{Mechtley2012} for a study of the host galaxy of the $z\sim6.4$ quasar J1148+5251]{Kunth1998, Verhamme2006, Gronwall2007, Atek2008, Sobral2019}.
The cumulative effect can be quantified by the so--called \lya\ escape fraction \citep[$f_{\rm esc, Ly\alpha}$, e.g.,][]{Kennicutt2012}.
The median value of $f_{\rm esc, Ly\alpha}$ estimated for our sample is $\lesssim$1\%.
This is an order of magnitude lower than typically reported for $z\sim6$ LAEs \citep[e.g.,][]{Ono2010, Hayes2011} but consistent with the most massive, highly star--forming galaxies observed in the  3D--\textit{HST}/CANDELS survey \citep[][]{Oyarzun2017}.
However, this value should be considered with some caution. 
The precise estimate of $f_{\rm esc, Ly\alpha}$ is strongly affected by different properties of the host galaxy \citep[e.g., neutral hydrogen column density, neutral fraction, geometry, gas--to--dust ratio, etc., see][; and references therein]{Draine2011, Hennawi2013} and by the \textit{patchiness} of the dust cocoon \citep[e.g.,][]{Casey2014}.
In addition, other mechanisms could contribute to the observed \lya\ emission, in particular the presence of the strong radiation field generated by the quasar \citep[e.g.,][; see also \autoref{sec:cgm}]{Cantalupo2005}.

\subsection{The kinematics of the gas}\label{sec:kine}

In \autoref{fig:mom} we presented the two dimensional flux--weighted maps of the velocity centroid and dispersion distribution of the extended \lya\ emission (see \autoref{sec:mom} for details).
In this section, we investigate these resolved kinematics maps in order to identify signatures of ordered motion, in/outflows, rotations, etc.
We remind the reader that these maps were computed in a non--parametric way in the regions identified by ${\rm MASK_{x,y,\lambda}}$, and that velocity shifts are not relative to the quasar's systemic redshift.

\subsubsection{Velocity fields}

The relatively low signal--to--noise of the first moment maps (see \autoref{fig:mom}) makes hard to infer the potential presence of ordered motion in the gas.
In addition, due to the resonant nature of the \lya\ line, the signature of coherent motion could be hindered by radiative transfer effects \citep[e.g.,][]{Cantalupo2005}.
Indeed, the majority of the halos identified in the \REQUIEM\ survey do not show evidence of rotation, as it was reported in extended \lya\ halos at $z\sim2-4$ \citep[][]{Borisova2016, Arrigoni2019, Cai2019}.

A noticeable exception is the nebular emission around the quasar P231$-$20.
A velocity gradient can be seen ranging from $-200$ to $+800$\,km\,s$^{-1}$ East to West (see \autoref{fig:pvel}).
Intriguingly, two \ciimu--bright companions located within $\lesssim14$\,kpc from the quasar host galaxy have been discovered \citep[][]{Decarli2017, Neeleman2019}.
Both the velocity shear and the presence of a rich environment are reminiscent of the enormous \lya\ nebular emission observed around the quasar SDSS~J1020$+$1040 at $z=3.2$ by \citet[][]{Arrigoni2018} (albeit on a smaller scale).
This system is considered a prototype to investigate the feasibility of inspiraling accretion onto a massive galaxy at $z\sim3$.
Indeed, simulations predict that baryons assemble in rotational structures, gaining angular momentum from their dark matters halos \citep[e.g.,][]{Hoyle1951, Fall1980, Mo1998} and from accretion streams \citep[e.g.,][]{Chen2003, Keres2009a, Keres2009b, Brook2011, Stewart2017}.
In this scenario the cool accreting gas should be able to shape the central galaxy, delivering both fuel for star formation and angular momentum to the central regions \citep[e.g.,][]{Sales2012, Bouche2013}. High resolution ($\sim0\farcs35$) observations of the \ciimu\ emission of the quasar host galaxy of P231$-$20 does not show a strong signature of rotation \citep[][]{Neeleman2019}.
This supports the idea that the system recently underwent a merger event with the close companion galaxy \citep[located at a separation of 9\,kpc and $-$135\,km\,s$^{-1}$,][]{Decarli2017, Neeleman2019} that perturbed the gas distribution.

If one assumes that the gas in the extended halo of P231$-$20 is (at first order) gravitationally bound, it is possible to approximate the dynamical mass of the system as $M_{\rm dyn}=1.16\times10^{5}\,v_{\rm circ}\,d$ where $d$ is the diameter of the nebular emission in pkpc and $v_{\rm circ}$ its circular velocity in km\,s$^{-1}$.
Given that the gas shows ordered motion, $v_{\rm circ}$ can be expressed as $v_{\rm circ}=0.75\,{\rm FWHM}_{\rm Ly\alpha}/\sin{i}$, with the (unknown) inclination typically considered to be $i=55^{\circ}$.
This gives us a dynamical mass of $M_{\rm dyn}\sim1.5\times10^{13}$\,M$_\odot$.
Albeit the large uncertainties associated with this measurement, the estimated dynamical mass is remarkably similar to the mass predicted for the \lya\ emission around the $z=2.28$ radio--quiet quasar UM287 \citep[][]{Cantalupo2014}\footnote{Estimates of the cool gas mass are strongly dependent on the assumed physical conditions of the gas. Combining photoionization models with sensitive searches for \heii\ and \civ\ extended emission around UM287 \citet{Arrigoni2015HeII} derived extreme gas clumping factors (and thus higher densities) and much lower mass of cool gas present in this nebula: $M_{\rm cool}\lesssim6.5\times10^{10}$\,M$_\odot$.}.
\citet{Martin2015, Martin2019} interpreted this emission, which extends out to $\sim$500\,kpc, as a large proto--galactic disc.
Deeper MUSE observations are however necessary to fully capture the complex kinematics of this system.

The low incidence of clearly rotating structures in our sample is broadly in agreement with results by \citet{Dubois2012} who re--simulated two massive ($0.5$ and $2.5\times$10$^{12}$\,M$_\odot$) $z\sim6$ halos. They show that a significant fraction of the gas in the halo can fall almost radially towards the center.
The reduced angular momentum inside the virial radius \citep[mostly due to the isotropic distribution of the 0 and to gravitational instabilities and mergers, see also][]{Prieto2015} allows for efficient funnelling of gas to the central regions of the halos, potentially sustaining the rapid growth of the first supermassive black holes.

\begin{figure}[ht]
\begin{center}
\includegraphics[width=0.97\columnwidth]{{{posVelP231m203Pan}}}
\caption{Velocity field of the extended \lya\ nebular emission detected around the quasar P231$-$20 convolved with a 3D Gaussian kernel with $\sigma_{\rm spat}=0\farcs2$ in the spatial and $\sigma_{\rm spec}=2.50$\,\AA\ in the spectral direction.
Zero velocity is set to the \ciimu\ redshift of the quasar ($z_{\rm sys}=6.5864\pm0.0005$, while in \autoref{fig:mom} velocity differences are relative to the flux--weighted centroid of the extended emission).
The red circle in the center of each image marks the location of the quasar.
Only voxels that are part of ${\rm MASK_{x,y,\lambda}}$ are shown.
$\Delta X$ goes from East to West, $\Delta Y$ from South to North.
The bulk of the emission is redshifted with respect to the quasar's systemic redshift, and the nebula shows a velocity gradient going from East to West.
}\label{fig:pvel}
\end{center}
\end{figure}

\subsubsection{Velocity dispersion}

The second--moment maps presented in \autoref{fig:mom} show that the detected extended \lya\ halos have average flux--weighted velocity dispersions ($\langle\sigma_{\rm Ly\alpha}\rangle$) spanning from $\langle\sigma_{\rm Ly\alpha}\rangle\sim30$\,km\,s$^{-1}$ to $460$\,km\,s$^{-1}$ with an average of $320\pm120$\,km\,s$^{-1}$.
Note that these values have been corrected for the limited spectral resolution of MUSE according to:
$\left(\sigma_{\rm Ly\alpha}^{\rm corr}\right)^2 = \left(\sigma_{\rm Ly\alpha}^{\rm meas.}\right)^2-\left(\sigma_{\rm Res.}^{\lambda}\right)^{2}$, where $\sigma_{\rm Res.}^{\lambda}\sim35$\,km\,s$^{-1}$ at the wavelengths explored in our sample.
The relatively quiescent $\langle\sigma_{\rm Ly\alpha}\rangle$ values are consistent with measurements reported by \citet{Borisova2016}, \citet{Arrigoni2019}, and \citet{Cai2019} around bright $z\sim2-4$ quasars.

At $z\sim3.7$ \citet{Borisova2016} reported a larger velocity dispersion ($\langle\sigma_{\rm ly\alpha}\rangle>400$\,km\,s$^{-1}$) for a halo around a radio--loud quasar than for halos around radio--quiet quasars.
In the \REQUIEM\ sample there is one radio--loud quasar, J2228$+$0110.
It shows a flux--weighted velocity dispersion of ($350\pm90$)\,km\,s$^{-1}$, in agreement with the rest of our (radio--quiet) sample.
This dispersion is consistent with \citet{Arrigoni2019}, who derived similar kinematics for nebulae around radio--loud and radio--quiet quasars at $z\sim3.2$ \footnote{It is worth noting that the radio--loud quasars in both our and \citeauthor{Arrigoni2019} samples are few magnitudes fainter than the one sampled by \citet{Borisova2016}.}

\citet{Ginolfi2018} suggested that the high velocity dispersions observed for the \lya\ extended emission around the broad--absorption line quasar J1605$-$0112 at $z=4.9$ could be linked to an outflow of material escaping the central black hole.
Our sample contains only one broad--absorption line quasar (J2216$-$0016) that is 2.9\,mag fainter than J1605$-$0112.
For this object we do not detect the presence of any significant extended emission.
Given the generally quiescent motion of the nebulae in our sample, it is unlikely we are probing fast outflows driven by the quasar \citep[expected to be of the order of $\gtrsim1000$\,km\,s$^{-1}$, e.g.,][]{Tremonti2007, VillarMartin2007, Greene2012}.
Nonetheless, we cannot exclude this scenario for the halos associated with the quasars P065$-$26 and P340$-$18, where the observed gas velocities are of the order ${\rm FWHM}\gtrsim1500$\,km\,s$^{-1}$.
Simulations of luminous ($>{\rm few}\times10^{46}$\,erg\,s$^{-1}$) quasars indeed predict AGN--driven winds with such large velocities, however these could happen at different scales, from less than 1\,kpc to several tens \citep[e.g.,][]{Costa2015, Bieri2017}.
However, given the current spatial resolution of our data, we are not sensitive to the presence of extreme kinematics on scales $\lesssim5$\,kpc (where the such an emission would be diluted by the flux of the central AGN).

\textit{Is the gas gravitationally bound to the halo?}
Recent observations (both in absorption and in emission) of the gas in the circum--galactic medium of quasars have revealed velocity dispersions consistent with the gravitational motion within dark matter halos with masses ${\rm M}_{\rm DM}\gtrsim10^{12.5}$\,M$_\odot$ \citep[e.g.,][]{Prochaska2009, Lau2018, Arrigoni2019}.
These are typical masses of halos hosting quasars, derived from strong quasar--quasar and quasar--galaxy clustering observed out to $z\sim4$ \citep[e.g.,][]{Shen2007, White2012, Eftekharzadeh2015, Garcia2017, Timlin2018, He2018}.
At $z\sim6$, however, such a direct measurement still eludes us.
Nevertheless, we can gain some insight by comparing the number density of bright quasars and massive dark matter halos \citep[e.g.,][]{Shankar2010}, under the assumption that there is a correlation between the luminosity (mass) of a quasar and the mass of the dark matter halo it is embed in \citep[see e.g.,][]{Volonteri2011}.
By integrating the \citet[][]{Kashikawa2015} luminosity function at $z=6.3$ (i.e., the average redshift of our survey), we can expect a number density of $\phi\left({\rm M}_{\rm 1450}<-25.25\right)=2.5\times10^{-9}$\,Mpc$^{-3}$ for quasars brighter than ${\rm M}_{\rm 1450}=-25.25$\,mag\footnote{Note that at these redshifts the quasar luminosity function is not well constrained. For instance, using the luminosity function inferred from a sample of 52 SDSS quasars from \citet{Jiang2016}, the number density of $z=6.3$ quasars is $\phi\left({\rm M}_{\rm 1450}<-25.25\right)=2.6\times10^{-10}$\,Mpc$^{-3}$}.
If we assume a high duty cycle of $f_{\rm duty}=0.9$ \citep[as predicted by][]{Shankar2010}, we can infer that the integral of the halo mass function from \citet{Behroozi2013} matches the integral of the luminosity function for masses ${\rm M}_{\rm DM}\sim10^{12.8}$\,M$_\odot$.

\begin{figure}[tb]
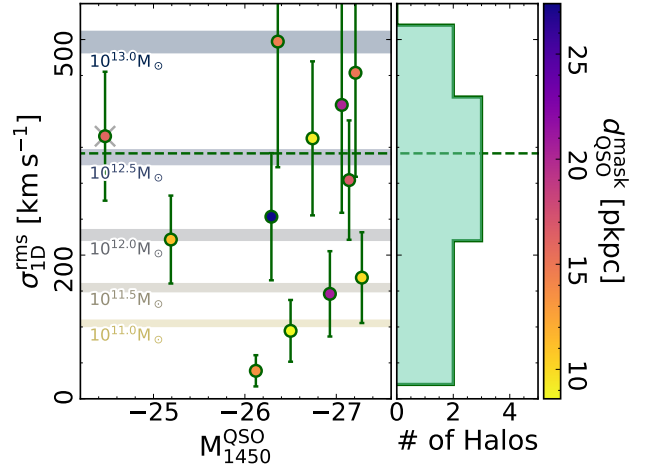

\begin{center}
\includegraphics[width=0.98\columnwidth]{{{sigma}}}
\caption{1D root--mean--square velocity dispersion measured in each detected nebulosity ($\sigma^{\rm rms}_{\rm 1D}$) versus the monochromatic luminosity of the quasars ($M_{\rm 1450}^{\rm QSO}$).
Points are color--coded by the size of the recovered halo ($d_{\rm QSO}^{\rm mask}$).
Horizontal bands are velocity dispersions expected for NFW dark matter halos with masses ranging from 10$^{11}$ to 10$^{13}$\,M$_{\odot}$.
Lower and upper limits of each band correspond to estimates at $z=5.9$ and $z=6.6$, respectively.
Despite the large scatter, the average velocity dispersion of the nebulae (plotted as a green dashed line) is consistent with values expected in a $z\sim6$ dark matter halo of ${\rm M}_{\rm DM}\sim10^{12.5}$\,M$_\odot$.
}\label{fig:mvir}
\end{center}
\end{figure}

We can now compare this value to the masses derived from the velocity dispersions observed in the detected halos.
Indeed, if we assume an NFW \citep[][]{Navarro1997} density profile and the concentration--mass relation presented in \citet{Dutton2014}\footnote{The Planck cosmology \citep{Planck2014} used in \citet{Dutton2014} is different from the one considered in this paper. However, effects of this discrepancy are negligible in the context of our calculations.}, the 1D root--mean--square velocity dispersion ($\sigma^{\rm rms}_{\rm 1D}$) can be directly related with the maximum circular velocity ($V^{\rm Max}_{\rm circ}$) as: $\sigma_{\rm 1D}=V^{\rm Max}_{\rm circ} / \sqrt{2}$ \citep[][]{Tormen1997}.
The average $\sigma^{\rm rms}_{\rm 1D}$ in the \REQUIEM\ sample is $\langle\sigma^{\rm rms}_{\rm 1D}\rangle=\left(340\pm125\right)$\,km\,s$^{-1}$, consistent with the gravitational motion in a ${\rm M_{\rm DM}}\sim10^{12.5}$\,M$_\odot$ halo at $z=6$ (see \autoref{fig:mvir}).
Although most of the detected nebulae are associated with quasars confined to a narrow luminosity range (i.e., between $\M1450\sim-26$\,mag and $-27.5$\,mag), no clear dependency between the velocity dispersion of the nebulae and $\M1450$ is observed (see \autoref{fig:mvir}).
This suggests that the mechanisms responsible for the broadening of the \lya\ line do not depend on the rest--frame UV emission of the central super--massive black hole.

\subsection{The powering mechanism(s) of the extended halos}\label{sec:cgm}

The currently favored mechanism to explain the extended emission observed around quasars is \textit{\lya\ fluorescence}, i.e., the recombination emission following photoionization of cool ($T\sim10^4$\,K) gas by the strong quasar radiation field \citep[e.g.,][]{Hennawi2013, Arrigoni2016, Arrigoni2019, Cantalupo2017}.
In general, if we assume that quasars are surrounded by a population of cool spherical gas clouds, we can directly infer the surface brightness of the fluorescence emission in two limiting regimes:
\begin{itemize}
\item[\textit{(i)}] The gas in the clouds is optically--thick (i.e., with $N_{\rm HI}\gg10^{17.2}$\,cm$^{-2}$).
In this case it is able to self--shield from the quasar's radiation and the \lya\ emission originates from a thin, highly ionized envelope around each individual cloud;
\item[\textit{(ii)}] The gas is optically thin (i.e., with $N_{\rm HI}\ll10^{17.2}$\,cm$^{-2}$) and it is maintained in a highly ionized state by the quasar radiation.
In this case, the \lya\ emission originates from the entire volume of each cloud.
\end{itemize}
In the following we will exploit the formalism presented in \citet{Hennawi2013} to gain insight into the physical status of the gas surrounding the first quasars.

\subsubsection{Optically thick scenario}

\begin{figure*}[tb]
\begin{center}
\includegraphics[width=0.91\columnwidth]{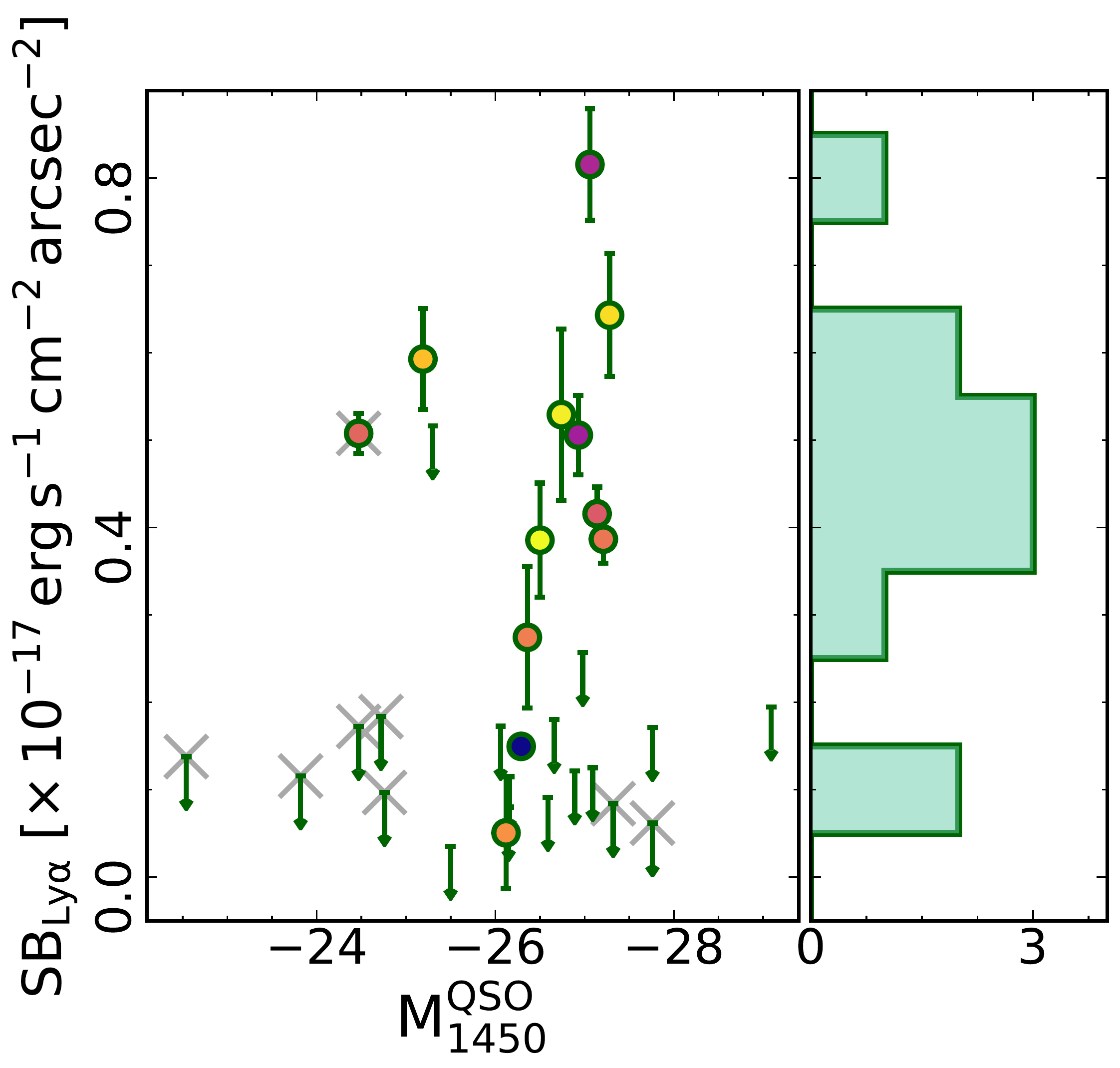}\qquad
\includegraphics[width=1.07\columnwidth]{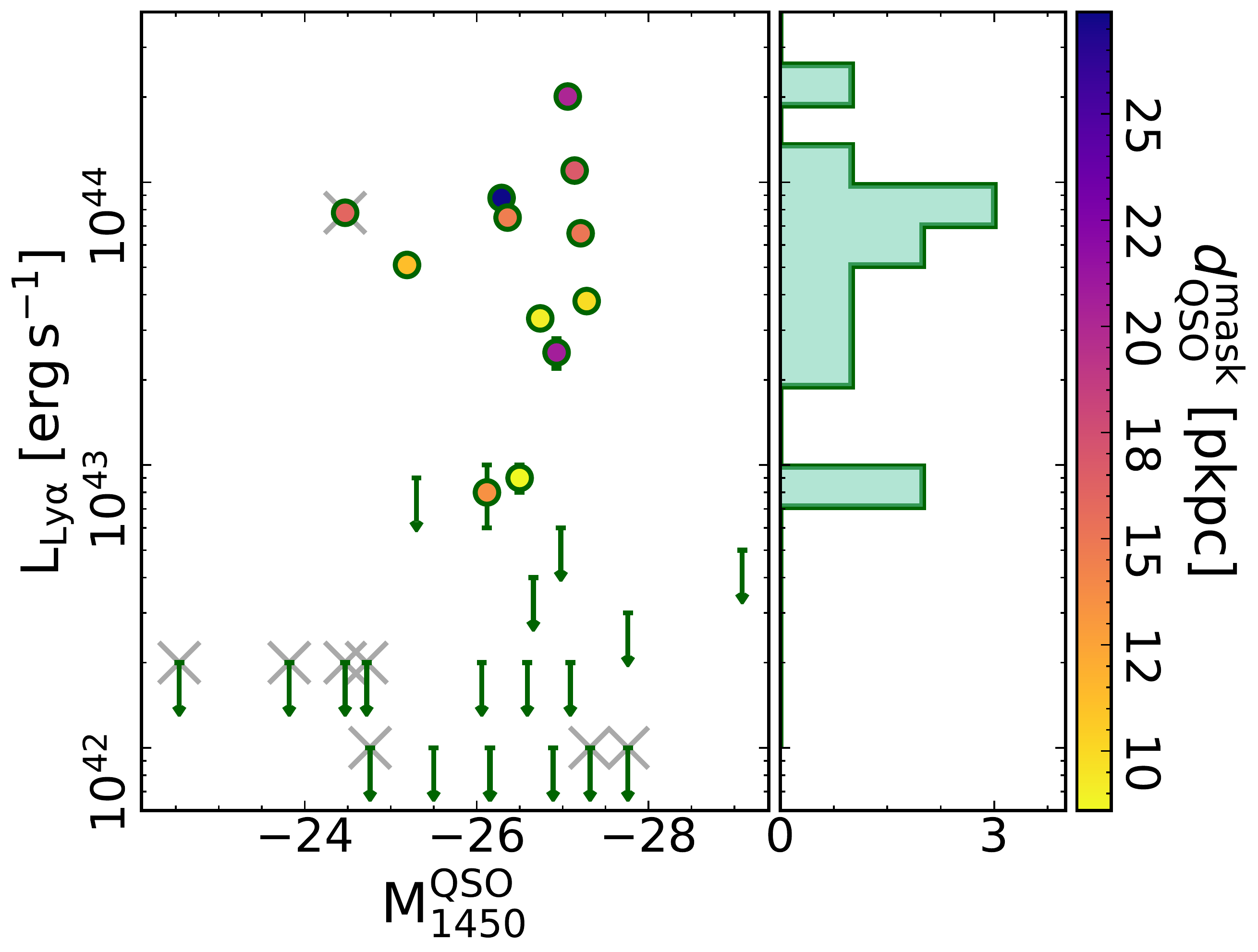}
\caption{\textit{Left Panel} --- Average surface brightness versus $\M1450$.
$3-\sigma$ upper--limits for non--detected nebulae are shown as downward arrows.
The colormap of  the points is the same as in \autoref{fig:sfr}.
For consistency with \autoref{sec:sb}, here the average ${\rm SB}_{\rm Ly\alpha}$ is calculated using a circular aperture of radius $d_{\rm QSO}^{\rm mask}$ (or 20\,pkpc if the halo was not detected) on pseudo--narrow--band images obtained collapsing the datacube between $\pm500$\,km\,s$^{-1}$ relative to the quasar's systemic redshift.
Targets not part of our \textit{core sample} are marked with gray crosses (see \autoref{sec:sample}).
\textit{Right Panel ---}
Same as the left panel but with the total \lya\ luminosity of the extended halo as $y$--axis. 
}\label{fig:sb}
\end{center}
\end{figure*}

If the gas is optically thick, the  \lya\ surface brightness of the extended emission is expected to be proportional to the flux of ionizing photons coming from the central AGN ($\Phi$), to the covering fraction of optically thick clouds ($f^{\rm thick}_{\rm C}$), and to the fraction of incident photons converted into \lya\ by the cloud's envelope \citep[$\eta_{\rm thick}$, see also][]{Hennawi2015, Farina2017, Cantalupo2017}:
\begin{equation}
{\rm SB^{\rm thick}_{\lya}}=\frac{\eta_{\rm thick}h\nu_{\rm \lya}}{4\pi\left(1+z\right)^{4}}f^{\rm thick}_{\rm C}\Phi\left(\frac{R}{\sqrt{3}}\right),
\end{equation}
where we considered the cool gas clouds to be spatially uniformly distributed in a spherical halo of radius $R$.
$\Phi$ can be expressed as a function of the luminosity of the quasar as:
\begin{equation}
\Phi\left(r\right)=\int_{\nu_{\rm LL}}^{\infty}\frac{F_{\nu}}{h\nu}d\nu=\frac{L_{\nu_{\rm LL}}}{4\pi r^2}\int_{\nu_{\rm LL}}^{\infty}\frac{1}{h\nu}\left(\frac{\nu}{\nu_{\rm LL}}\right)^{\alpha_{\rm UV}}d\nu,
\end{equation}
where we considered that, blueward of the Lyman limit ($\nu_{\rm LL}$) the quasar spectral energy distribution has the form $L_{\nu}=L_{\nu_{\rm LL}}\left(\nu/\nu_{\rm LL}\right)^{\alpha_{\rm UV}}$ with $\alpha_{\rm UV}=-1.7$. 
The luminosity at the Lyman edge ($L_{\nu_{\rm LL}}$) can be directly derived from $\M1450$ as: ${\rm M}_{912}=\M1450+0.33$ \citep[see][]{Lusso2015}.
Considering $\eta_{\rm thick}=0.66$, we can thus write:
\begin{multline}\label{eq:sbthick}
\frac{{\rm SB}_{\rm Ly\alpha}^{\rm thick}}{10^{-17}{\rm erg\,s^{-1}\,cm^{-2}\,arcsec^{-2}}}=21.0\left(\frac{1+z}{1+6.2}\right)^{-4}\times\\
\times\left(\frac{f_{\rm C}^{\rm thick}}{0.5}\right)\left(\frac{R}{20\,{\rm pkpc}}\right)^{-2}\left(\frac{L_{\nu_{\rm LL}}}{10^{31}\,{\rm erg\,s^{-1}\,Hz^{-1}}}\right).
\end{multline}
Considering that our \textit{core sample} has an average luminosity at the Lyman edge of $\langle L_{\nu_{\rm LL}}\rangle=2\times10^{31}$\,erg\,s$^{-1}$\,Hz, the optically thick scenario predicts a surface brightness of ${\rm SB}_{\rm Ly\alpha}^{\rm thick}\sim40\times10^{-17}$\,erg\,s$^{-1}$\,cm$^{-2}$\,arcsec$^{-2}$, i.e. $\gtrsim2$ orders of magnitude higher than observed (see \autoref{fig:sb}).
Despite the presence of unknowns such as the geometry of the quasar emission or the covering fraction of optically thick clouds \citep[that may be of the order of 60\% within a projected distance of 200\,pkpc from $z\sim2-3$ quasars, e.g.][]{Prochaska2013a}, this discrepancy points to a different scenario for the origin of the extended \lya\ emission.
The optically thick regime is also disfavored by the absence of a clear correlation between ${\rm SB}_{\rm Ly\alpha}$ and $L_{\nu_{\rm LL}}$ (and thus $\M1450$) as expected from \autoref{eq:sbthick} (see \autoref{fig:sb}).

\subsubsection{Optically thin scenario}

If the quasar radiation is sufficiently intense to keep the gas highly ionized (i.e. if the neutral fraction $x_{\rm HI}=n_{\rm HI}/n_{\rm H}\ll1$), the expected average surface brightness arising from these optically thin clouds is independent of the quasar luminosity and can be expressed as:
\begin{multline}\label{eq:sbthin}
\frac{{\rm SB}_{\rm Ly\alpha}^{\rm thin}}{10^{-19}{\rm erg\,s^{-1}\,cm^{-2}\,arcsec^{-2}}}=3.6\left(\frac{1+z}{1+6.2}\right)^{-4}\times\\
\times\left(\frac{f_{\rm C}^{\rm thin}}{0.5}\right)\left(\frac{n_{\rm H}}{1\,{\rm cm^{-3}}}\right)\left(\frac{N_{\rm H}}{10^{20.5}\,{\rm cm^{-2}}}\right),
\end{multline}
where $f^{\rm thin}_{\rm C}$ is the covering fraction of optically thin clouds, and $n_{\rm H}$ and $N_{\rm H}$ are the cloud's hydrogen volume and column densities, respectively \citep[see][ for further details]{Osterbrock2006, Gould1996, Hennawi2013}.
Assuming photoionization equilibrium allows us to express the neutral column density averaged over the area of the halo ($\langle N_{\rm HI}\rangle$) in terms of $L_{\nu_{\rm LL}}$ and $L_{\rm Ly\alpha}$:
\begin{multline}
\frac{\langle N_{\rm HI}\rangle}{10^{17.2}\,{\rm cm}^{-2}}=0.1\left(\frac{L_{\rm Ly\alpha}}{10^{44}\,{\rm erg\,s^{-1}}}\right)\times\\
\times\left(\frac{L_{\nu_{\rm LL}}}{10^{31}\,{\rm erg\,s^{-1}\,Hz^{-1}}}\right)^{-1}.
\end{multline}
Given the observed luminosities of the nebulae in our core sample ($L_{\rm Ly\alpha}\lesssim10^{44}$\,erg\,s$^{-1}$, see \autoref{tab:spechalo}) we obtain $\langle N_{\rm HI}\rangle\ll10^{17.2}$\,cm$^{-2}$, consistent with the optically thin regime.
However, we stress that $\langle N_{\rm HI}\rangle$ is obtained by averaging over the whole area of the halo. So, while $\langle N_{\rm HI}\rangle\gg10^{17.2}$\,cm$^{-2}$ definitively determines the clouds to be optically thick, a small value of $\langle N_{\rm HI}\rangle$ does not provide the same clear result, since individual clouds may still be optically thick while being surrounded by a thinner medium.

Under the assumption that the clouds are optically thin, it is of interest to use \autoref{eq:sbthin} to derive constraints on the gas volume density ($n_{\rm H}$).
Studies of absorption systems associated with gas surrounding $z\sim2$ quasars suggest that $N_{\rm H}$ is almost constant within an impact parameter 200\,pkpc at a median value of $N_{\rm H}=10^{20.5}$\,cm$^{-2}$ \citep[e.g.][]{Lau2016}.
If $z\sim6$ quasars are embedded in halos with similar hydrogen column densities, our observations imply $n_{\rm H}>1$\,cm$^{-3}$.
Intriguingly, similarly high gas densities have been invoked to explain the \lya\ emission in giant nebulae discovered around $z\sim2-3$ quasars \citep[][]{Cantalupo2014, Hennawi2015, Arrigoni2015HeII, Arrigoni2018, Cai2018}.

\subsubsection{Other possibilities}

In addition, other mechanisms have been proposed to explain the presence of extended \lya\ nebulae including gravitational cooling radiation \citep[e.g.,][]{Haiman2000, Fardal2001, Furlanetto2005, Dijkstra2009}, shocks powered by outflows \citep[e.g.,][]{Taniguchi2000, Mori2004}, or resonant scattering of \lya\ photons \citep[e.g.][]{Gould1996, Dijkstra2008}.

However, \lya\ emission coefficients for collisional excitation are exponentially dependent on the temperature \citep[][]{Osterbrock2006}.
The concurrence of a very narrow density and temperature range for all the gas in every observed \lya\ nebula would thus be necessary to validate this.
Instead, recombination radiation has a much weaker dependence on temperature \citep[][]{Osterbrock2006}, providing a more natural explanation for the \lya\ extended emission in the presence of a strong ionizing flux \citep[e.g.,][]{Borisova2016}.
In addition, the relatively quiescent motion of the gas in the detected halos (see \autoref{sec:kine}) is not easily reconciled with shock--powered emission \citep[see also discussion in][]{Arrigoni2019}.

On the other hand, resonant scattering of \lya\ photons from the central AGNs and from young stars in the host galaxies can provide a relevant contribution to the emission, if the gas is optically thick at the \lya\ transition ($N_{\rm HI}\gtrsim10^{14}$\,cm$^{-2}$).
This was proposed as the main process powering the extended \lya\ emission detected around $3\lesssim z\lesssim6$ \lya\ emitters by \citet{Wisotzki2016}.
\citet{Hennawi2013} showed that the surface brightness of extended \lya\ emission produced via resonant scattering by neutral gas in the CGM (${\rm SB}_{\rm Ly\alpha}^{\rm scatt.}$) is expected to be directly proportional to the flux of ionizing photons emitted close to the \lya\ resonance ($L_{\rm \nu Ly\alpha}$).
Given that the peak of the \lya\ line of $z\sim6$ quasars is typically absorbed by neutral hydrogen, there is no direct way to test for the presence of such a correlation in the \REQUIEM\ survey.
In any case, \citet{Arrigoni2019} reported the lack of significant correlation between the surface brightness of \lya\ halos and the luminosity of the peak of the \lya\ line of $z\sim3$ quasars.
In addition, we do not detect clear signals of the characteristic double--peaked profiles expected for resonantly trapped \lya\ photons \citep[e.g.,][]{Dijkstra2017}.
However, a detailed analysis of the \lya\ line shape performed on high signal--to--noise, high spectral resolution spectra is required to properly test this scenario.

We stress that all the aforementioned mechanisms can be in place at the same time and contribute at different levels to the observed emission.
Additional factors can also modulate the total luminosity of the halos. For instance, the presence of dust on scales larger than 20\,kpc \citep[e.g.][]{Roussel2010, Menard2010} can destroy \lya\ photons, and/or the variability of the quasar emission \citep[e.g.,][]{MacLeod2012, Yang2019} can be faster than the response of the halo (with a strong dependence on $n_{\rm H}$) and wash out some of the expected correlations.
Future observations of non--resonant lines such as \heii\ or \ha\ will be instrumental in disentangling different emission mechanisms \citep[e.g.][]{Arrigoni2015HeII, Leibler2018, Cantalupo2019}.
This is particularly challenging at $z>6$, where only space--based observations will have the sensitivity necessary to provide additional information about the gas surrounding the first quasars.

\subsection{\lya\ nebulae and galaxy overdensities}

Several giant \lya\ nebulae extending on scales $\gg100$\,kpc have recently been reported in the literature \citep[][]{Cantalupo2014, Hennawi2015, Cai2018, Arrigoni2018, Arrigoni2019Filament, Lusso2019}.
The incidence of such large nebulae has been estimated to be of the order of few percent at $z\sim2-3$ \citep[][]{Hennawi2013, Arrigoni2016, Arrigoni2019}.
A larger sample of $z\gtrsim6$ quasars is necessary to assess if this low occurrence holds at high redshifts.
In any case, all $z<4$ giant nebulae appear to be invariably associated with overdensities of AGN and galaxies, suggesting a connection between proto--cluster structures and extremely extended emission \citep[e.g.,][, but see \citealt{Buadescu2017} for examples of large \lya\ blobs located at the outskirts of high--density regions]{Hennawi2015, Arrigoni2018Mammoth}.

We can qualitatively test this scenario at $z>6$, by searching for peculiarities in the nebulae associated with quasars for which deep ALMA observations have revealed the presence of bright \ciimu\ companions \citep[i.e., J0305$-$3150, P231$-$20, P308$-$21, and J2100$-$1715; see \autoref{sec:sample} and][]{Decarli2017, Willott2017, Venemans2019}\footnote{We note that for all these quasars MUSE observations have been gathered with integration times $>3\times$ longer than the median of our sample (see \autoref{tab:sample}).}.
J2100$-$1715 has a \ciimu\ companion located at a separation of $\sim60$\,kpc \citep[][]{Decarli2017, Neeleman2019} but does not show the presence of any significant extended \lya\ emission.
J0305$-$3150, although located in an overdensity with three \ciimu\ and one LAE emitter \citep[][]{Farina2017, Venemans2019}, shows a really faint halo.
Finally, P231$-$20 and P308$-$21 host among the brightest and most extended halos in the \REQUIEM\ survey and both are in the middle of gravitational interactions with their companions \citep[][]{Decarli2019, Neeleman2019}.

The variety of the \lya\ emission observed in this (small) sample of quasars with companions suggests that at $z>6$, bright and extended \lya\ halos may be associated with ongoing merger events.
If this is the case, P323$+$12, that exhibits the brightest nebular emission in our sample, is likely to be in a gravitational interaction.
\citet{Mazzucchelli2017} presented low resolution \textit{NOEMA} observations on the \ciimu\ emission line of this source, without detecting a merger.
However testing the merger scenario requires a higher sensitivity and better spatial resolution, for example with \textit{ALMA} or \textit{NOEMA}.

\subsection{The average surface--brightness profile}\label{sec:avesbprofile}

In this section we will infer the average surface brightness profile of the \lya\ emission around quasars in the \REQUIEM\ survey and we will compare its shape with lower redshift studies.
To avoid selection effects, we will focus only on our \textit{core sample}.
We remind the reader that this consists of 23~radio--quiet quasars at an average redshift of $\langle z \rangle=6.27$ and absolute magnitude ranging from $\M1450=-25.2$\,mag to $-29.1$\,mag, with an average of $\langle \M1450 \rangle=-27.1$\,mag (see \autoref{sec:sample}).
As a lower redshift comparison we will use the following studies (see \autoref{fig:comparison}):
\textit{(i)} \citet{Cai2019}, who investigated with KCWI 16~quasars at $2.20<z<2.38$ (with an average of $\langle z \rangle=2.27$) and absolute magnitude between $\M1450=-29.0$\,mag and $-26.1$\,mag (with $\langle \M1450 \rangle=-27.3$\,mag);
\textit{(ii)} \textit{QSO~MUSEUM} \citep{Arrigoni2019}, a MUSE investigation for extended \lya\ emission around a sample of 61~quasars at $3.03<z<3.46$ (with $\langle z \rangle=3.21$) with absolute magnitudes in the range $-28.3\,{\rm mag}<\M1450<-25.6$\,mag (with $\langle \M1450 \rangle=-27.2$\,mag); and
\textit{(iii)} \citet{Borisova2016}, who explored with MUSE the vicinity of 19 bright ($-29.0\,{\rm mag}<M1450<-26.8\,{\rm mag}$, and  $\langle \M1450 \rangle=-28.0$\,mag) quasars at $3.02<z<3.91$ (with $\langle z \rangle=3.36$)\footnote{The average surface brightness profile for the \citeauthor{Borisova2016} sample is presented in \citet{Marino2019}.}.

\begin{figure}[tb]
\begin{center}
\includegraphics[width=0.98\columnwidth]{{{zmALL}}}
\caption{
Redshift vs. $\M1450$ distribution of the quasars part of the surveys from \citet[][, 16 quasars, magenta circles and histograms]{Cai2019}, \citet[][, 61 quasars, light blue down--pointing triangles and histograms]{Arrigoni2019}, and \citet[][, green right--pointing triangles and histograms]{Borisova2016}.
The 23 quasars part of our \textit{core sample} are shown as orange diamonds and histograms.
Histograms are normalized by the total number of targets and by the bin size (with steps $0.3$ in redshift and of $0.3$\,mag in absolute magnitude).
The average absolute magnitude of the \citeauthor{Cai2019}, \citeauthor{Arrigoni2019}, and our \textit{core} sample are nearly identical, while the quasars studies in \citeauthor{Borisova2016} are, in average, $\sim0.7$\,mag brighter.
}\label{fig:comparison}
\end{center}
\end{figure}

\begin{figure*}[tb]
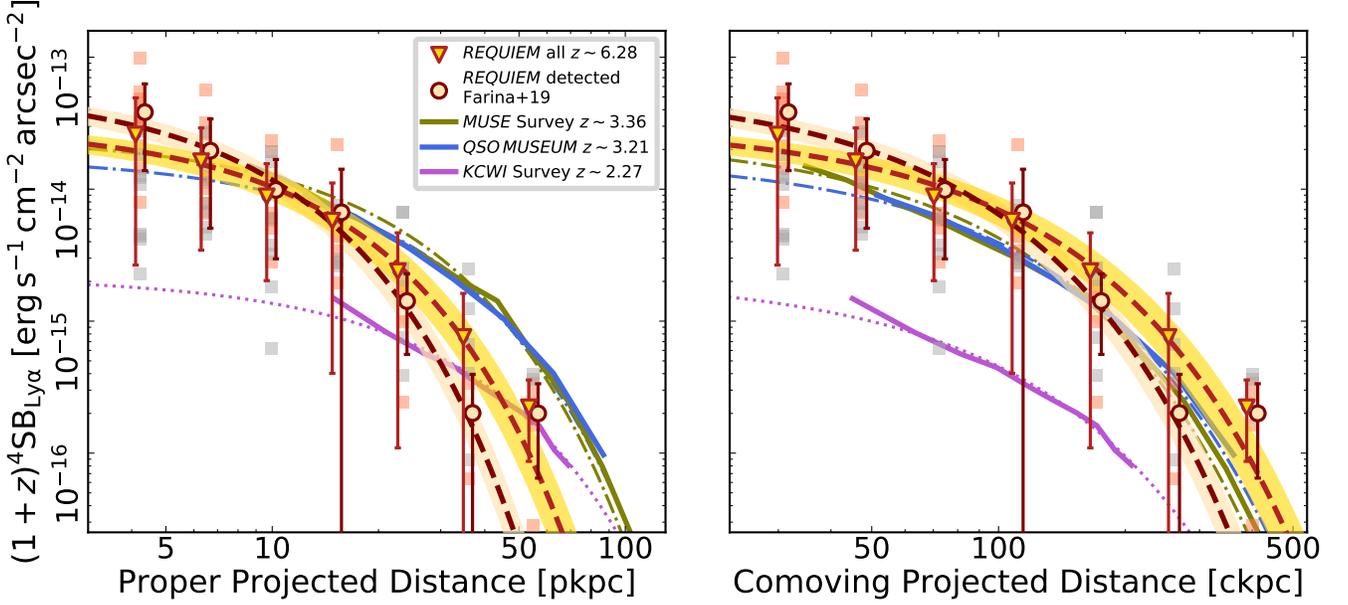

\begin{center}
\includegraphics[width=0.98\textwidth]{{{sbProfile}}}
\caption{
\textit{Left Panel --} Average \lya\ surface brightness profile of the \textit{core sample} of our \REQUIEM\ survey.
We show the individual circularly--averaged profiles extracted around each quasar as orange squares for detected halos and gray squares for non--detections.
Sum--averaged profiles considering all quasars (orange triangles) or only significantly detected halos (purple circles) are also shown, together with their respective best exponential fit (orange and purple dashed lines, shaded regions provide 1--$\sigma$ uncertainties in the fit).
Data--points have been slightly shifted along the $x$--axis for the sake of clarity.
The average surface brightness profile of the extended emission around quasars at $z\sim2.27$ \citep[magenta line,][]{Cai2019}, $z\sim3.21$ \citep[light blue line,][]{Arrigoni2019}, and $z\sim3.36$ \citep[olive green line,][]{Marino2019} are also plotted.
The corresponding exponential best fits are shown in the same color palette.
Note that all measurements are corrected for the $(1+z)^4$ factor to compensate for cosmological dimming.
\textit{Right Panel --} Same of Left Panel, but in comoving units.
}\label{fig:sbprofile}
\end{center}
\end{figure*}

The extended nebulae detected in the \REQUIEM\ survey appear to have complex morphologies and clear asymmetries (see \autoref{fig:psfsub}).
We proceeded following the standard approach in the literature and we obtained the surface brightness profiles averaging over circular apertures centered on the location of the quasars.
As explained in \autoref{sec:sb}, single profiles were extracted from pseudo--narrow band images created by collapsing the datacubes over 30\,\AA\ centered at the location of the \lya\ line, redshifted to the quasar's systemic redshift.
To create the stacked profile, we first correct these profiles for cosmological dimming 
(i.e., by a factor $(1+z)^4$) and then we average over them with equal weights.
This prevents the introduction of biases towards deeper exposures and/or brighter objects
(the marginal variations caused by the use of the median to combine the different radial profiles is discussed in \autoref{sec:median}).
We also create the stacked profile only using the sub--sample of quasars for which an extended emission has been detected with significance.
The results of this procedure are plotted in \autoref{fig:sbprofile}, where the average surface brightness profile obtained for all quasars is shown as orange triangles and the one from quasars embedded in halos as purple circles (the average radial profile for the entire \textit{core sample} is tabulated in \autoref{tab:sbprofile} in \autoref{sec:median}).
For comparison, the average profile from \citet{Cai2019}, \citet{Arrigoni2019}, and \citet{Marino2019} are displayed as magenta, light blue, and olive solid lines, respectively.

In order to extract information from the stacked profiles, we perform a fit with an exponential function: $(1+z)^4\,{\rm SB}_{\rm Ly\alpha}\left(r\right)=C\exp{\left(-r/r_h\right)}$, where $C$ is the normalization and $r_h$ is the scale length of the profile.
The resulting parameters are $C=(3.0\pm0.4)\times10^{-14}$\,erg\,s$^{-1}$\,cm$^{-2}$\,arcsec$^{-2}$ and $r_h=(9.4\pm0.8)$\,kpc for the full sample and $C=(5.6\pm0.8)\times10^{-14}$\,erg\,s$^{-1}$\,cm$^{-2}$\,arcsec$^{-2}$ and $r_h=(6.4\pm0.3)$\,kpc for the stack quasars with detected halos.
As expected, while the two profile match within the errors, the latter appears slightly more concentrated due to the stronger signal in the central $\sim20$\,kpc.
In the following, we will keep showing both profiles, however in the discussion we will focus solely on the one that includes all quasars as it is more representative of the full high--$z$ quasar population.

The scale length derived for $z\sim6.2$ quasars is a factor of $\sim2\times$ smaller than the $r_{h}=(15.5\pm0.5)$\,pkpc and $r_{h}=(21.1\pm0.9)$\,pkpc measured for radio--quiet quasars at $z\sim3.2$ \citep{Arrigoni2019} and at $z\sim2.3$ \citep{Cai2019}, suggesting that extended halos are more compact at higher redshift.
For comparison, the sample of \lya\ emitters in the \textit{Hubble} Ultra Deep Field shows a much milder evolution of halo scale length, increasing from $r_h=(3.8\pm1.3)$\,pkpc at $5< z\lesssim6$ to $r_h=(4.4\pm1.5)$\,pkpc at $3\lesssim z<4$
\citep{Leclercq2017}.
However, we should note that, given the difference in apparent brightness of the quasars, the cosmological evolution of the angular diameter distance, and the factor $\sim8\times$ in sensitivity due to redshift dimming, our observations are more sensitive to regions closer to the quasar while \citet{Cai2019}, \citet{Arrigoni2019}, and \citet{Borisova2016} are more sensitive to extended emission at larger scales.

\citet{Arrigoni2019} reported a strong evolution of the average properties of the extended emission with cosmic time.
This was based on the comparison of the average surface brightness radial profiles of their quasars split into a $z\sim3.3$ and a $z\sim3.1$ sub--samples and the results obtained from a narrow--band survey of bright radio--quiet quasars at $z\sim2.2$ \citep[][]{Arrigoni2016}.
Recently, \citet{Cai2019} showed that studies based on narrow--band imaging underestimated the total nebular emission of an order of magnitude \citep[see also Discussion in][]{Borisova2016}.
The new IFU observations revealed a less pronounced evolution, with halos surrounding quasars at $z\sim2$ being $\sim0.4$\,dex fainter than at $z\sim3$ \citep{Cai2019}.
In the following, we will test if this trend holds out to the redshifts provided by the \REQUIEM\ survey.

In the optically thin scenario, the \lya\ surface brightness scales as ${\rm SB_{\lya}} \propto n_{\rm H}N_{\rm H}$ (see \autoref{eq:sbthin}).
If the gas clouds are bound to the dark matter halo hosting the quasar (see \autoref{sec:kine}), it can be shown that $N_{\rm H}\propto n_{\rm H}R_{\rm vir}$, where $R_{\rm vir}$ is the virial radius \citep[see also ][, for a similar argument applied to \mgii\ absorbers in the CGM of $z<1$ galaxies]{Churchill2013Vir}.
In addition, given the inferred high densities ($n_{\rm H}>1$\,cm$^{-3}$), the emitting gas is not likely to trace the evolution of the cosmic mean density.
We thus expect the size of the nebular emission to scale with the growth of $R_{\rm vir}$ with cosmic times.
For the sake of simplicity, we will consider that quasars are hosted by massive halos with $M_{\rm DM}=10^{12.5}$\,M$_\odot$ independent of their redshift \citep[see, e.g., discussion in][]{Shen2007, He2018}. 
Thus the virial radius $R_{\rm vir}\left(z\right)=\left[3M_{\rm DM}/4\pi\rho_{\rm c}\left(z\right)\right]^\frac{1}{3}$ depends only on the critical density of the Universe as a determined redshift [$\rho_{\rm c}\left(z\right)$].
For the considered halo, the virial radius calculated at $z=6.28$ increases of a factor $\sim1.6\times$ down to $z=3.36$ and of a factor $\sim2.2\times$ down to $z=2.27$.

\begin{figure}[tb]
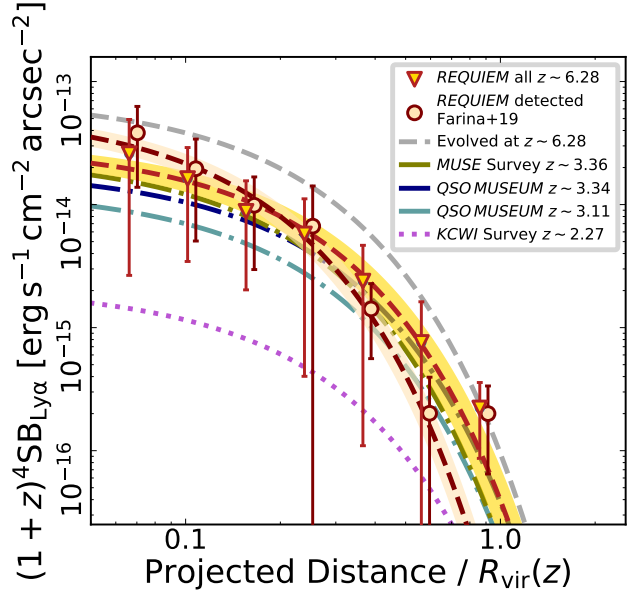

\begin{center}
\includegraphics[width=0.98\columnwidth]{{{sbProfileEvolution}}}
\caption{
Average, cosmological dimming corrected, surface brightness profiles of the extended emission around quasars with the radius normalized to the virial radius R$_{vir}\left(z\right)$ of a $10^{12.5}$\,M$_\odot$ dark matter halo located at different redshifts.
The color code of data--points and curves is the same as in \autoref{fig:sbprofile}.
In addition, following \citet[][]{Arrigoni2019}, we split quasars part of the \textit{QSO~MUSEUM} in two sub--samples with median redshifts $z=3.11$ and $z=3.34$.
The corresponding exponential fit of the surface brightness profiles are plot as light and dark blue dashed--dotted line, respectively.
The gray dashed line shows the expected profile if the evolutionary trend observed between $z=2.27$ and $z=3.36$ is extrapolated linearly in redshift to $z=6.28$.
}\label{fig:sbevolution}
\end{center}
\end{figure}

In \autoref{fig:sbevolution} we show the average surface brightness profiles at $z=6.28$, $3.36$, $3.34$, $3.11$, and $2.27$ normalized by the virial radius.
At all the these redshifts, the emitting gas appears to be located well within $R_{\rm vir}$ and the average profile becomes brighter at higher redshifts (with the $z\sim3$ and $z\sim6$ profiles being consistent within the scatter).
If we model the profiles normalized by the virial radius with an exponential function: $(1+z)^4\,{\rm SB}_{\rm Ly\alpha}\left(r\right)=C_h\exp{\left(-x/x_h\right)}$ with $C_h$ and $x_h$ as free parameters and $x=r/R_{\rm vir}$, we obtain: $C_h=(3.0\pm0.4)\times10^{-14}$\,erg\,s$^{-1}$\,cm$^{-2}$\,arcsec$^{-2}$ and $x_h=(0.15\pm0.01)$ for the full quasar sample and $C_h=(5.6\pm0.8)\times10^{-14}$\,erg\,s$^{-1}$\,cm$^{-2}$\,arcsec$^{-2}$ and $x_h=(0.13\pm0.01)$\,kpc for the sub--sample of detected halos.
The value of $x_h$ measured at $z\sim6.28$ matches the $x_h=0.14$, $0.16$, $0.15$, and $0.16$ estimated for quasars at $z\sim3.36$, $z\sim3.34$, $z\sim3.11$, and $z\sim2.27$ respectively.
This suggests a scenario where the (average properties of the) extended \lya\ emission mirrors the cosmic evolution of the dark matter halos they reside in (see also \autoref{sec:kine}).
On the other hand, $C_h$ rapidly increases with redshift from $z\sim2$ to $z\sim3$ and grows much gradually between $z\sim3$ and $z\sim6$.
This behavior is described by the gray dashed line in \autoref{fig:sbevolution}.
This is the expected average profile of the extended emission if the increase of the normalization observed between $z\sim2.27$ and $z\sim3.36$ would keep its pace linearly with redshift up to $z\sim6.28$.
The observed profile from the \REQUIEM\ survey lies $\sim0.4$\,dex below this prediction.

Intriguingly, hydrodynamical cosmological simulations show that high--$z$ galaxies in dark matter halos of $M_{\rm DM}\gtrsim10^{12}$\,M$_\odot$ are mainly fed by cool gas streams (co--existing with a hot, shocked medium) down to redshift $z\sim2-3$.
Below this ``critical'' redshift, these cool streams are not able to balance the virial shock--heating and are suppressed \citep[e.g.,][]{Dekel2006, Dekel2009, Dekel2019}.
This theoretical picture would naturally explain the observed evolution of the surface brightness profile. 
Under the assumption that the emission arises from optically thin clouds of gas, the surface brightness is expected to scale as: ${\rm SB_{\lya}}\propto(1+z)^4f_{\rm C}^{\rm thin}n_{\rm H}N_{\rm H}$ and $N_{\rm H}$ is proportional to the total mass in cool gas $M_{\rm cool}$ \citep[see \autoref{sec:cgm} and][]{Hennawi2013, Arrigoni2019}.
Thus, the small variation in \lya\ surface brightness reported between $z\sim6$ and $z\sim3$ suggests that cool streams are able to replenish the CGM with gas, permitting $M_{\rm cool}$ to keep pace with $M_{\rm DM}$.
The consequent heating of massive halos at $z\lesssim3$ may be responsible for the drop of cool gas in the CGM and, thus, of the average \lya\ emission.
However, this picture clashes with the large amount of cool gas revealed by absorption studies of the CGM around $0.5\lesssim z\lesssim 2$ quasars \citep[e.g.,][]{Bowen2006, Farina2013, Farina2014, Prochaska2014, Johnson2015}.

\subsection{Extended emission and quasar near zones}\label{sec:nearzones}

Near the end of the cosmic reionization epoch, the large fraction of neutral hydrogen present in the Universe suppresses virtually all the emission blueward of the \lya\ line \citep[e.g.,][]{Gunn1965}.
However, the intense radiation of a luminous quasar is able to ionize the surrounding gas, creating a bubble of enhanced transmission in the \lya\ forest in its immediate vicinity known as the proximity (or near) zone \citep[e.g.,][]{Cen2000, Madau2000, Haiman2001}.

The size of the proximity zone ($R_{\rm p}$) is traditionally defined as the distance out to which the Ly$\alpha$ transmission, smoothed on a scale of 20{\AA}, first falls below 10\% \citep{Fan2006}.
Absorption studies of the first quasars revealed that proximity zones can extend to several proper megaparsecs \citep[][]{Fan2006, Carilli2010, Venemans2015, Eilers2017}, i.e. on scales much larger than the one probed by the extended \lya\ emission around quasars (typically $\lesssim$100\,kpc).
Proximity zones are thus expected to be more sensitive to the state of the IGM than to the quasar's CGM \citep[e.g.,][]{Fan2006}.

The interpretation of the magnitude of $R_{\rm p}$ depends on the neutral fraction of the surrounding IGM.
If the gas is close to fully neutral, the size of the resulting proximity zone will reflect the ionized bubble carved out by the quasar:
\begin{equation}\label{eq:near}
R_{\rm p}=\left(\frac{3\dot{N}_{\rm ion}t_{\rm QSO}}{4 \pi n_{\rm H} x_{\rm HI}}\right)^{\frac{1}{3}},
\end{equation}
where $\dot{N}_{\rm ion}$ is the emission rate of the ionizing photons and $t_{\rm QSO}$ is the age of the quasars.
Alternatively, if the gas is already mostly ionized, the proximity zone size instead reflects the distance out to which the ionizing flux from the quasar is enough to keep the \lya\ forest sufficiently transparent \citep[][]{Bolton2007}.
The size of the proximity zone can still vary with the age of the quasar if the quasar is sufficiently young, such that the IGM gas has not yet reached photoionization equilibrium with the newly enhanced ionizing flux \citep[][]{Eilers2017, Eilers2018, Davies2019}.

Once the dependence on the quasar luminosity is removed, the typical size for the ``corrected'' near zone of a $z\sim6$ quasar is $R_{\rm p}^{\rm corr}=R_{\rm p}\times10^{-0.4(-27.0-\M1450)/2.35}\sim5$\,pMpc.
Deviations below $R_{\rm p,corr}\sim3$\,pMpc should be very rare unless non--equilibrium photoionization is at play \citep{Eilers2017, Davies2019}.
Thus, the discovery of quasars with exceptionally small proximity zones and no evidence for significantly neutral gas by \citet{Eilers2017} implies that these objects must have been shining for $t_{\rm QSO}\lesssim10^5$\,yr (see \citealt{Eilers2018} and \citealt{Davies2019} for further details).

Interestingly, this value is comparable to the light crossing time given the size of the nebulae observed in our survey: $t_{\rm cross}=d_{\rm QSO}/c\sim3\times10^4$\,yr for $d_{\rm QSO}=10$\,pkpc.
This suggests that if a quasar in our sample has a peculiarly small $R_{\rm p}^{\rm corr}$, then its extended emission should also be small or not--existent \citep[see Discussion in][]{Eilers2018}.
We remind the reader that the recombination time scales as $t_{\rm rec}=1/n_e\alpha_A$ where the electron density can be calculated as $n_e=n_{\rm H}\left(1+Y/2X\right)$ (assuming that all helium is doubly ionized) and the case~A recombination coefficient evaluated at $T=10^{5}$\,K is $\alpha_A=4.2\times10^{-13}$\,cm$^{3}$\,s$^{-1}$ \citep{Osterbrock2006}.
Thus, for the volume densities inferred in \autoref{sec:cgm}, we obtain values of $t_{\rm rec}\lesssim10^{5}$\,year.
This means that, in case of intermittent quasar activity, the extended \lya\ halo would disappear if the time scale in which the quasar is inactive is $t_{\rm off}>t_{\rm rec}$.

We can test this scenario within our \REQUIEM\ survey.
Indeed, our targets overlap with the \citet{Eilers2017} sample in six quasars, two of which have $R_{\rm p,corr}\lesssim3$\,pMpc:
J2229$+$1457 with $R_{\rm p}^{\rm corr}=(1.07\pm0.33)$\,pMpc, and
J0100$+$2802 with $R_{\rm p}^{\rm corr}=(3.09\pm0.06)$\,pMpc\footnote{
The proximity zone sizes are calculated from ESI/\textit{Keck\,II} spectra considering the \citet{Planck2014} cosmology.
$R_{\rm p}^{\rm corr}$ values for the remainder quasars part of both the \citeauthor{Eilers2017} and the \REQUIEM\ survey samples are:
J0210$-$0456 with $R_{\rm p}^{\rm corr}=(3.47\pm0.34)$\,pMpc,
J2329$-$0301 with $R_{\rm p}^{\rm corr}=(4.86\pm0.70)$\,pMpc,
J1030$+$0524 with $R_{\rm p}^{\rm corr}=(5.95\pm0.36)$\,pMpc,
J2054$-$0005 with $R_{\rm p}^{\rm corr}=(4.32\pm0.19)$\,pMpc.
}.
It is alluring that none of these young quasar candidates show an extended \lya\ halo.
Deeper MUSE observations complemented with sensitive NIR spectroscopy aimed to confirm the true nature of these small zones \citep[e.g.,][]{Eilers2018} will provide new information on the nature of this class of objects (Eilers et al.\ in prep.).

\subsection{Is the halo around P323$+$12 lensed?}\label{sec:lens}

\begin{figure}[tb]
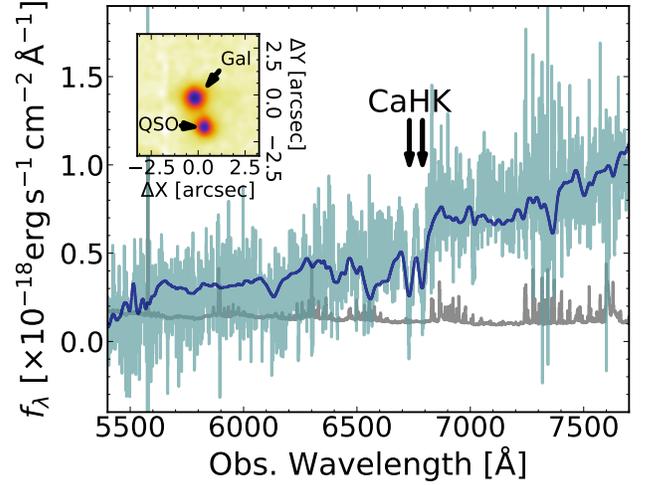

\begin{center}
\includegraphics[width=0.98\columnwidth]{{{LensingGalaxy}}}
\caption{Spectrum of the foreground galaxy located 1\farcs57 North North--East of the optical position of the quasar P323$+$12 (light blue).
The elliptical galaxy template from \citet{Mannucci2001} redshifted to $z_{\rm gal}=0.711$ and rescaled to the flux observed in the Pan-STARRS $i$--band $i_{\rm PS1,\,gal}=23.29$\,mag is shown as a dark blue line (see \autoref{sec:lens} for details).
The inset plots the location of the galaxy with respect to the quasar in the pseudo--broad--band image obtained by collapsing the MUSE cube between 8200\AA\ and 9200\AA.
}\label{fig:lens}
\end{center}
\end{figure}

The procedure to find and remove low--redshift contaminants described in \autoref{sec:psfsubtraction} revealed the presence of a galaxy located within the bright halo detected in association with the quasar P323$+$12, i.e. 1\farcs6 NNE from the quasar at RA$_{\rm J2000}$=21:32:33.22 and Dec.$_{\rm J2000}$=+12:17:56.8 (see \autoref{fig:lens}).
This galaxy is spatially resolved in deep near--infrared imaging collected with \textit{LBT}/LUCI2$+$ARGOS (see \autoref{sec:luci}).
The detection of the Ca\,H\&K~$\lambda\lambda$3969,3934 (hereafter Ca\,HK, see \autoref{fig:lens}) in the galaxy spectrum determined its redshift at $z_{\rm gal}=0.711\pm0.001$.
In the following we check the possibility that this galaxy could act as a lens and thus enhance the total luminosity observed for this halo.

The expected radius of the Einstein ring ($\theta_{\rm E}$) can be estimated by assuming the potential well of the galaxy is well described by a singular isothermal sphere (SIS).
This allows us to directly relate $\theta_{\rm E}$ to the velocity dispersion of the SIS ($\sigma_{\rm SIS}$), to the angular diameter distance between the halo and the lens ($D_{\rm LH}$), and to the angular diameter distance between the halo and the observer \citep[$D_{\rm H}$, e.g,][]{Narayan1990, Peacock1999, Chieregato2007}:
\begin{equation}\label{eq:lens}
\theta_{\rm E} = 4\pi\left(\frac{\sigma_{\rm SIS}}{c}\right)^{2}\frac{D_{\rm LH}}{D_{\rm H}}
\end{equation}
Given the relatively low $S/N$ per pixel and spectral resolution (i.e. $R=\lambda/\Delta\lambda\sim2700$ at $\sim6800$\,\AA) of the spectrum in our data, we assume that $\sigma_{\rm SIS}=\sigma_\star$ (i.e. the velocity dispersion of the galaxy) and we infer $\sigma_\star$ from the Faber--Jackson relation \citep[][]{Faber1976}.
Using the updated relation from \citet{Nigoche2010}, a galaxy with an $r$--band absolute magnitude of $M_r=-20.44$\,mag has a $\sigma_\star=100$\,km\,s$^{-1}$.
Plugging these values in \autoref{eq:lens}, we obtain $\theta_{\rm E}\sim0\farcs2$.
This is well below the current spatial resolution of our MUSE observations (see \autoref{tab:sample}) and thus our measurements are not significantly biased by lensing.

Given the estimated size of the Einstein ring, the new AO system \textit{GALACSI} on \textit{MUSE} \citep[][]{Stuik2006} should be able to resolve it.
Future high spatial resolution observations of this system will allow us to investigate the extended halo of a $z\sim6.6$ quasar in unprecedented detail.

\section{SUMMARY AND CONCLUSIONS}\label{sec:conclusions}

We conducted a sensitive search for extended \lya\ emission around a sample of 31 $5.7<z<6.6$ quasars spanning absolute magnitudes from $\M1450=-22.5$\,mag to $\M1450=-29.1$\,mag.
This ongoing \textit{VLT}/MUSE effort represents the first statistical study of the circum--galactic medium of quasars during the epoch of reionization (see \autoref{fig:zLum}).
After subtracting the contribution of the central AGN, we unveil the presence of significant extended \lya\ emission around 12 targets.
The detected nebulosities extend out to $\sim30$\,kpc from the quasars and show a variety of morphologies and physical properties.
The study of these systems reveals that:

\begin{figure}[tb]
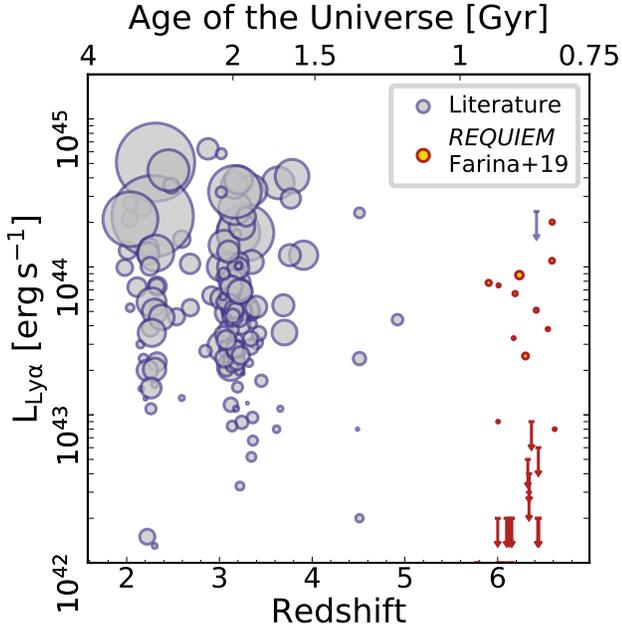

\begin{center}
\includegraphics[width=0.98\columnwidth]{{{zLum}}}
\caption{Redshift vs. total luminosity of all extended \lya\ nebulae associated with QSOs known to date.
Gray and orange points are data from the literature and from the \REQUIEM\ survey, respectively.
The size of the point is proportional to the area (in pkpc$^{2}$) covered by each halo.
Downward arrows are 3--$\sigma$ upper limits estimated integrating the nominal surface brightness limits over circular apertures with radius 20\,pkpc.
All values are uniformed to the concordance cosmology used in this paper (see \autoref{tab:allhalos} in \autoref{sec:allhalos}).
However, effects of different sensitivities and observing technique are not taken into account.
}\label{fig:zLum}
\end{center}
\end{figure}

\begin{enumerate}
\item[(i)] The redshift of the extended emission well aligns with the systemic redshift of the quasar host--galaxies traced by the \ciimu\ line with an average shift of $\langle \Delta V_{\rm sys} \rangle=(69\pm36)$\,km\,s$^{-1}$.
\item[(ii)] The luminosities of the halos appear to be independent of the amount of star formation in the host galaxy and of the UV luminosity of the central AGN.
\item[(iii)] The velocity dispersion of the gas in the halos is consistent with gravitational motion in dark matter halos of $M_{\rm DM}\lesssim10^{13}$\,M$_\odot$ at $z\sim6$.
\item[(iv)] For most of our objects, we do not find clear evidence of rotation or ordered motion.
However, the extended emission around the quasar P231$+$20 shows indications of a rotational pattern.
\item[(v)] The surface brightness of the detected halos is consistent with the emission expected from optically thin clouds illuminated by the quasars.
However, this requires high volume densities of the order $n_{\rm H}\gtrsim1$\,cm$^{-3}$.
\item[(vi)] The average surface brightness profile of the halos is well fit by an exponential curve.
After correcting for redshift dimming and scaling distances by the virial radius of a halo of $10^{12.5}$\,M$_{\odot}$, we observe no strong evolution of the profile between $z\sim6$ and $z\sim3$, followed by a decline in surface brightness down to $z\sim2$.
\item[(vii)] The two quasars that have peculiarly small near zones ($R^{\rm corr}_{\rm p}\lesssim3$\,pMpc) do not show evidence for extended emission.
This is consistent with a scenario where these quasars that have been shining for less than $\sim10^5$\,yr.
\end{enumerate}

We can ask ourselves if the reservoirs of cool gas observed around the first quasars are sufficient to sustain the enormous star formation rates of the host galaxies (with depletion time of $t_{\rm dep}\sim10-100$\,Myr) and fast growth of the central supermassive black holes.
The little evolution in the characteristics of the extended \lya\ halos observed between $z\sim6$ and $z\sim3$ suggests that the emitting clouds retain similar properties within this redshift range.
In this idealized model and assuming a spherical distribution for the clouds, the total mass in cool gas can be calculated as:
\begin{equation}
M_{\rm cool}=\pi R f_c N_{\rm H} \frac{m_p}{X}.
\end{equation}
Given that the extended emissions occur on scales of $10-30$\,pkpc, we can roughly estimate $M_{\rm cool}\gtrsim10^9$\,M$_\odot$ around the first quasars.
In general, hydrodynamical cosmological simulation are necessary to follow the complex journey of this gas from the IGM down to the host galaxy.
However, given the observed kinematics, we can assume that the angular momentum has little impact in the accretion process and consider the free--fall time ($t_{\rm ff}=\sqrt{3\pi/32\,G\,\rho}\sim50$\,Myr) as the minimum time--scale over which gas inflows.
This implies that, potentially, the rate of gas supply (of the order of ${\dot M}_{\rm cool}\sim10-100$\,M$_\odot$\,yr$^{-1}$) is compatible with the SFR estimated for the quasar host galaxies.
Further investigations of the detected nebulae are necessary to fully capture the physical status of the emitting material, however our \REQUIEM\ survey suggest that the halos of the first quasars contain sufficient fuel to maintain the observed high--rate of gas consumption.

As the first IFS study aimed at mapping the \lya\ emission around a statistical sample of $z>6$ quasars, the \REQUIEM\ survey demonstrates that direct detection of the CGM of the first massive galaxies is possible in 1--10\,hours of \textit{VLT}/MUSE per target.
The detected nebulae are unique targets for future multi--band follow--up observations to characterize the distribution of the gas and to constrain its physical conditions.
In the near future, we will exploit the rich dataset provided by the \REQUIEM\ survey to study the clustering of galaxies around these quasars (Paper~II), to search for UV--bright counterparts of high--$z$ absorption selected galaxies (Paper~III), and to perform a detailed comparison between the dynamics of the host--galaxies and the properties of the extended emission (Paper~IV).\par

\acknowledgments

EPF, ABD, MN, and FW acknowledge support from the ERC Advanced Grant 740246 (Cosmic Gas).
EPF is grateful to S.\ Vegetti and G.\ Kauffmann fur useful discussions and comments on the manuscript and to Z.\ Cai for sharing information on $z\sim2$ \lya\ nebulae.
We thank the members of the ENIGMA group\footnote{\url{http://enigma.physics.ucsb.edu/}} at UCSB for helpful discussions.
EPF is thankful to V.\ Springel (and all collegues at the MPA) for the hospitality while writing this manuscript.
For access to the data and codes used in this work, please contact the authors or visit \url{https://emastro.github.io/requiem/index.html}.

\facilities{ESO--\textit{VLT}/MUSE \& \textit{LBT}/LUCI2+ARGOS}

\software{This research made use of \textsc{Astropy}, a community--developed core \textsc{Python} package for Astronomy \citep{Astropy2013, Astropy2018} and~of~\textsc{IRAF}\footnote{\textsc{IRAF} \citep{Tody1986, Tody1993}, is distributed by the National Optical Astronomy Observatories, which are operated by the Association of Universities for Research in Astronomy, Inc., under cooperative agreement with the National Science Foundation.}.}

\clearpage

\appendix

\section{Atlas of the quasars part of the \REQUIEM\ survey}\label{app:allspec}

In \autoref{fig:allspec} we show the RGB postage stamps of the quasar vicinity created by combining three 2000\,km\,s$^{-1}$ wide pseudo--narrow--band images: one located 16000\,km\,s$^{-1}$ blueward, one 5000\,km\,s$^{-1}$ redward, and one at the redshifted \lya\ wavelength.
The spectra of the quasars extracted over apertures with a radius two times larger than the seeing are also shown.

\begin{figure*}[b]
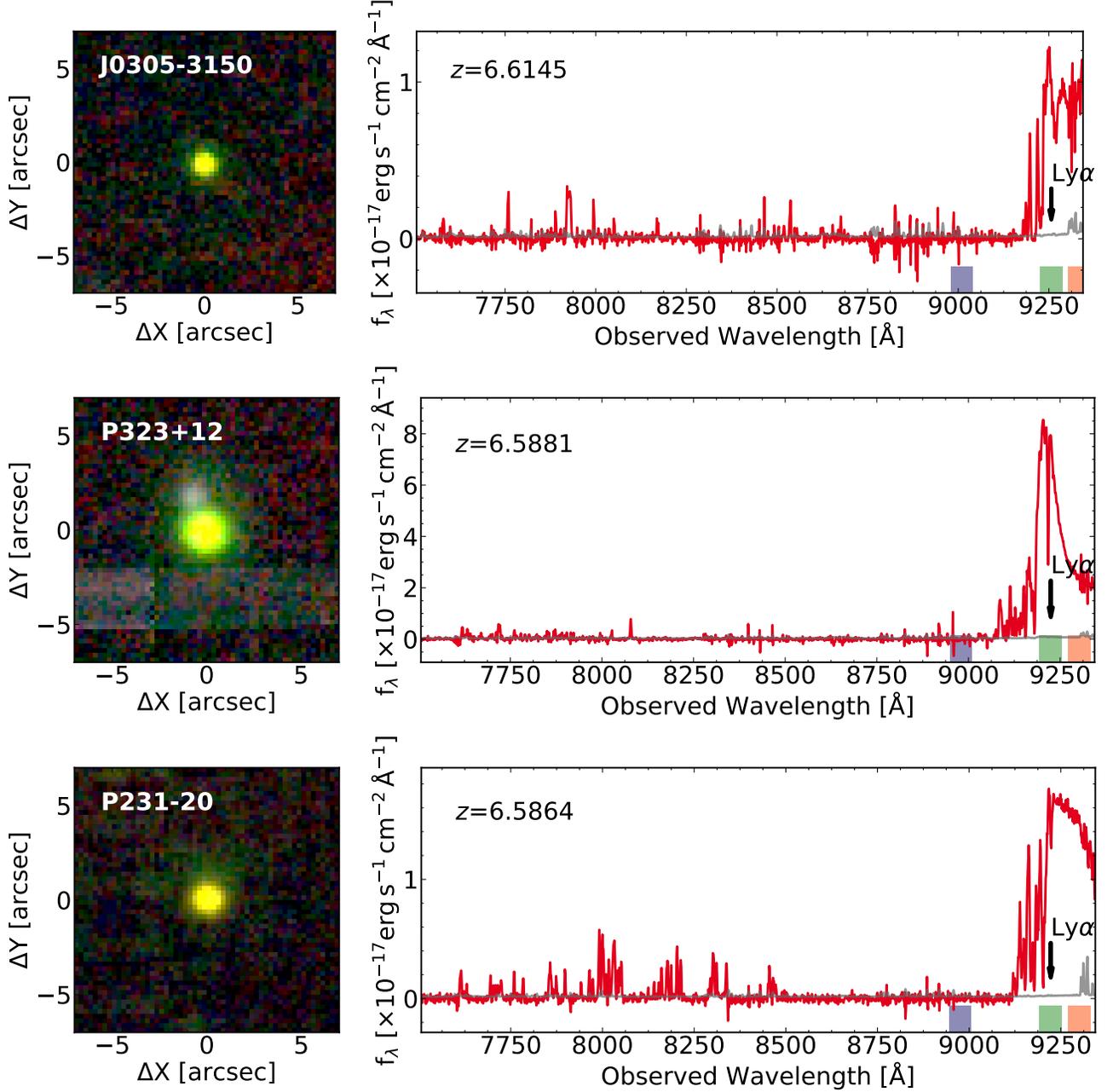

\begin{center}
\includegraphics[width=0.95\textwidth]{{{fullJ0305m3150_z6.6145}}}
\includegraphics[width=0.95\textwidth]{{{fullP323p12_z6.5881}}}
\includegraphics[width=0.95\textwidth]{{{fullP231m20_z6.5864}}}
\caption{
RGB images (left) and spectra (right) of the high--redshift quasars targeted in the \REQUIEM\ survey (ordered by decreasing redshift).
The wavelength ranges used to create the RGB images are highlighted with red, green, and blue boxes in the right panels.
The wavelength of the \lya\ line redshifted at $z_{\rm sys}$ is marked with a black arrow.
}\label{fig:allspec}
\end{center}
\end{figure*}

\addtocounter{figure}{-1}

\begin{figure*}[p]
\begin{center}
\includegraphics[width=0.95\textwidth]{{{fullP036p03_z6.5412}}}
\includegraphics[width=0.95\textwidth]{{{fullJ2318m3113_z6.4435}}}
\includegraphics[width=0.95\textwidth]{{{fullP183p05_z6.4386}}}
\includegraphics[width=0.95\textwidth]{{{fullJ0210m0456_z6.4323}}}
\caption{continued.}
\end{center}
\end{figure*}

\addtocounter{figure}{-1}

\begin{figure*}[p]
\begin{center}
\includegraphics[width=0.95\textwidth]{{{fullJ2329m0301_z6.4164}}}
\includegraphics[width=0.95\textwidth]{{{fullJ1152p0055_z6.3637}}}
\includegraphics[width=0.95\textwidth]{{{fullJ2211m3206_z6.3394}}}
\includegraphics[width=0.95\textwidth]{{{fullJ0142m3327_z6.3379}}}
\caption{continued.}
\end{center}
\end{figure*}

\addtocounter{figure}{-1}

\begin{figure*}[p]
\begin{center}
\includegraphics[width=0.95\textwidth]{{{fullJ0100p2802_z6.3258}}}
\includegraphics[width=0.95\textwidth]{{{fullJ1030p0524_z6.3}}}
\includegraphics[width=0.95\textwidth]{{{fullP308m21_z6.2341}}}
\includegraphics[width=0.95\textwidth]{{{fullP065m26_z6.1877}}}
\caption{continued.}
\end{center}
\end{figure*}

\addtocounter{figure}{-1}

\begin{figure*}[p]
\begin{center}
\includegraphics[width=0.95\textwidth]{{{fullP359m06_z6.1722}}}
\includegraphics[width=0.95\textwidth]{{{fullJ2229p1457_z6.1517}}}
\includegraphics[width=0.95\textwidth]{{{fullP217m16_z6.1498}}}
\includegraphics[width=0.95\textwidth]{{{fullJ2219p0102_z6.1492}}}
\caption{continued.}
\end{center}
\end{figure*}

\addtocounter{figure}{-1}

\begin{figure*}[p]
\begin{center}
\includegraphics[width=0.95\textwidth]{{{fullJ2318m3029_z6.1458}}}
\includegraphics[width=0.95\textwidth]{{{fullJ1509m1749_z6.1225}}}
\includegraphics[width=0.95\textwidth]{{{fullJ2216m0016_z6.0962}}}
\includegraphics[width=0.95\textwidth]{{{fullJ2100m1715_z6.0812}}}
\caption{continued.}
\end{center}
\end{figure*}

\addtocounter{figure}{-1}

\begin{figure*}[tbp]
\begin{center}
\includegraphics[width=0.95\textwidth]{{{fullJ2054m0005_z6.0391}}}
\includegraphics[width=0.95\textwidth]{{{fullP009m10_z6.0039}}}
\includegraphics[width=0.95\textwidth]{{{fullP340m18_z6.01}}}
\includegraphics[width=0.95\textwidth]{{{fullJ0055p0146_z6.006}}}
\caption{continued.}
\end{center}
\end{figure*}

\addtocounter{figure}{-1}

\begin{figure*}[t]
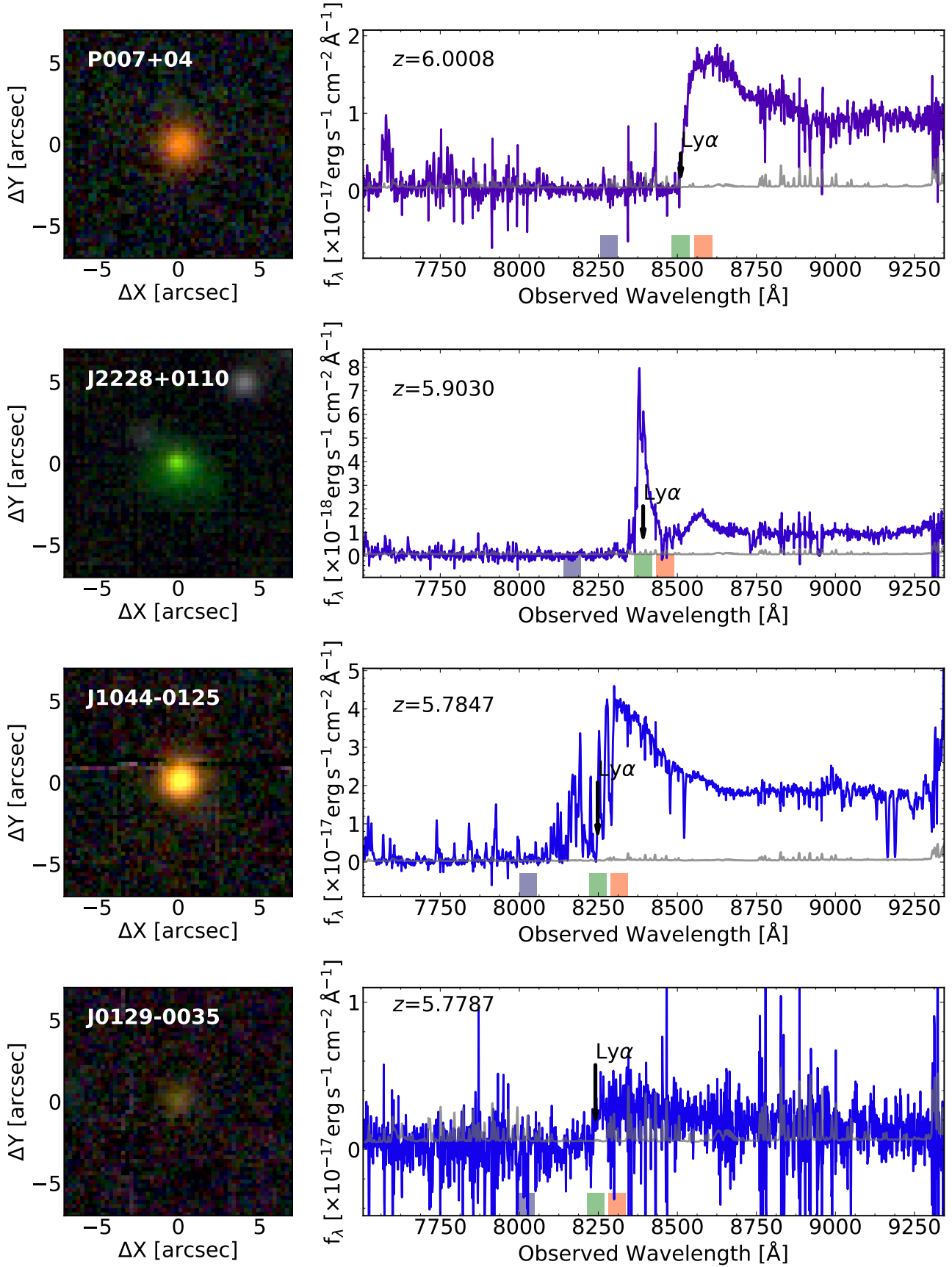

\begin{center}
\includegraphics[width=0.95\textwidth]{{{fullP007p04_z6.0008}}}
\includegraphics[width=0.95\textwidth]{{{fullJ2228p0110_z5.903}}}
\includegraphics[width=0.95\textwidth]{{{fullJ1044m0125_z5.7847}}}
\includegraphics[width=0.95\textwidth]{{{fullJ0129m0035_z5.7787}}}
\caption{continued.}
\end{center}
\end{figure*}

\section{The spectrum of P009$-$10}

The spectrum of P009$-$10 plotted in \autoref{fig:allspec} shows a deviation from the typical blue slope of quasars at $\lambda>8700$\,\AA.
This behavior appears to be independent from the spectrophotometric star used for flux calibration and from the frames used to correct for flat field.
We argue that this is most probably due to imperfect illumination correction due to the rapid variation of the sky conditions occurred during the observation of the target during the night of August 3rd, 2018.
This is supported by the strong variation on the background in the red side of different MUSE IFUs.
However, given that the \lya\ line is redshifted at $\lambda\sim8500$\,\AA, this has no impact on the current analysis.

\section{Analysis of the \textit{LBT}/LUCI2 + ARGOS images of P323$+$12}\label{sec:luci}

\begin{figure*}[tb]
\begin{center}
\includegraphics[width=0.99\textwidth]{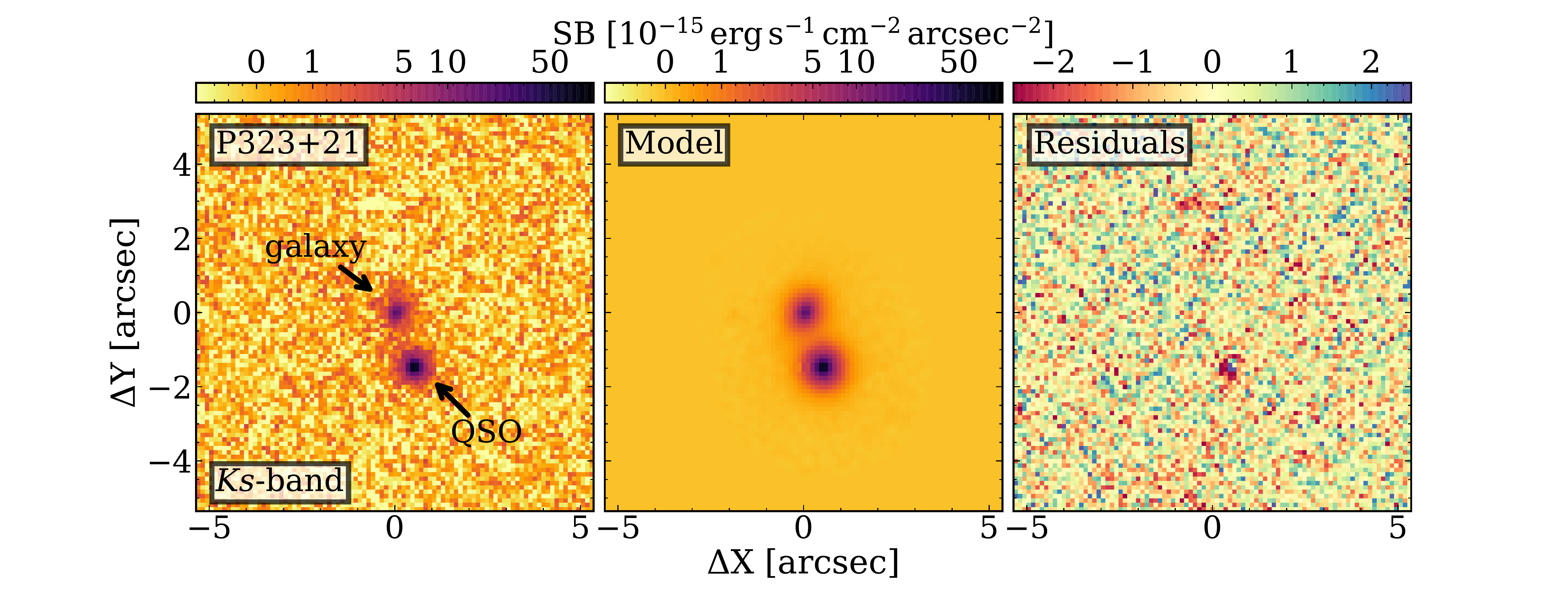}
\caption{
Results from the modelling of the quasar and close--by galaxy on the $Ks$--band images obtained with \textit{LBT}/LUCI\,2$+$ARGOS.
Different panels show, from left to right: zoom--in on the 10\arcsec$\times$10\arcsec region centered on the galaxy; model of the quasar and galaxy emission; residuals after model subtraction (see \autoref{sec:luci} for details).
In all panels, North is up and West is right.
}\label{fig:argos}
\end{center}
\end{figure*}

High--resolution $Ks$--band images of the quasar P323$+$12 have been collected with the Large Binocular Telescope \citep[\textit{LBT},][]{Hill2004, Hill2012} high with the Advanced Rayleigh guided Ground layer adaptive Optics System \citep[ARGOS;][]{Rabien2010, Rabien2019} coupled with LUCI\,2 \citep[i.e., \textit{LBT} Utility Camera in the Infrared;][]{Seifert2003, Ageorges2010}.
Data where collected on 25$^{\rm th}$~October~2017 during an ARGOS commissioning run.
The total time on targets was 660\,s, divided in 263 individual 2.51\,s exposures.
The data reduction has been performed with standard \textsc{IRAF} routines following the procedure described in \citet{Farina2018} and \citet{Georgiev2019}.
We registered the image to the WCS using the \textsc{Astrometry.net} software \citep[][]{Lang2010}.
The absolute flux calibration was achieved by matching sources with the 2MASS catalogue \citep[][]{Cutri2003} and considering a Vega to AB conversion in the $K_s$--band of ${\rm m}_{\rm AB}-{\rm m}_{\rm Vega}=1.85$\,mag.
Uncertainty in the zeropoint is of the order of 0.1\,mag.
During the observations, the DIMM seeing was 1\farcs34 in the optical.
The three green light (532\,nm) lasers focused at 12\,km used by ARGOS to correct for the ground layer turbulence, allowed us to enhance the $Ks$--band image quality to 0\farcs27 (FWHM of an unresolved source) of the entire LUCI\,2 field--of--view.
The 5--$\sigma$ detection limit for a point source (estimated from the \textsc{rms} of the sky counts integrated over the radius of an unresolved source) is $Ks_{\rm lim}=23.9$\,mag\footnote{The final image in \textsc{FITS} format is available at: \url{https://github.com/EmAstro/LBT_ARGOS}.}.

We exploit this data to look for the possible presence of multiple lensed images of the quasar generated by the presence of the $z=0.711$ elliptical galaxy located 1\farcs6 NNE from the quasar (see \autoref{sec:lens}).
First, we construct a spatially variable PSF model and we evaluated it at the quasar location \citep[for further details see][]{Farina2018, Georgiev2019}.
Then, we use this model to subtract both the emission from both the quasar and the close--by galaxy using the \textsc{GALFIT} v3.0.5 package \citep[][, see \autoref{fig:argos}]{Peng2010, Peng2011}.
The galaxy emission is well represented by a S{\'e}rsic profile \citep{Sersic1963} with magnitude $Ks_{\rm gal}=(20.26\pm0.16)$\,mag, effective radius $R_e=(1.1\pm0.3)$\,arcsec, and S{\'e}rsic index $n=(4.9\pm2.1)$.
The quasar is unresolved, with apparent magnitude $Ks_{\rm QSO}=(19.33\pm0.11)$\,mag (see \autoref{fig:argos}).
This implies that the host--galaxy is either compact (with radius $<1.5$\,pkpc) or its emission is below a surface brightness of $\mu_{Ks, {\rm host}}>22.7$\,mag\,arcsec$^{-1}$ 5--$\sigma$ limit over a 1\,arcsec$^{2}$ aperture).
These limits are slightly looser, but consistent with those obtained by \citet{Mechtley2012} on the host galaxy of the $z=6.42$ quasar J1148$+$5251.

In the residual image we do not detect any source in the close proximity of the galaxy (down to a 2--$\sigma$ surface brightness limit of $\mu_{Ks, {\rm lim}}>23.7$\,mag\,arcsec$^{-1}$ over an aperture of 1\,arcsec$^{2}$) that could be interpreted as multiple images of the quasar.
This supports our simple model presented in \autoref{sec:lens} where we showed that the Einstein ring is expected to be smaller than the separation between the quasar and the galaxy (i.e., $<1\farcs6$).

\section{The median surface--brightness profile}\label{sec:median}

In \autoref{fig:sbprofilemedian} we show the surface brightness profile computed by median combining the profiles extracted from the pseudo--narrow band images of quasars part of the \textit{core sample} of our \REQUIEM\ survey (see \autoref{sec:sb}).
Given the relatively small number of $z>5.5$ quasars observed with MUSE, we are not able to estimate the incidence of outliners \citep[such as the nabulae with sizes $\gg100$\,pkpc observed around $\sim1\%$ of intermediate redshift quasars, e.g.,][]{Hennawi2015, Arrigoni2019} in our sample.
We thus consider the average surface brightness profile presented in \autoref{sec:avesbprofile} to be a more befitting depiction of the diffuse \lya\ emission around $z\sim6$ quasars.
We notice, however, that our results do not depend on the type of profile chosen.
A fit with the exponential function $(1+z)^4\,{\rm SB}_{\rm Ly\alpha}\left(r\right)=C\exp{\left(-r/r_h\right)}$ shows that the median profile has a slightly fainter normalization [$C=(2.2\pm0.3)\times10^{-14}$\,erg\,s$^{-1}$\,cm$^{-2}$\,arcsec$^{-2}$] and a similar scale length [$r_h=(8.9\pm0.6)$\,kpc] with respect to the average profile (see \autoref{sec:avesbprofile} and \autoref{fig:sbprofilemedian}).

Measured average and median surface brightness radial profiles of the extended \lya\ emission around $z\sim6$ quasars are tabulated in \autoref{tab:sbprofile}. 

\begin{figure*}[t]
\begin{center}
\includegraphics[width=0.98\textwidth]{{{sbProfileSingleMedianSimple}}}
\caption{
\textit{Left Panel --} Median \lya\ surface brightness profiles around quasars at different redshifts.
Measurements are from the \textit{core sample} of our \REQUIEM\ survey (dark orange line), \citet[][, olive green line]{Marino2019},
\citet[][, light blue line]{Arrigoni2019}, and \citet[][, magenta line]{Cai2019}.
All data are corrected for cosmological dimming.
Shaded regions represents the 25th--75th percentiles in each quasar sample.
Exponential best fits are shown in the same color palette.
\textit{Right Panel --} Same of Left Panel, but in comoving units.
}\label{fig:sbprofilemedian}
\end{center}
\end{figure*}

\begin{deluxetable*}{cccccc}
\tablecaption{Average and median \lya\ surface brightness profile around quasars in the \textit{core sample} of the \REQUIEM\ survey \label{tab:sbprofile}}
\tablecolumns{6}
\tablewidth{0pt}
\tablehead{
\colhead{Radius}                                 &
\colhead{average \lya\ SB}                       &
\colhead{RMS}                                    &
\colhead{median \lya\ SB}                        &
\colhead{25$^{\rm th}$ percentile}               &
\colhead{75$^{\rm th}$ percentile}               \\
\colhead{(pkpc)}                                 &
\multicolumn{5}{c}{($10^{-16}$\,erg\,s$^{-1}$\,cm$^{-2}$\,arcsec$^{-2}$)}  
}
\startdata
\phm{0}4.2 &       259.2 &       232.7 &       198.7 &       101.3 &       345.7 \\
\phm{0}6.5 &       162.7 &       128.2 &       120.8 & \phm{0}77.9 &       175.5 \\
\phm{0}9.9 & \phm{0}88.1 & \phm{0}67.9 & \phm{0}60.5 & \phm{0}37.6 &       106.7 \\
      15.3 & \phm{0}57.7 & \phm{0}53.7 & \phm{0}39.2 & \phm{0}28.0 & \phm{0}65.5 \\
      23.4 & \phm{0}23.9 & \phm{0}22.8 & \phm{0}18.5 & \phm{00}8.8 & \phm{0}25.3 \\
      35.9 & \phm{00}7.6 & \phm{00}8.7 & \phm{00}4.0 & \phm{00}2.1 & \phm{00}9.6 \\
      55.0 & \phm{00}2.2 & \phm{00}1.4 & \phm{00}2.1 & \phm{00}1.6 & \phm{00}3.4 \\
\enddata
\tablecomments{All measurements are corrected for cosmological surface brightness dimming.
}
\end{deluxetable*}

\begin{widetext}

\section{List of known \lya\ nebulae associated to quasars}\label{sec:allhalos}

\autoref{tab:allhalos} lists sizes and luminosities of the extended \lya\ nebulae associated to quasars known as of the end of October 2019.
Data are homogenized to the cosmology used in this paper.
However, no attempt has been made to correct for the different sensitivities or for the diverse observing techniques employed in the listed studies.

\startlongtable
\begin{deluxetable*}{lccccl}
\tablecaption{\lya\ nebulae associated to quasars from the literature.\label{tab:allhalos}}
\tablecolumns{6}
\tablewidth{0pt}
\tablehead{
\colhead{ID}                                     &
\colhead{Type}                                   &
\colhead{$z$}                                    &
\colhead{d$_{{\rm Ly}\alpha}$}                   &
\colhead{L$_{{\rm Ly}\alpha}$}                   &
\colhead{Ref.}                                   \\
\colhead{}                                       &
\colhead{}                                       &
\colhead{}                                       &
\colhead{(pkpc)}                                 &
\colhead{($10^{43}$\,erg\,s$^{-1}$)}             &
\colhead{}                                       
}
\startdata
Q1658$+$575              & QSO--RL        & 1.979 & \phm{0}89\phm{\tablenotemark{o}}   & \phm{0}9.9\phm{\tablenotemark{o}} & \citet{Heckman1991b}                                                       \\
Q0017$+$154              & QSO--RL        & 2.012 & \phm{0}97\phm{\tablenotemark{o}}   & 12.9\phm{\tablenotemark{o}}       & \citet{Heckman1991b}                                                       \\
Q1354$+$258              & QSO--RL        & 2.032 & \phm{0}88\phm{\tablenotemark{o}}   & 21.1\phm{\tablenotemark{o}}       & \citet{Heckman1991b}                                                       \\
Q0225$-$014              & QSO--RL        & 2.037 & \phm{0}44\phm{\tablenotemark{o}}   & \phm{0}5.3\phm{\tablenotemark{o}} & \citet{Heckman1991b}                                                       \\
Q1345$+$258              & QSO--RL        & 2.039 & \phm{0}79\phm{\tablenotemark{o}}   & 26.8\phm{\tablenotemark{o}}       & \citet{Heckman1991b}                                                       \\
Jackpot                  & 4$\times$QSO   & 2.040 &       310\phm{\tablenotemark{o}}   & 21.0\phm{\tablenotemark{o}}       & \citet{Hennawi2015}                                                        \\
Q0445$+$097              & QSO--RL        & 2.113 &       106\phm{\tablenotemark{o}}   & \phm{0}7.3\phm{\tablenotemark{o}} & \citet{Heckman1991a, Heckman1991b}                                         \\
J0112$-$0048             & QSO            & 2.149 & \phm{0}42\phm{\tablenotemark{o}}   & \phm{0}3.0\phm{\tablenotemark{o}} & \citet{Fathivavsari2016}                                                   \\
Q0109$+$176              & QSO--RL        & 2.157 & \phm{0}26\phm{\tablenotemark{o}}   & \phm{0}1.5\phm{\tablenotemark{o}} & \citet{Heckman1991b}                                                       \\
Q1318$+$113              & QSO--RL        & 2.176 & \phm{0}96\phm{\tablenotemark{o}}   & 25.0\phm{\tablenotemark{o}}       & \citet{Heckman1991a, Heckman1991b}                                         \\
J1154$-$0215             & QSO            & 2.181 & \phm{0}50\phm{\tablenotemark{o}}   & \phm{0}2.4\phm{\tablenotemark{o}} & \citet{Fathivavsari2016}                                                   \\
Q2125$+$0112             & QSO            & 2.203 & \phm{0}19\phm{\tablenotemark{o}}   & \phm{0}1.3\phm{\tablenotemark{o}} & \citet{Cai2019}                                                            \\
Q0050$+$0051             & QSO            & 2.222 &       116\phm{\tablenotemark{o}}   & \phm{0}2.0\phm{\tablenotemark{o}} & \citet{Cai2019}                                                            \\
Q0814$+$3250             & QSO            & 2.222 & \phm{0}85\phm{\tablenotemark{o}}   & \phm{0}0.2\phm{\tablenotemark{o}} & \citet{Cai2019}                                                            \\
Q1228$+$3128             & QSO            & 2.231 &       124\phm{\tablenotemark{o}}   & 12.3\phm{\tablenotemark{o}}       & \citet{Cai2019}                                                            \\
J2233$-$606              & QSO            & 2.238 &       105\phm{\tablenotemark{o}}   & 13.6\phm{\tablenotemark{o}}       & \citet{Bergeron1999}                                                       \\
Q1444$+$3904             & QSO            & 2.250 &       101\phm{\tablenotemark{o}}   & 10.1\phm{\tablenotemark{o}}       & \citet{Cai2019}                                                            \\
Q1426$+$2555             & QSO            & 2.256 & \phm{0}96\phm{\tablenotemark{o}}   & \phm{0}3.9\phm{\tablenotemark{o}} & \citet{Cai2019}                                                            \\
Q2127$+$0049             & QSO            & 2.261 & \phm{0}58\phm{\tablenotemark{o}}   & \phm{0}1.1\phm{\tablenotemark{o}} & \citet{Cai2019}                                                            \\
Q0107$+$0314             & QSO            & 2.262 &       114\phm{\tablenotemark{o}}   & \phm{0}1.5\phm{\tablenotemark{o}} & \citet{Cai2019}                                                            \\
Q1227$+$2848             & QSO            & 2.268 &       164\phm{\tablenotemark{o}}   & \phm{0}5.8\phm{\tablenotemark{o}} & \citet{Cai2019}                                                            \\
Q2123$-$0050             & QSO            & 2.271 &       154\phm{\tablenotemark{o}}   & \phm{0}3.6\phm{\tablenotemark{o}} & \citet{Cai2019}                                                            \\
Slug                     & 2$\times$QSO   & 2.279 &       460\phm{\tablenotemark{o}}   & 22.0\phm{\tablenotemark{o}}       & \hspace{-0.765cm}\makecell[l]{\citet{Cantalupo2014}; \\ \citet{Martin2015, Martin2019}}     \\
Q0052$+$0140             & QSO            & 2.300 &       127\phm{\tablenotemark{o}}   & \phm{0}2.0\phm{\tablenotemark{o}} & \citet{Cai2019}                                                            \\
Q1416$+$2649             & QSO            & 2.301 &       141\phm{\tablenotemark{o}}   & \phm{0}5.0\phm{\tablenotemark{o}} & \citet{Cai2019}                                                            \\
J1058$+$0315             & QSO            & 2.302 & \phm{0}34\phm{\tablenotemark{o}}   & \phm{0}1.6\phm{\tablenotemark{o}} & \citet{Fathivavsari2016}                                                   \\
Q0848$-$0114             & QSO            & 2.302 & \phm{0}28\phm{\tablenotemark{o}}   & \phm{0}0.1\phm{\tablenotemark{o}} & \citet{Cai2019}                                                            \\
MAMMOTH-1                & QSO            & 2.311 &       442\phm{\tablenotemark{o}}   & 51.0\phm{\tablenotemark{o}}       & \citet{Cai2017}                                                            \\
Q1230$+$3320             & QSO            & 2.313 &       204\phm{\tablenotemark{o}}   & 12.4\phm{\tablenotemark{o}}       & \citet{Cai2019}                                                            \\
Q2150$+$053              & QSO--RL        & 2.323 & \phm{0}87\phm{\tablenotemark{o}}   & \phm{0}7.4\phm{\tablenotemark{o}} & \citet{Heckman1991b}                                                       \\
Q0048$+$0056             & QSO            & 2.327 &       104\phm{\tablenotemark{o}}   & \phm{0}2.3\phm{\tablenotemark{o}} & \citet{Cai2019}                                                            \\
Q2222$+$051              & QSO--RL        & 2.328 &       104\phm{\tablenotemark{o}}   & \phm{0}7.4\phm{\tablenotemark{o}} & \citet{Heckman1991b}                                                       \\
NDFWS~J143725$+$351048   & QSO            & 2.332 & \phm{0}80\tablenotemark{a}         & \phm{0}5.3\phm{\tablenotemark{o}} & \citet{Yang2009}                                                           \\
Q2121$+$0052             & QSO            & 2.377 &       141\phm{\tablenotemark{o}}   & \phm{0}4.6\phm{\tablenotemark{o}} & \citet{Cai2019}                                                            \\
ELAN0101$+$020           & 2$\times$QSO   & 2.450 &       232\phm{\tablenotemark{o}}   & 45.0\phm{\tablenotemark{o}}       & \citet{Cai2018}                                                            \\
J0049$+$3510             & QSO            & 2.480 & \phm{0}85\phm{\tablenotemark{o}}   & 35.4\phm{\tablenotemark{o}}       & \citet{Barrio2008}                                                         \\
TXS~1436$+$157           & QSO--RL        & 2.537 & \phm{0}92\phm{\tablenotemark{o}}   & \phm{0}4.6\phm{\tablenotemark{o}} & \hspace{-0.765cm}\makecell[l]{\citet{Roettgering1997}; \\ \citet{vanOjik1997}; \\ \citet{Humphrey2013}}\\
Q2206$-$199              & QSO            & 2.577 & \phm{0}80\phm{\tablenotemark{o}}   & \nodata\tablenotemark{b}          & \citet{Moller2000}                                                         \\
J0953$+$0349             & QSO            & 2.594 & \phm{0}29\phm{\tablenotemark{o}}   & \phm{0}1.3\phm{\tablenotemark{o}} & \citet{Fathivavsari2016}                                                   \\
Q2338$+$042              & QSO--RL        & 2.594 & \phm{0}94\phm{\tablenotemark{o}}   & 15.4\phm{\tablenotemark{o}}       & \hspace{-0.765cm}\makecell[l]{\citet{Heckman1991b}; \\ \citet{Lehnert1998}}                 \\
Q0758$+$097              & QSO--RL        & 2.683 &       110\phm{\tablenotemark{o}}   & 10.5\phm{\tablenotemark{o}}       & \citet{Heckman1991a, Heckman1991b}                                         \\
Q0730$+$257              & QSO--RL        & 2.686 & \phm{0}93\phm{\tablenotemark{o}}   & \phm{0}5.3\phm{\tablenotemark{o}} & \citet{Heckman1991b}                                                       \\
AMS05                    & QSO            & 2.850 & \phm{0}66\phm{\tablenotemark{o}}   & \phm{0}2.7\phm{\tablenotemark{o}} & \citet{Smith2009}                                                          \\
Q0805$+$046              & QSO--RL        & 2.877 &       116\phm{\tablenotemark{o}}   & 62.8\phm{\tablenotemark{o}}       & \citet{Heckman1991a, Heckman1991b}                                         \\
Q0941$+$261              & QSO--RL        & 2.913 & \phm{0}99\phm{\tablenotemark{o}}   & \phm{0}6.4\phm{\tablenotemark{o}} & \citet{Heckman1991a, Heckman1991b}                                         \\
J1253$+$1007             & QSO            & 3.015 & \phm{0}49\phm{\tablenotemark{o}}   & \phm{0}8.5\phm{\tablenotemark{o}} & \citet{Fathivavsari2016}                                                   \\
CTS~A31.05               & QSO            & 3.020 &       120\phm{\tablenotemark{o}}   & \phm{0}6.1\phm{\tablenotemark{o}} & \citet{Borisova2016}                                                       \\
J1135$-$0221             & 2$\times$QSO   & 3.020 & \phm{0}60\phm{\tablenotemark{o}}   & 32.0\phm{\tablenotemark{o}}       & \citet{Arrigoni2019Filament}                                               \\
UM669                    & QSO            & 3.021 &       160\phm{\tablenotemark{o}}   & 10.0\phm{\tablenotemark{o}}       & \citet{Borisova2016}                                                       \\
J0952$+$0114             & QSO            & 3.020 & \phm{0}58\tablenotemark{a}         & 58.3\phm{\tablenotemark{o}}       & \citet{Marino2019}                                                         \\
Q0041$-$2638             & QSO            & 3.036 &       170\phm{\tablenotemark{o}}   & \phm{0}2.9\phm{\tablenotemark{o}} & \citet{Borisova2016}                                                       \\
SDSSJ0219$-$0215         & QSO            & 3.036 & \phm{0}87\tablenotemark{a}         & \phm{0}3.5\phm{\tablenotemark{o}} & \citet{Arrigoni2019}                                                       \\
Q1205$-$30               & QSO            & 3.047 & \phm{0}81\tablenotemark{a}         & \phm{0}5.8\phm{\tablenotemark{o}} & \citet{Arrigoni2019}                                                       \\
Q1759$+$7539             & QSO            & 3.049 & \phm{0}65\phm{\tablenotemark{o}}   & \phm{0}9.0\phm{\tablenotemark{o}} & \citet{Christensen2006}                                                    \\
Q1205$-$30               & QSO            & 3.040 & \phm{0}40\phm{\tablenotemark{o}}   & \phm{0}6.3\phm{\tablenotemark{o}} & \hspace{-0.765cm}\makecell[l]{\citet{Weidinger2004, Weidinger2005}; \\  \citet{Fynbo2000}}  \\
HE0940$-$1050            & QSO            & 3.050 &       170\phm{\tablenotemark{o}}   & 14.0\phm{\tablenotemark{o}}       & \citet{Borisova2016}                                                       \\
SDSSJ1342$+$1702         & QSO            & 3.053 &       100\tablenotemark{a}         & \phm{0}2.4\phm{\tablenotemark{o}} & \citet{Arrigoni2019}                                                       \\
SDSSJ0947$+$1421         & QSO            & 3.073 & \phm{0}80\tablenotemark{a}         & \phm{0}2.0\phm{\tablenotemark{o}} & \citet{Arrigoni2019}                                                       \\
LBQS1209$+$1524          & QSO            & 3.075 &       108\tablenotemark{a}         & \phm{0}3.2\phm{\tablenotemark{o}} & \citet{Arrigoni2019}                                                       \\
AWL~11                   & QSO            & 3.079 &       130\phm{\tablenotemark{o}}   & \phm{0}4.9\phm{\tablenotemark{o}} & \citet{Borisova2016}                                                       \\
SDSSJ0100$+$2105         & QSO            & 3.097 & \phm{0}67\tablenotemark{a}         & \phm{0}5.0\phm{\tablenotemark{o}} & \citet{Arrigoni2019}                                                       \\
TEX1033$+$137            & QSO--RL        & 3.097 &       122\tablenotemark{a}         & 12.7\phm{\tablenotemark{o}}       & \citet{Arrigoni2019}                                                       \\
Q-N1097.1                & QSO            & 3.099 & \phm{0}87\tablenotemark{a}         & \phm{0}3.0\phm{\tablenotemark{o}} & \citet{Arrigoni2019}                                                       \\
Q0058$-$292              & QSO            & 3.101 &       109\tablenotemark{a}         & \phm{0}3.8\phm{\tablenotemark{o}} & \citet{Arrigoni2019}                                                       \\
S3~1013$+$20             & QSO--RL        & 3.108 &       110\tablenotemark{a}         & \phm{0}5.6\phm{\tablenotemark{o}} & \citet{Arrigoni2019}                                                       \\
SDSSJ1240$+$1455         & QSO            & 3.113 & \phm{0}40\phm{\tablenotemark{o}}   & \phm{0}4.2\phm{\tablenotemark{o}} & \citet{Matsuda2011}                                                        \\
CTS~A11.09               & QSO            & 3.121 &       150\phm{\tablenotemark{o}}   & \phm{0}2.1\phm{\tablenotemark{o}} & \citet{Borisova2016}                                                       \\
J0525$-$233              & QSO--RL        & 3.123 & \phm{0}77\tablenotemark{a}         & \phm{0}1.2\phm{\tablenotemark{o}} & \citet{Arrigoni2019}                                                       \\
CT~656                   & QSO            & 3.125 &       130\phm{\tablenotemark{o}}   & \phm{0}2.8\phm{\tablenotemark{o}} & \citet{Borisova2016}                                                       \\
SDSSJ1209$+$1138         & QSO            & 3.126 & \phm{0}83\tablenotemark{a}         & \phm{0}2.3\phm{\tablenotemark{o}} & \citet{Arrigoni2019}                                                       \\
CTSH22.05                & QSO            & 3.127 &       123\tablenotemark{a}         & \phm{0}7.4\phm{\tablenotemark{o}} & \citet{Arrigoni2019}                                                       \\
UM672                    & QSO--RL        & 3.127 & \phm{0}93\tablenotemark{a}         & \phm{0}6.5\phm{\tablenotemark{o}} & \citet{Arrigoni2019}                                                       \\
CTS~B27.07               & QSO            & 3.132 &       160\phm{\tablenotemark{o}}   & 10.0\phm{\tablenotemark{o}}       & \citet{Borisova2016}                                                       \\
Q0140$-$306              & QSO--RL        & 3.132 &       113\tablenotemark{a}         & \phm{0}5.1\phm{\tablenotemark{o}} & \citet{Arrigoni2019}                                                       \\
UM683                    & QSO            & 3.132 &       142\tablenotemark{a}         & \phm{0}7.9\phm{\tablenotemark{o}} & \citet{Arrigoni2019}                                                       \\
SDSSJ2100$-$0641         & QSO            & 3.136 & \phm{0}68\tablenotemark{a}         & \phm{0}2.1\phm{\tablenotemark{o}} & \citet{Arrigoni2019}                                                       \\
SDSSJ0814$+$1950         & QSO--RL        & 3.137 & \phm{0}50\tablenotemark{a}         & \phm{0}2.6\phm{\tablenotemark{o}} & \citet{Arrigoni2019}                                                       \\
SDSSJ0827$+$0300         & QSO--RL        & 3.137 & \phm{0}57\tablenotemark{a}         & \phm{0}0.8\phm{\tablenotemark{o}} & \citet{Arrigoni2019}                                                       \\
PKS0537$-$286            & QSO--RL        & 3.141 &       112\tablenotemark{a}         & \phm{0}4.6\phm{\tablenotemark{o}} & \citet{Arrigoni2019}                                                       \\
Q2138$-$4427             & QSO            & 3.142 & \phm{0}82\tablenotemark{a}         & \phm{0}4.7\phm{\tablenotemark{o}} & \citet{Arrigoni2019}                                                       \\
SDSSJ1550$+$0537         & QSO            & 3.147 & \phm{0}99\tablenotemark{a}         & \phm{0}5.6\phm{\tablenotemark{o}} & \citet{Arrigoni2019}                                                       \\
LBQS1244$+$1129          & QSO            & 3.157 &       101\tablenotemark{a}         & \phm{0}5.2\phm{\tablenotemark{o}} & \citet{Arrigoni2019}                                                       \\
6dF~J0032$-$0414         & QSO--RL        & 3.162 &       149\tablenotemark{a}         & 35.7\phm{\tablenotemark{o}}       & \citet{Arrigoni2019}                                                       \\
UM24                     & QSO            & 3.163 &       107\tablenotemark{a}         & \phm{0}2.6\phm{\tablenotemark{o}} & \citet{Arrigoni2019}                                                       \\
SDSSJ0905$+$0410         & QSO--RL        & 3.165 & \phm{0}98\tablenotemark{a}         & \phm{0}2.8\phm{\tablenotemark{o}} & \citet{Arrigoni2019}                                                       \\
PKS1017$+$109            & QSO            & 3.167 &       185\tablenotemark{a}         & 24.9\phm{\tablenotemark{o}}       & \citet{Arrigoni2019}                                                       \\
SDSSJ1020$+$1040         & 3$\times$QSO   & 3.167 &       297\phm{\tablenotemark{o}}   & 32.2\phm{\tablenotemark{o}}       & \citet{Arrigoni2018}                                                       \\
SDSSJ2319$-$1040         & QSO            & 3.172 & \phm{0}86\tablenotemark{a}         & \phm{0}3.2\phm{\tablenotemark{o}} & \citet{Arrigoni2019}                                                       \\
SDSSJ1243$+$0720         & QSO--RL        & 3.178 & \phm{0}89\tablenotemark{a}         & \phm{0}5.1\phm{\tablenotemark{o}} & \citet{Arrigoni2019}                                                       \\
Q2204$-$408              & QSO            & 3.179 & \phm{0}30\tablenotemark{a}         & \phm{0}1.1\phm{\tablenotemark{o}} & \citet{Arrigoni2019}                                                       \\
J0823$+$0529             & QSO            & 3.188 & \phm{0}45\phm{\tablenotemark{o}}   & 16.8\phm{\tablenotemark{o}}       & \citet{Fathivavsari2016}                                                   \\
UM678                    & QSO            & 3.188 &       150\phm{\tablenotemark{o}}   & \phm{0}7.8\phm{\tablenotemark{o}} & \citet{Borisova2016}                                                       \\
SDSSJ2348$-$1041         & QSO            & 3.190 & \phm{0}92\tablenotemark{a}         & \phm{0}2.3\phm{\tablenotemark{o}} & \citet{Arrigoni2019}                                                       \\
SDSSJ1032$+$1206         & QSO--RL        & 3.195 & \phm{0}58\tablenotemark{a}         & \phm{0}1.5\phm{\tablenotemark{o}} & \citet{Arrigoni2019}                                                       \\
Q0052$-$3901A            & QSO--RL        & 3.203 &       120\tablenotemark{a}         & \phm{0}9.8\phm{\tablenotemark{o}} & \citet{Arrigoni2019}                                                       \\
UM670                    & QSO            & 3.203 & \phm{0}92\tablenotemark{a}         & \phm{0}2.8\phm{\tablenotemark{o}} & \citet{Arrigoni2019}                                                       \\
Q1425$+$606              & QSO            & 3.204 & \phm{0}37\phm{\tablenotemark{o}}   & 10.1\phm{\tablenotemark{o}}       & \citet{Christensen2006}                                                    \\
SDSSJ0819$+$0823         & QSO            & 3.205 &       158\tablenotemark{a}         & 38.7\phm{\tablenotemark{o}}       & \citet{Arrigoni2019}                                                       \\
CTS~G18.01               & QSO            & 3.207 &       240\phm{\tablenotemark{o}}   & 17.0\phm{\tablenotemark{o}}       & \citet{Borisova2016}                                                       \\
PKS1614$+$051            & QSO            & 3.215 & \phm{0}50\phm{\tablenotemark{o}}   & \phm{0}2.0\phm{\tablenotemark{o}} & \citet{Husband2015}                                                        \\
UM679                    & QSO            & 3.215 & \phm{0}94\tablenotemark{a}         & \phm{0}4.5\phm{\tablenotemark{o}} & \citet{Arrigoni2019}                                                       \\
PKS1614$+$051            & QSO--RL        & 3.217 & \phm{0}66\phm{\tablenotemark{o}}   & \phm{0}7.3\phm{\tablenotemark{o}} & \citet{Hu1987}                                                             \\
CT-669                   & QSO            & 3.218 & \phm{0}97\tablenotemark{a}         & 11.0\phm{\tablenotemark{o}}       & \citet{Arrigoni2019}                                                       \\
PKS1614$+$051            & QSO            & 3.210 & \phm{0}40\phm{\tablenotemark{o}}   & 10.3\phm{\tablenotemark{o}}       & \citet{Matsuda2011}                                                        \\
Q0115$-$30               & QSO            & 3.221 & \phm{0}46\tablenotemark{a}         & \phm{0}0.3\phm{\tablenotemark{o}} & \citet{Arrigoni2019}                                                       \\
Q2139$-$4434             & QSO            & 3.221 &       140\tablenotemark{a}         & \phm{0}6.8\phm{\tablenotemark{o}} & \hspace{-0.765cm}\makecell[l]{\citet{Arrigoni2019}; \\ \citet{Lusso2019}}   \\
Q2139$-$4433             & QSO            & 3.229 &       100\tablenotemark{a}         & \phm{0}2.5\phm{\tablenotemark{o}} & \citet{Lusso2019}                                                          \\
SDSSJ1307$+$1230         & QSO            & 3.229 &       117\tablenotemark{a}         & \phm{0}7.0\phm{\tablenotemark{o}} & \citet{Arrigoni2019}                                                       \\
Q0347$-$383              & QSO            & 3.230 &       113\tablenotemark{a}         & \phm{0}4.1\phm{\tablenotemark{o}} & \citet{Arrigoni2019}                                                       \\
SDSSJ2321$+$1558         & QSO            & 3.241 & \phm{0}71\tablenotemark{a}         & \phm{0}0.9\phm{\tablenotemark{o}} & \citet{Arrigoni2019}                                                       \\
SDSSJ1025$+$0452         & QSO            & 3.243 &       144\tablenotemark{a}         & 18.7\phm{\tablenotemark{o}}       & \citet{Arrigoni2019}                                                       \\
CTS~C22.31               & QSO            & 3.246 &       101\tablenotemark{a}         & \phm{0}5.0\phm{\tablenotemark{o}} & \citet{Arrigoni2019}                                                       \\
Q0057$-$3948             & QSO            & 3.251 &       107\tablenotemark{a}         & \phm{0}5.1\phm{\tablenotemark{o}} & \citet{Arrigoni2019}                                                       \\
Q1451$+$122              & QSO            & 3.253 & \phm{0}16\phm{\tablenotemark{o}}   & \phm{0}1.9\phm{\tablenotemark{o}} & \citet{Christensen2006}                                                    \\
Q0042$-$2627             & QSO            & 3.280 &       320\phm{\tablenotemark{o}}   & 17.0\phm{\tablenotemark{o}}       & \citet{Borisova2016}                                                       \\
SDSSJ1557$+$1540         & QSO            & 3.288 &       107\tablenotemark{a}         & 21.8\phm{\tablenotemark{o}}       & \citet{Arrigoni2019}                                                       \\
Q2233$+$131              & QSO            & 3.301 & \phm{0}11\phm{\tablenotemark{o}}   & \phm{0}1.2\phm{\tablenotemark{o}} & \citet{Christensen2006}                                                    \\
Q0956$+$122              & QSO            & 3.309 & \phm{0}90\phm{\tablenotemark{o}}   & \phm{0}5.6\phm{\tablenotemark{o}} & \citet{Fumagalli2016}                                                      \\
Q0956$+$1217             & QSO            & 3.316 &       106\tablenotemark{a}         & \phm{0}5.3\phm{\tablenotemark{o}} & \citet{Arrigoni2019}                                                       \\
SDSSJ0125$-$1027         & QSO            & 3.319 & \phm{0}43\tablenotemark{a}         & \phm{0}2.3\phm{\tablenotemark{o}} & \citet{Arrigoni2019}                                                       \\
Q2348$-$4025             & QSO            & 3.334 &       103\tablenotemark{a}         & \phm{0}5.1\phm{\tablenotemark{o}} & \citet{Arrigoni2019}                                                       \\
SDSSJ0250$-$0757         & QSO            & 3.336 & \phm{0}86\tablenotemark{a}         & \phm{0}2.9\phm{\tablenotemark{o}} & \citet{Arrigoni2019}                                                       \\
SDSSJ0817$+$1053         & QSO            & 3.336 &       102\tablenotemark{a}         & \phm{0}5.0\phm{\tablenotemark{o}} & \citet{Arrigoni2019}                                                       \\
SDSSJ0154$-$0730         & QSO            & 3.337 & \phm{0}83\tablenotemark{a}         & \phm{0}3.3\phm{\tablenotemark{o}} & \citet{Arrigoni2019}                                                       \\
SDSSJ1337$+$0218         & QSO            & 3.344 & \phm{0}51\tablenotemark{a}         & \phm{0}0.5\phm{\tablenotemark{o}} & \citet{Arrigoni2019}                                                       \\
SDSSJ0001$-$0956         & QSO--RL        & 3.348 &       131\tablenotemark{a}         & 10.8\phm{\tablenotemark{o}}       & \citet{Arrigoni2019}                                                       \\
CTS~R07.04               & QSO            & 3.351 &       170\phm{\tablenotemark{o}}   & 33.0\phm{\tablenotemark{o}}       & \citet{Borisova2016}                                                       \\
SDSSJ1427$-$0029         & QSO            & 3.354 & \phm{0}49\tablenotemark{a}         & \phm{0}4.1\phm{\tablenotemark{o}} & \citet{Arrigoni2019}                                                       \\
Q0042$-$269              & QSO            & 3.357 & \phm{0}61\tablenotemark{a}         & \phm{0}1.0\phm{\tablenotemark{o}} & \citet{Arrigoni2019}                                                       \\
FBQSJ2334$-$0908         & QSO--RL        & 3.361 & \phm{0}52\tablenotemark{a}         & \phm{0}0.7\phm{\tablenotemark{o}} & \citet{Arrigoni2019}                                                       \\
Q2355p0108               & QSO            & 3.395 &       121\tablenotemark{a}         & \phm{0}5.4\phm{\tablenotemark{o}} & \citet{Arrigoni2019}                                                       \\
SDSSJ1019$+$0254         & QSO            & 3.395 & \phm{0}73\tablenotemark{a}         & \phm{0}3.1\phm{\tablenotemark{o}} & \citet{Arrigoni2019}                                                       \\
SDSSJ1429$-$0145         & QSO            & 3.425 & \phm{0}79\tablenotemark{a}         & \phm{0}3.6\phm{\tablenotemark{o}} & \citet{Arrigoni2019}                                                       \\
SDSSJ1057$-$0139         & QSO--RL        & 3.452 & \phm{0}65\tablenotemark{a}         & \phm{0}1.7\phm{\tablenotemark{o}} & \citet{Arrigoni2019}                                                       \\
0054$-$284               & QSO            & 3.616 & \phm{0}38\phm{\tablenotemark{o}}   & \phm{0}0.8\phm{\tablenotemark{o}} & \citet{Bremer1992}                                                         \\
Q0055$-$269              & QSO            & 3.634 &       180\phm{\tablenotemark{o}}   & 37.0\phm{\tablenotemark{o}}       & \citet{Borisova2016}                                                       \\
0055$-$264               & QSO            & 3.656 & \phm{0}30\phm{\tablenotemark{o}}   & 1.1 \phm{\tablenotemark{o}}       & \citet{Bremer1992}                                                         \\
Q1621$-$0042             & QSO            & 3.689 &       120\phm{\tablenotemark{o}}   & \phm{0}5.5\phm{\tablenotemark{o}} & \citet{Borisova2016}                                                       \\
Q1317$-$0507             & QSO            & 3.701 &       140\phm{\tablenotemark{o}}   & \phm{0}3.6\phm{\tablenotemark{o}} & \citet{Borisova2016}                                                       \\
QB2000$-$330             & QSO--RL        & 3.759 &       120\phm{\tablenotemark{o}}   & 12.0\phm{\tablenotemark{o}}       & \citet{Borisova2016}                                                       \\
PKS1937$-$101            & QSO--RL        & 3.769 &       110\phm{\tablenotemark{o}}   & 29.0\phm{\tablenotemark{o}}       & \citet{Borisova2016}                                                       \\
J0124$+$0044             & QSO            & 3.783 &       190\phm{\tablenotemark{o}}   & 41.0\phm{\tablenotemark{o}}       & \citet{Borisova2016}                                                       \\
BRI1108$-$07             & QSO            & 3.907 &       160\phm{\tablenotemark{o}}   & 12.0\phm{\tablenotemark{o}}       & \citet{Borisova2016}                                                       \\
Q0953$+$4749             & QSO            & 4.489 & \phm{0}14\phm{\tablenotemark{o}}   & \phm{0}0.8\phm{\tablenotemark{o}} & \hspace{-0.765cm}\makecell[l]{\citet{Christensen2006}; \\ \citet{Bunker2003}}               \\
BR1033$-$0327            & QSO            & 4.510 & \phm{0}70\phm{\tablenotemark{o}}   & \phm{0}2.4\phm{\tablenotemark{o}} & \hspace{-0.765cm}\makecell[l]{\citet{North2012}; \\ \citet{Courbin2008}}    \\
SDSSJ14472$+$0401        & QSO            & 4.510 & \phm{0}42\phm{\tablenotemark{o}}   & \phm{0}0.2\phm{\tablenotemark{o}} & \citet{North2012}                                                          \\
SDSSJ21474$-$0838        & QSO            & 4.510 & \phm{0}56\phm{\tablenotemark{o}}   & 23.2\phm{\tablenotemark{o}}       & \citet{North2012}                                                          \\
1605$-$0112              & QSO            & 4.920 & \phm{0}60\phm{\tablenotemark{o}}   & \phm{0}4.4\phm{\tablenotemark{o}} & \citet{Ginolfi2018}                                                        \\
J2228$+$0110             & QSO--RL        & 5.903 & \phm{0}30\phm{\tablenotemark{o}}   & \phm{0}7.8\phm{\tablenotemark{o}} & \hspace{-0.765cm}\makecell[l]{\citet{Roche2014}; \\ \citet{Drake2019}; \\ \citet{Farina2019}}          \\
P009$-$10                & QSO            & 6.004 & \phm{0}15\phm{\tablenotemark{o}}   & \phm{0}0.9\phm{\tablenotemark{o}} & \citet{Farina2019}                                                         \\
P340$-$18                & QSO            & 6.010 & \phm{0}18\phm{\tablenotemark{o}}   & \phm{0}7.5\phm{\tablenotemark{o}} & \citet{Farina2019}                                                         \\
P359$-$06                & QSO            & 6.172 & \phm{0}17\phm{\tablenotemark{o}}   & \phm{0}3.3\phm{\tablenotemark{o}} & \citet{Farina2019}                                                         \\
P065$-$26                & QSO            & 6.188 & \phm{0}25\phm{\tablenotemark{o}}   & \phm{0}6.6\phm{\tablenotemark{o}} & \citet{Farina2019}                                                         \\
P308$-$21                & QSO            & 6.234 & \phm{0}43\phm{\tablenotemark{o}}   & \phm{0}8.8\phm{\tablenotemark{o}} & \citet{Farina2019}                                                         \\
J1030$+$0524             & QSO            & 6.300 & \phm{0}34\phm{\tablenotemark{o}}   & \phm{0}2.5\phm{\tablenotemark{o}} & \hspace{-0.765cm}\makecell[l]{\citet{Decarli2012}; \\ \citet{Drake2019}; \\ \citet{Farina2019}}        \\
J2329$-$0301             & QSO            & 6.416 & \phm{0}22\phm{\tablenotemark{o}}   & \phm{0}5.1\phm{\tablenotemark{o}} & \citet{Farina2019}                                                         \\
J2329$-$0301             & QSO            & 6.416 & \phm{0}22\phm{\tablenotemark{o}}   & \phm{0}5.1\phm{\tablenotemark{o}} & \hspace{-0.765cm}\makecell[l]{\citet{Goto2009, Goto2012}; \\ \citet{Willott2011}; \\ \citet{Momose2018}; \\ \citet{Drake2019}; \\ \citet{Farina2019}} \\
P036$+$03                & QSO            & 6.541 & \phm{0}19\phm{\tablenotemark{o}}   & \phm{0}3.8\phm{\tablenotemark{o}} & \citet{Farina2019}                                                         \\
P231$-$20                & QSO            & 6.586 & \phm{0}28\phm{\tablenotemark{o}}   & 11.0\phm{\tablenotemark{o}}       & \hspace{-0.765cm}\makecell[l]{\citet{Drake2019}; \\ \citet{Farina2019}}     \\
P323$+$12                & QSO            & 6.588 & \phm{0}25\phm{\tablenotemark{o}}   & 20.1\phm{\tablenotemark{o}}       & \citet{Farina2019}                                                         \\
J0305$-$3150             & QSO            & 6.615 & \phm{0}17\phm{\tablenotemark{o}}   & \phm{0}0.8\phm{\tablenotemark{o}} & \citet{Farina2017, Farina2019}                                             \\
\enddata
\tablecomments{The `Type' indicates if the nebula is associated with a radio--loud quasar (QSO--RL), a radio--quiet quasar (QSO), or with a system of multiple quasars ($N\times$QSO).
The size is the maximum diameter distance of the \lya\ emission.
The table is published in its entirety in the machine-readable format.
A portion is shown here for guidance regarding its form and content.
\tablenotetext{a}{Value calculated assuming circular source, i.e., d$_{{\rm Ly}\alpha}$ is the diameter of a circle with area equal to the area of the source.}
\tablenotetext{b}{Published spectrum not flux--calibrated.}
}
\end{deluxetable*}

\end{widetext}

\end{document}